\documentclass[twocolumn,showpacs,nofootinbib]{revtex4-1}%
\usepackage{amsmath}
\usepackage{amssymb}
\usepackage{graphicx}
\usepackage{amsfonts}
\usepackage{txfonts}
\usepackage[colorlinks,linkcolor=blue,anchorcolor=blue,citecolor=blue]{hyperref}
\begin{document}
\title{Quantum many-body theory for electron spin decoherence in nanoscale nuclear
spin baths}
\author{Wen Yang}
\affiliation{Beijing Computational Science Research Center, Beijing 100193, China}
\author{Wen-Long Ma}
\affiliation{Department of Physics, The Chinese University of Hong Kong, Shatin, N. T.,
Hong Kong, China}
\affiliation{Centre for Quantum Coherence, The Chinese University of Hong Kong, Shatin, N.
T., Hong Kong, China}
\author{Ren-Bao Liu}
\thanks{Corresponding author. rbliu@phy.cuhk.edu.hk}
\affiliation{Department of Physics, The Chinese University of Hong Kong, Shatin, N. T.,
Hong Kong, China}
\affiliation{Centre for Quantum Coherence, The Chinese University of Hong Kong, Shatin, N.
T., Hong Kong, China}
\affiliation{Institute of Theoretical Physics, The Chinese University of Hong Kong, Shatin,
N. T., Hong Kong, China}

\begin{abstract}
Decoherence of electron spins in nanoscale systems is important to quantum
technologies such as quantum information processing and magnetometry. It is
also an ideal model problem for studying the crossover between quantum and
classical phenomena. At low temperatures or in light-element materials where
the spin-orbit coupling is weak, the phonon scattering in nanostructures is
less important and the fluctuations of nuclear spins become the dominant
decoherence mechanism for electron spins. Since 1950s, semiclassical noise
theories have been developed for understanding electron spin decoherence. In
spin-based solid-state quantum technologies, the relevant systems are in the
nanometer scale and the nuclear spin baths are quantum objects which require a
quantum description. Recently, quantum pictures have been established to
understand the decoherence and quantum many-body theories have been developed
to quantitatively describe this phenomenon. Anomalous quantum effects have
been predicted and some have been experimentally confirmed. A systematically
truncated cluster correlation expansion theory has been developed to account
for the many-body correlations in nanoscale nuclear spin baths that are built
up during the electron spin decoherence. The theory has successfully predicted
and explained a number of experimental results in a wide range of physical
systems. In this review, we will cover these recent progresses. The
limitations of the present quantum many-body theories and possible directions
for future development will also be discussed.

\end{abstract}
\maketitle
\tableofcontents

\section{Introduction}

A quantum object can be in a superposition of states. An isolated quantum
object can be in a pure state with full \textit{quantum coherence}, a state in
which each component of the superposition has a deterministic coefficient up
to a global phase factor. Quantum coherence gives rise to a series of
non-classical phenomena such as interference and entanglement. It is also the
basis of quantum technologies
\cite{BenioffJSP1980,FeynmanIJTP1982,DeutschPRSLA1985}, such as quantum
cryptography \cite{BennetPICC1984,GisinRMP2002}, quantum-enhanced imaging and
sensing \cite{CavesPRD1981,BudkerNatPhys2007,GiovannettiNatPhoton2011}, and
quantum computers \cite{DiVincenzoScience1995,LaddNature2010}.

Realistic quantum systems are always coupled to environments, thus the quantum
coherence is destroyed by the environmental noise
\cite{LeggettRMP1987,StampRPP2000,ZurekRMP2003}. On the one hand, such
decoherence processes prevent quantum interference, restore classical
behaviors, and pose a critical challenge to quantum technologies. On the other
hand, decoherence could be utilized to reveal information about the
environments. This prospect has been persued for a long history in magnetic
resonance spectroscopy \cite{AbragamBook1961}, where the decoherence of a
large number of electronic or nuclear spins are used to reveal the
interactions and motions of atoms in bulk materials. In recent years, the
progresses in active control and measurement of single spins have allowed
single spins to be used as ultrasensitive quantum sensors to reveal the
structures and dynamics of the environments with nanoscale resolution
\cite{ColeNanotechnol2009,ZhaoNatNano2011,ZhaoNatNano2012} (see Ref.
\cite{RondinRPP2014} for a review).

Additionally, a great diversity of physical systems have been proposed for
spin-based quantum technologies and quantum sensing. In particular, the spins
of individual electrons and atomic nuclei offer a promising combination of
environmental isolation and controllability, thus they can serve as the basic
units of quantum machines: the qubits. Electronic and nuclear spins in
semiconductors have distinct technical advantages such as scalability and
compatability with modern semiconductor technology \cite{MooreElectronics1965}%
, tunable spin properties by energy band and wavefunction engineering, and the
ability to manipulate the spins by using the well established electron spin
resonance and nuclear magnetic resonance techniques as well as optical and
electrical approaches \cite{HansonNature2008}. Here we concentrate on
semiconductor quantum dots (QDs) \cite{HansonRMP2007} and impurity/defect
centers such as phosphorus and bismuth donors in silicon
\cite{MortonNature2011} and nitrogen-vacancy (NV) centers in diamond
\cite{AwschalomSA2007}. In these nanoscale systems, a few electronic or
nuclear spins (referred to as \textit{central spins} for clarity) can be
addressed, so they are used as qubits, while the many unresolved nuclear spins
form a magnetic environment that causes decoherence of the central spins. In
addition, the central spins are directly coupled to nearby electronic
spins from impurities and defects and are also influenced by charge and voltage fluctuations (e.g.,
from lattice vibration and nearby electron/hole gases and trapped charges) via
spin-orbit coupling. However, these environmental noises can be suppressed, e.g., by
careful material and device engineering to remove parasitic charge and spin
defects, lowering the temperature to suppress phonon scattering, or using
light-element materials to suppress the spin-orbit coupling. Therefore, the
most relevant noise sources for the central spins in quantum technologies are
the nuclear spins.

\begin{figure}[ptb]
\includegraphics[width=0.9\columnwidth]{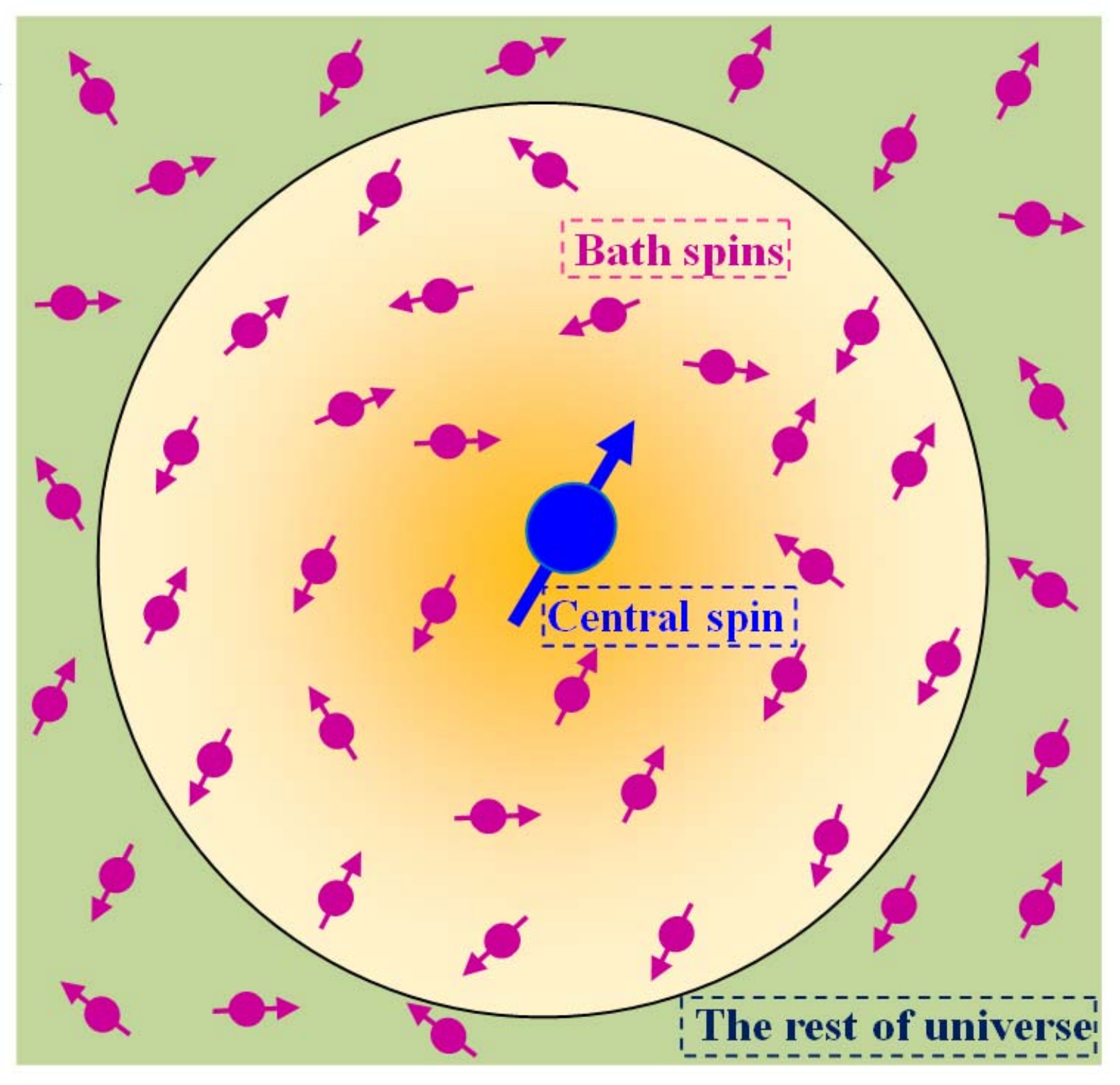}
\caption{A central spin
and the nanoscale spin bath evolve as a closed system in the relevant
timescales. The thermal distribution of the spin bath causes a static thermal
noise, and the quantum evolution of the spin bath induces a dynamical quantum
noise. The rest of universe indicates the larger environment beyond the spin
bath, which induces classical noises (being static or dynamical).}%
\label{G_SPINBATH}%
\end{figure}

Since 1950s, the semiclassical picture of \textit{spectral diffusion} has been
adopted to study the central spin decoherence in spin
baths\cite{AndersonRMP1953,AndersonJPSJ1954,KuboJPSJ1954}. The semiclassical
theory treats the spin bath as a source of classical magnetic noise. In modern
quantum nanodevices, the wave function of the central spin is localized, so
the nuclear spins coupled to the central spin form a nanoscale spin bath. The
central spin and the nanoscale spin bath form a closed system in the time
scale of interest (see Fig. \ref{G_SPINBATH}) and the quantum nature of the
spin bath becomes important. In recent years, quantum pictures have been
established to understand the central spin decoherence. By the quantum theory,
anomalous quantum effects have been predicted, some of which have been
experimentally confirmed. To quantitatively describe central spin decoherence,
a variety of quantum many-body theories have been developed, including the
pair-correlation approximation \cite{YaoPRB2006,LiuNJP2007,YaoPRL2007},
cluster expansion \cite{WitzelPRB2005,WitzelPRB2006}, linked-cluster expansion
\cite{SaikinPRB2007}, cluster-correlation expansion~(CCE)
\cite{YangPRB2008,YangPRB2009}, disjoint cluster approximation
\cite{MazePRB2008,HallPRB2014}, and ring diagram approximation
\cite{CywinskiPRL2009,CywinskiPRB2009}. In particular, the CCE theory \cite{YangPRB2008,YangPRB2009} provides a systematic account
for the many-body correlations in nanoscale spin baths that lead to central
spin decoherence. The CCE method has successfully predicted and explained a
number of experimental results in a wide range of solid state systems. In this
review, we will provide a pedagogical review on the basic concepts of
coherence and decoherence, the recent quantum many-body theories, their
relationships, limitations, and possible directions for future development.

The organization of this review is as follows. In the first three sections, we
introduce the basic concepts (Sec. \ref{SEC_CONCEPT}), decoherence theory
(Sec. \ref{SEC_SEMICLASSICAL}) and coherence protection (Sec.
\ref{SEC_DD_FILTER}) based on the semi-classical noise model. Then we
introduce, in Sec. \ref{SEC_QUANTUM_NOISE}, the concept of quantum noise and,
in Sec. \ref{SEC_QUANTUM_PICTURE}, a full quantum picture of central spin
decoherence. In Sec. \ref{SEC_SYSTEM}, we introduce the coupling of the
central spin to the phonon and nuclear spin baths and experimental
measurements in paradigmatic solid-state physical systems that identifies the
nuclear spin bath as the most relevant decohering environment. Next we review
the microscopic quantum many-body theories for central spin decoherence in
nuclear spin baths (Sec. \ref{SEC_MANYBODY_THEORY}) and discuss a series of
quantum decoherence effects (Sec. \ref{SEC_EFFECTS}). Finally, the possible directions for future development are discussed in Sec.
\ref{SEC_OUTLOOK}. For convenience, we take $\hbar=1$ throughout this review.

\section{Basic concepts of spin decoherence}

\label{SEC_CONCEPT}

In this section, we introduce the basic concepts for the environmental noise
induced decoherence of a central spin-1/2, including quantum coherence and
decoherence, density matrix and ensembles, classification of central spin
decoherence and their geometric representation with Bloch vectors, and
description of central spin decoherence caused by the simplest environmental
noises: rapidly fluctuating noise and static noise.

\begin{figure}[ptb]
\includegraphics[width=\columnwidth,clip]{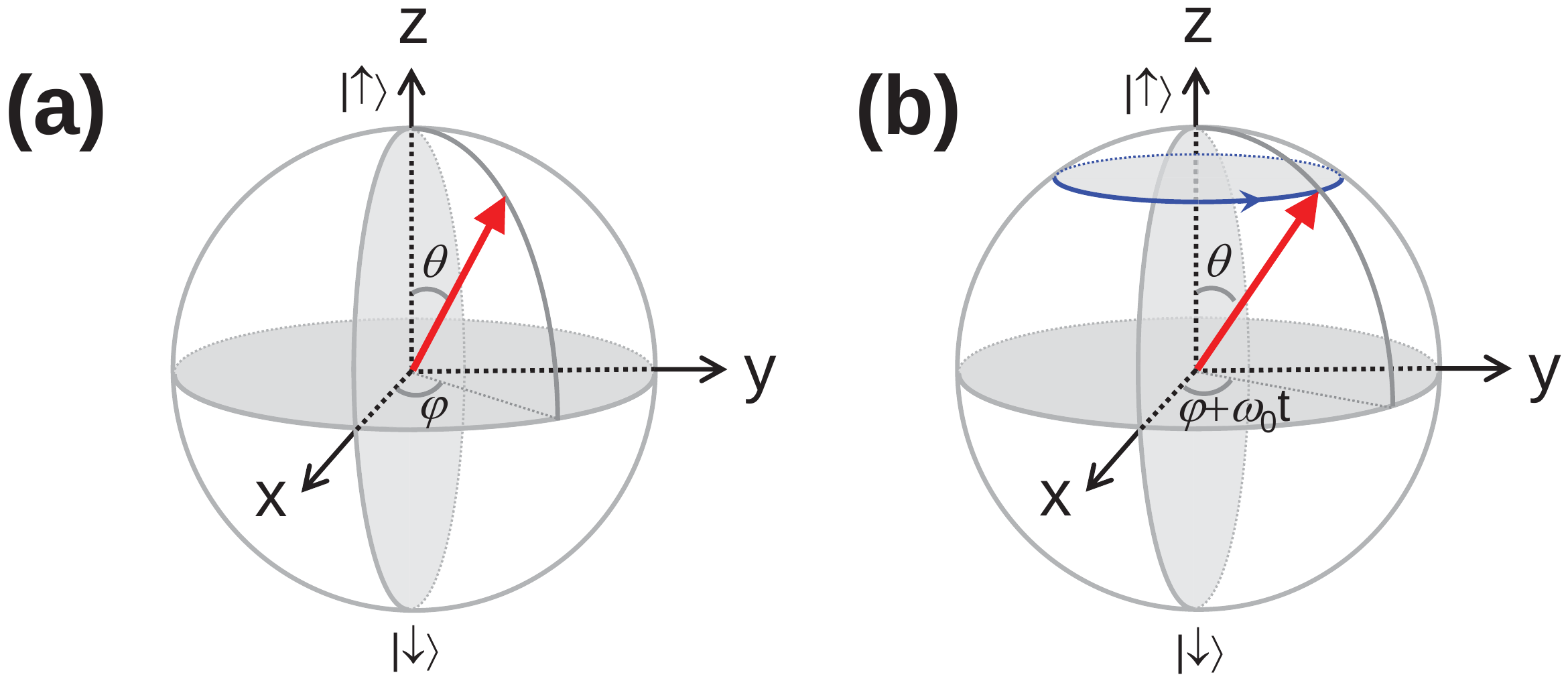}
 \caption{A central
spin quantized in a magnetic field along the $z$ axis:\ (a) geometric
representation of a general pure state as a Bloch vector, and (b) central spin
evolution as the precession of the Bloch vector around the magnetic field.}%
\label{G_BLOCHSPHERE}%
\end{figure}

Under an external magnetic field, the central spin is quantized along the
magnetic field (defined as the $z$ axis) and its evolution is governed by the
Zeeman Hamiltonian
\begin{equation}
\hat{H}_{0}=\omega_{0}\hat{S}_{z}, \label{H0}%
\end{equation}
with two energy eigenstates $|\uparrow\rangle$ (spin up) and $|\downarrow
\rangle$ (spin down). A general pure superposition state of a spin-1/2 can be
parametrized by two real numbers $\theta$ and $\varphi$ as
\begin{equation}
|\theta,\varphi\rangle\equiv\cos\frac{\theta}{2}|\uparrow\rangle+\sin
\frac{\theta}{2}e^{i\varphi}|\downarrow\rangle. \label{THETA_PHI}%
\end{equation}
Quantum coherence is fully preserved when the central spin is isolated from
the environment and undergoes unitary evolution according to its own,
deterministic Hamiltonian. For example, the Zeeman Hamiltonian in Eq. (\ref{H0})
leads to the coherent evolution $|\theta,\varphi\rangle\rightarrow
e^{-i\hat{H}_{0}t}|\theta,\varphi\rangle=|\theta,\varphi+\omega_{0}t\rangle$.
The couplings of the central spin to the environment amounts to measurement of
the central spin by the environment (with the results unknown to any observers
though). As a result, the central spin undergoes random collapses from a fully
coherent pure state into an incoherent mixture (i.e., a statistical ensemble)
of distinct pure states, i.e., quantum coherence breaking or decoherence in short.

\subsection{Temporal ensembles and spatial ensembles}

A quantum system in a pure state $|\psi\rangle$ is described by the density
operator $\hat{\rho}=|\psi\rangle\langle\psi|$, while a quantum system that is
found in the $k$th distinct pure state $|\psi_{k}\rangle$ with probability
$p_{k}$ ($k=1,2,\cdots$) is described by the density operator $\hat{\rho}%
=\sum_{k}p_{k}|\psi_{k}\rangle\langle\psi_{k}|$. In the energy eigenstates
$|\uparrow\rangle$ and $|\downarrow\rangle$ of the central spin, the density
operator becomes a 2$\times$2 \textit{density matrix} as
\[
\hat{\rho}=%
\begin{bmatrix}
\rho_{\uparrow\uparrow} & \rho_{\uparrow\downarrow}\\
\rho_{\downarrow\uparrow} & \rho_{\downarrow\downarrow}%
\end{bmatrix}
,
\]
where the diagonal matrix elements $\rho_{\uparrow\uparrow}$ and
$\rho_{\downarrow\downarrow}$ describe the population of each energy
eigenstate, and the off-diagonal elements $\rho_{\downarrow\uparrow}%
=\rho_{\uparrow\downarrow}^{\ast}$ describe the phase correlation between
different energy eigenstates.

The density matrix $\hat{\rho}(t)$ describes the statistics of many identical
measurements over an \textit{ensemble} of central spins. In recent years,
single-shot measurement of a single central spin has been demonstrated in
various solid-state systems
\cite{ElzermanNature2004,HansonPRL2005,BarthelPRL2009,MorelloNature2010,NeumannScience2010,VamivakasNature2010,RobledoNature2011,DelteilPRL2014,WaldherrNature2014,ShulmanNatCommun2014}%
. For such single-spin measurements, one still need to repeat the measurement
cycle (i.e., initialization-evolution-measurement) many times to retrieve the
correct probabilities of different measurement outcomes. In this case, each
cycle corresponds to a \textit{sample} of the \textit{temporal ensemble}.
According to the characteristic time scale of the noise fluctuation (see Sec.
\ref{SEC_NOISE_AUTOCORRELATION} for more details), the environmental noises
fall into two categories: \textit{dynamical quantum noises} that change
randomly during the evolution of each sample and \textit{static thermal
noises} that remain invariant for each sample but change randomly from sample
to sample (see Sec. \ref{SEC_QUANTUM_NOISE} for discussions about the
difference between dynamical quantum noises and static thermal noises). Note
that \textquotedblleft noises\textquotedblright\ that remain invariant during
all repeated measurements just renormalize the external field and do not cause
decoherence, e.g., decoherence is suppressed under fast measurements
\cite{BarthelPRL2009,ShulmanNatCommun2014,DelbecqPRL2016}.

In traditional spin resonance measurements, a large number of spatially
separated central spins are simultaneously prepared, evolved, and measured. In
this case, each central spin is a \textit{sample} of the \textit{spatial
ensemble}. Since spatially separated spins may be subjected to different
static macroscopic conditions (e.g., due to inhomogeneous magnetic fields,
$g$-factors, and strains), this introduces additional static noises that could
qualitatively change the central spin dephasing \cite{DobrovitskiPRB2008}.
Nevertheless, since static noises are just static inhomogeneities of the
environments for different samples, they can be eliminated by techniques that
remove these inhomogeneities, such as spin echo
\cite{HahnPR1950,JelezkoPRL2004}. It is also possible to employ environmental
engineering to suppress quasi-static noises. For example, to combat electron spin decoherence in nuclear spin baths, a widely pursued approach is
to narrow the distribution of the quasi-static noise by polarizing the bath
\cite{CoishPRB2004,LondonPRL2013,LiuNanoscale2014}, quantum measurements of
the bath
\cite{GiedkePRA2006,KlauserPRB2006,StepanenkoPRL2006,CappellaroPRA2012,ShulmanNatCommun2014}%
, and nonlinear feedback between the electron spins and the nuclear spin baths
\cite{GreilichScience2007,XuNature2009,SunPRL2012,LattaNatPhys2009,BluhmPRL2010,ToganNature2011}
(see Ref. \cite{YangPRB2013} for the theories about the nonlinear feedback).
Thus the dynamical quantum noises are the most relevant mechanism of central spin dephasing. Single-spin
and many-spin measurements would give similar statistics if the dynamical
quantum noise do not vary appreciably for spatially separated spins.

\subsection{Classification of decoherence processes}

\label{SEC_CLASSIFICATION}

The state of the central spin can be visualized by the \textit{Bloch vector}
defined as $2\langle\hat{\mathbf{S}}(t)\rangle\equiv2\operatorname*{Tr}%
[\hat{\mathbf{S}}\hat{\rho}(t)]$\ through the decomposition
\[
\hat{\rho}(t)=\frac{\hat{I}}{2}+\langle\hat{\mathbf{S}}(t)\rangle\cdot
\hat{\boldsymbol{\sigma}},
\]
where $\hat{I}$ is the identity matrix and $\hat{\boldsymbol{\sigma}}%
=(\hat{\sigma}_{x},\hat{\sigma}_{y},\hat{\sigma}_{z})^{T}$ are Pauli matrices
along the $x/y/z$ directions. The Bloch vector of the general pure state
$|\theta,\varphi\rangle$ in Eq. (\ref{THETA_PHI}) is a unit vector with polar
angle $\theta$ and azimuth angle $\varphi$ [Fig. \ref{G_BLOCHSPHERE}(a)]. The
unitary evolution transforms a pure state into another pure state with the
length of the Bloch vector preserved. For example, the coherent evolution
$|\theta,\varphi\rangle\rightarrow|\theta,\varphi+\omega_{0}t\rangle$ governed
by the Zeeman Hamiltonian in Eq. (\ref{H0}) is mapped to the Larmor precession of
the Bloch vector around the magnetic field ($z$ axis) [Fig.
\ref{G_BLOCHSPHERE}(b)]: $\langle\mathbf{\dot{S}}(t)\rangle=\omega
_{0}\mathbf{e}_{z}\times\langle\hat{\mathbf{S}}(t)\rangle$ or equivalently
$\langle\dot{S}_{z}(t)\rangle=0$ and $\langle\dot{S}_{-}(t)\rangle
=-i\omega_{0}\langle\hat{S}_{-}(t)\rangle$, where $S_{\pm}\equiv S_{x}\pm
iS_{y}$. By contrast, spin decoherence transforms, via non-unitary evolution,
a pure state into a mixed state, described by a Bloch vector with shrinking length.

\begin{figure}[ptb]
\includegraphics[width=\columnwidth,clip]{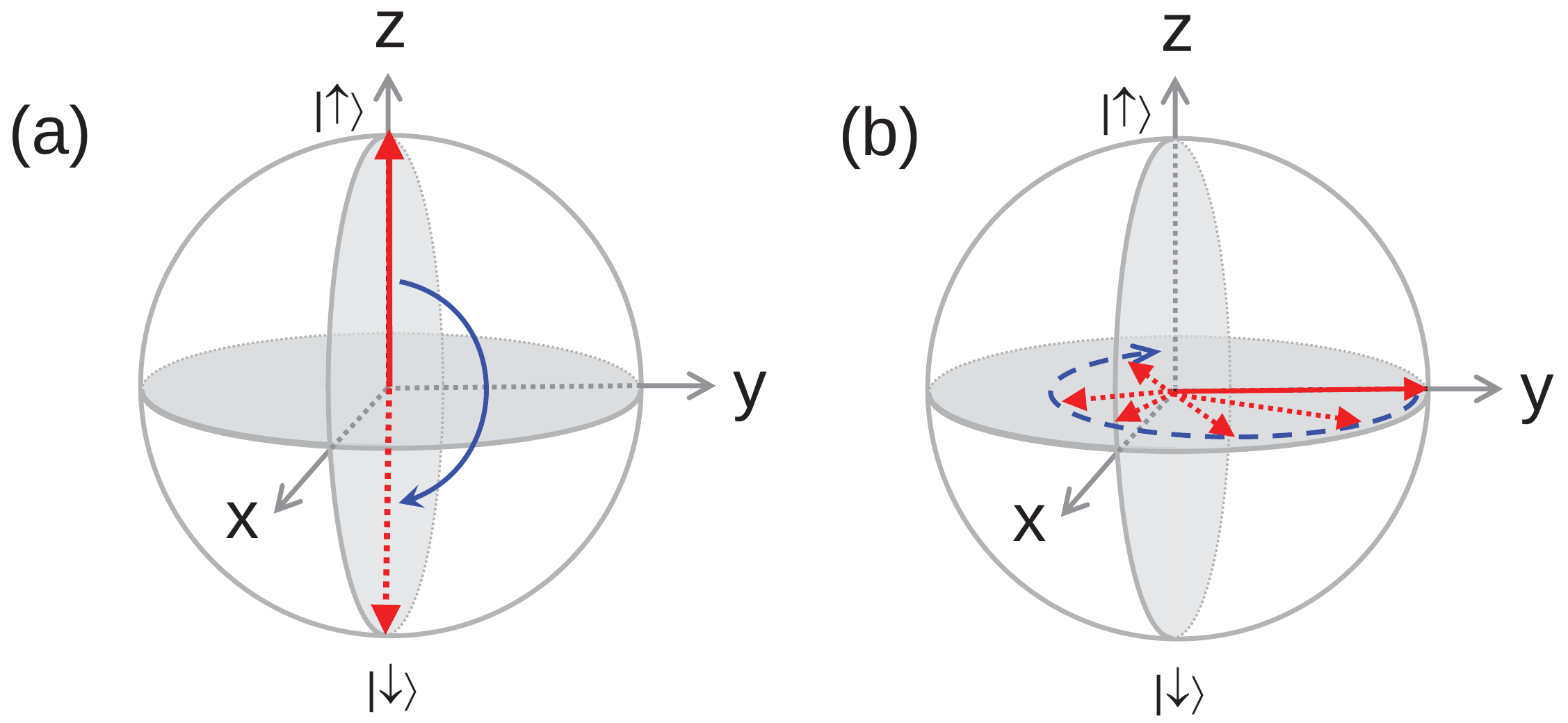}
\caption{Evolution of the
Bloch vector under (a) spin relaxation and (b) spin dephasing.}%
\label{G_T1T2}%
\end{figure}

The environmental noise induces two kinds of changes to the central spin state:

\leftmargini=3mm

\begin{enumerate}
\item Spin relaxation (also called longitudinal relaxation or $T_{1}$ process
in literature), which refers to the change of the diagonal populations
$\rho_{\uparrow\uparrow}(t)$ and $\rho_{\downarrow\downarrow}(t)$ or
equivalently the longitudinal component of the Bloch vector $2\langle\hat
{S}_{z}(t)\rangle=\rho_{\uparrow\uparrow}(t)-\rho_{\downarrow\downarrow}(t)$,
as shown in Fig. \ref{G_T1T2}(a).

\item Spin dephasing (also called transverse relaxation or $T_{2}$ process in
literature), which refers to the decay of the \textit{off-diagonal coherence}
\begin{equation}
L(t)\equiv\frac{\rho_{\uparrow\downarrow}(t)}{\rho_{\uparrow\downarrow}%
(0)}=\frac{\langle\hat{S}_{-}(t)\rangle}{\langle\hat{S}_{-}(0)\rangle}
\label{L_DEF}%
\end{equation}
or equivalently the transverse components $\langle\hat{S}_{x}(t)\rangle
=\operatorname{Re}\rho_{\uparrow\downarrow}(t)$ and $\langle\hat{S}%
_{y}(t)\rangle=-\operatorname{Im}\rho_{\uparrow\downarrow}(t)$ of the Bloch
vector, as shown in Fig. \ref{G_T1T2}(b).
\end{enumerate}

The spin relaxation ($T_{1}$ process) is always accompanied by spin dephasing
($T_{2}$ process), but there are two kinds of physical mechanisms that
contribute to \textit{pure dephasing} (i.e., without causing spin relaxation):
(1)\ dynamical quantum noises lead to \textit{\textquotedblleft
true\textquotedblright\ decoherence} $(T_{\varphi}$ process)
\cite{JoosBook2003}; (2) static thermal noises lead to \textit{inhomogeneous
dephasing} ($T_{2}^{\ast}$ process). For $T_{1}$ and $T_{\varphi}$ processes,
to provide an intuitive physical picture, we only consider noises that flucutate and hence lose memory much faster than
central spin decoherence.

\subsection{Spin relaxation (T$_{1}$ process)}

\label{SEC_T1}

When the environmental noise induces the central spin flip between
$|\uparrow\rangle$ and $|\downarrow\rangle$, the central spin energy changes
by an amount $\omega_{0}$, which is compensated by the environment to ensure
the conservation of energy. For noises that fluctuate rapidly and hence lose memory much faster
than the central spin relaxes, the random central spin flip is memoryless,
i.e., the central spin state at time $t$ completely determines its state at
the next instant. If $\hat{\rho}(t)=|\psi\rangle\langle\psi|$ is a pure
superposition $|\psi\rangle\equiv|\psi_{\uparrow}\rangle+|\psi_{\downarrow
}\rangle$ of the spin-up component $|\psi_{\uparrow}\rangle$ and spin-down
component $|\psi_{\downarrow}\rangle$ and the environment induces the random
jump $|\uparrow\rangle\rightarrow|\downarrow\rangle$ at a constant rate
$\gamma$ [Fig. \ref{G_T1T2}(a)], then during a small interval $dt$, the
component $|\psi_{\downarrow}\rangle$ remains intact, while $|\psi_{\uparrow
}\rangle$ has a probability $\gamma dt$ to \textit{incoherently }jump to
$\hat{S}_{-}|\psi_{\uparrow}\rangle=\hat{S}_{-}|\psi\rangle$. Therefore, the
central spin state at the next instant $t+dt$ is given by the density matrix
\[
\hat{\rho}(t+dt)=\hat{M}_{1}\hat{\rho}(t)\hat{M}_{1}^{\dagger}+\hat{M}_{0}%
\hat{\rho}(t)\hat{M}_{0}^{\dagger},
\]
which describes the incoherent mixture of the collapsed component
$\sqrt{\gamma dt}\hat{S}_{-}|\psi\rangle\equiv\hat{M}_{1}|\psi\rangle$ and the
non-collapsed component
\[
|\psi_{\downarrow}\rangle+\sqrt{1-\gamma dt}|\psi_{\uparrow}\rangle\approx
e^{-(\gamma dt/2)\hat{S}_{-}^{\dagger}\hat{S}_{-}}|\psi\rangle\equiv\hat
{M}_{0}|\psi\rangle.
\]
This evolution corresponds to a general binary-outcome weak measurement of the
central spin by the environment (with the results unknown to any observers):\ depending on the two possible outcomes, the
central spin collapses to $\hat{M}_{1}|\psi\rangle$ or $\hat{M}_{0}%
|\psi\rangle$. The central spin evolution due to the random jump
$|\uparrow\rangle\rightarrow|\downarrow\rangle$ is
\[
\lbrack\dot{\rho}(t)]_{|\uparrow\rangle\rightarrow|\downarrow\rangle}%
\equiv\frac{\hat{\rho}(t+dt)-\hat{\rho}(t)}{dt}=\gamma\mathcal{D}[\hat{S}%
_{-}]\hat{\rho}(t),
\]
where $\mathcal{D}[\hat{L}]\hat{\rho}\equiv\hat{L}\hat{\rho}\hat{L}^{\dagger
}-\{\hat{L}^{\dagger}\hat{L},\hat{\rho}\}/2$ is the standard Lindblad form for dissipation.

In general, an environment could not only induce $|\uparrow\rangle
\rightarrow|\downarrow\rangle$ by absorbing an energy quantum $\omega_{0}$
from the central spin, but also induce the reverse process $|\downarrow
\rangle\rightarrow|\uparrow\rangle$ by delivering an energy quantum
$\omega_{0}$ to the central spin. When the environment is in thermal
equilibrium with an inverse temperature $\beta\equiv1/(k_{B}T_{\mathrm{env}}%
)$, the latter process would be slower than the former process by a Boltzman
factor $e^{-\beta\omega_{0}}$, e.g., for $T_{\mathrm{env}}=0$, the environment
is in its ground state and hence cannot deliver the energy quantum $\omega
_{0}$, so the latter process is blocked. Including both processes, the
environment induced central spin evolution is described by
\begin{equation}
\lbrack\dot{\rho}(t)]_{\mathrm{T}_{1}}=\gamma(\mathcal{D}[\hat{S}%
_{-}]+e^{-\beta\omega_{0}}\mathcal{D}[\hat{S}_{+}])\hat{\rho}(t)=%
\begin{bmatrix}
-\frac{\rho_{\uparrow\uparrow}(t)-\rho_{\uparrow\uparrow}^{\mathrm{eq}}}%
{T_{1}} & -\frac{\rho_{\uparrow\downarrow}(t)}{2T_{1}}\\
-\frac{\rho_{\downarrow\uparrow}(t)}{2T_{1}} & -\frac{\rho_{\downarrow
\downarrow}(t)-\rho_{\downarrow\downarrow}^{\mathrm{eq}}}{T_{1}}%
\end{bmatrix}
, \label{DRHO_T1}%
\end{equation}
which is characterized by a single time constant $T_{1}\equiv\lbrack
(1+e^{\beta\omega_{0}})\gamma]^{-1}$ (so-called spin relaxation time) and
drives the central spin into thermal equilibrium with the environment:
\[
\hat{\rho}^{\mathrm{eq}}\equiv\frac{e^{-\beta\hat{H}_{0}}}{\operatorname*{Tr}%
e^{-\beta\hat{H}_{0}}}=%
\begin{bmatrix}
\frac{1}{1+e^{\beta\omega_{0}}} & 0\\
0 & \frac{e^{\beta\omega_{0}}}{1+e^{\beta\omega_{0}}}%
\end{bmatrix}
.
\]
During the spin relaxation process, both the populations and\ the off-diagonal
coherence of the central spin exponentially decay to their respective thermal
equilibrium values, with the decay rate of the latter being only half that of
the former.

\subsection{\textquotedblleft True\textquotedblright\ decoherence by dynamical
quantum noises ($T_{\varphi}$ process)}

\label{SEC_TPHI}

During pure dephasing, the environmental noise induces random jumps of the
relative phase between the energy eigenstates $|\uparrow\rangle$ and
$|\downarrow\rangle$ of the central spin. For noises that lose
memory much faster than the spin dephasing, the central spin evolution is
memoryless. Again we take $\hat{\rho}(t)=|\psi\rangle\langle\psi|$ and assume
that the environment induces the random phase jump $|\psi\rangle
\rightarrow\hat{\sigma}_{z}|\psi\rangle$ at a constant rate $\gamma_{\varphi}%
$. The central spin state at the next instant $t+dt$ is an incoherent mixture
of $\sqrt{\gamma_{\varphi}dt}\hat{\sigma}_{z}|\psi\rangle\equiv\hat
{M}_{\mathrm{c}}|\psi\rangle$ and
\[
\sqrt{1-\gamma_{\varphi}dt}|\psi\rangle\approx e^{-(\gamma_{\varphi}%
dt/2)\hat{\sigma}_{z}^{\dagger}\hat{\sigma}_{z}}|\psi\rangle\equiv\hat
{M}_{\mathrm{nc}}|\psi\rangle,
\]
described by the density matrix $\hat{\rho}(t+dt)=\hat{M}_{\mathrm{c}}%
\hat{\rho}(t)\hat{M}_{\mathrm{c}}^{\dagger}+\hat{M}_{\mathrm{nc}}\hat{\rho
}(t)\hat{M}_{\mathrm{nc}}^{\dagger}$. This incoherent collapse corresponds to
a general binary-outcome weak measurement of the central spin by the
environment (with the results unknown to any observers). The central spin evolution due to this process assumes the
standard Lindblad form%
\begin{equation}
\lbrack\dot{\rho}(t)]_{\mathrm{T}_{\varphi}}=\gamma_{\varphi}\mathcal{D}%
[\hat{\sigma}_{z}]\hat{\rho}(t)=%
\begin{bmatrix}
0 & -\frac{\rho_{\uparrow\downarrow}(t)}{T_{\varphi}}\\
-\frac{\rho_{\downarrow\uparrow}(t)}{T_{\varphi}} & 0
\end{bmatrix}
, \label{DRHO_TPHI}%
\end{equation}
which is characterized by a single time constant $T_{\varphi}\equiv
1/(2\gamma_{\varphi})$ (so-called pure dephasing time). During \textquotedblleft
true\textquotedblright\ decoherence, the longitudinal Bloch vector component
remain invariant, while the magnitude of the transverse components decay
exponentially on a time scale $T_{\varphi}$ [Fig. \ref{G_T1T2}(b)].

\subsection{Inhomogeneous dephasing by static thermal noises ($T_{2}^{\ast}$
process)}

\label{SEC_T2STAR}

\begin{figure}[ptb]
\includegraphics[width=\columnwidth,clip]{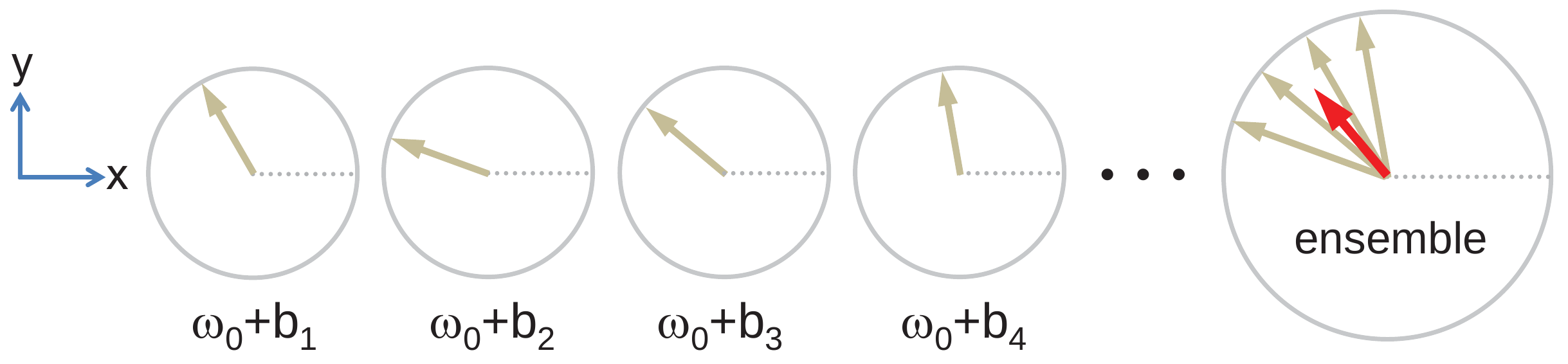}
\caption{Geometric
representation of inhomogeneous dephasing due to averaging over an ensemble of
coherently precessing samples.}%
\label{G_INHOMODEPHASING}%
\end{figure}

In the presence of static noises, the central spin evolution is governed by
the Hamiltonian $\hat{H}_{\mathbf{b}}\equiv\hat{H}_{0}+\mathbf{\tilde{b}}%
\cdot\hat{\mathbf{S}}$, where the local field $\mathbf{\tilde{b}}$ remains
static for each sample of the ensemble but fluctuates from sample to sample
according to a certain probability distribution $P_{\mathrm{inh}}(\mathbf{b}%
)$. For a sample subjected to the local field $\mathbf{b}$, the central spin
undergoes unitary evolution $\hat{\rho}_{\mathbf{b}}(t)=e^{-i\hat
{H}_{\mathbf{b}}t}\hat{\rho}(0)e^{i\hat{H}_{\mathbf{b}}t}$ and its Bloch
vector $\langle\hat{\mathbf{S}}(t)\rangle_{\mathbf{b}}\equiv\operatorname*{Tr}%
[\hat{\mathbf{S}}\hat{\rho}_{\mathbf{b}}(t)]$ undergoes coherent precession
$\langle\mathbf{\dot{S}}(t)\rangle_{\mathbf{b}}=(\omega_{0}\mathbf{e}%
_{z}+\mathbf{b})\times\langle\hat{\mathbf{S}}(t)\rangle_{\mathbf{b}}$ that
preserves its length. The density matrix that describes the ensemble,
\begin{equation}
\hat{\rho}(t)=\int\hat{\rho}_{\mathbf{b}}(t)P_{\mathrm{inh}}(\mathbf{b}%
)d\mathbf{b}, \label{RHO_ENSEMBLE}%
\end{equation}
and the Bloch vector%
\[
\langle\hat{\mathbf{S}}(t)\rangle=\int\langle\hat{\mathbf{S}}(t)\rangle
_{\mathbf{b}}P_{\mathrm{inh}}(\mathbf{b})d\mathbf{b}.
\]
In principle, the inhomogeneous distribution of the local field can result in
both spin relaxation and spin dephasing.

When the external field is much stronger than the noise field, the transverse
noises $\tilde{b}_{x},\tilde{b}_{y}$ can hardly tilt the precession axis away
from the $z$ axis. In this case, the longitudinal spin relaxation is
suppressed by the large energy splitting $\omega_{0}$ between the spin-up
$|\uparrow\rangle$ and spin-down $|\downarrow\rangle$ eigenstates, and only
pure dephasing by the longitudinal noise $\tilde{b}_{z}$ occurs (see Fig.
\ref{G_INHOMODEPHASING}): the different precession frequencies of different
samples lead to progressive spread out of their azimuth angles $\varphi
_{j}(t)=(\omega_{0}+b_{j})t$ and hence decay of the transverse Bloch vector
components $\langle\hat{S}_{-}(t)\rangle$. The off-diagonal coherence (the
intrinsic phase factor $e^{-i\omega_{0}t}$ removed)
\begin{equation}
L_{\mathrm{inh}}(t)=\int e^{-ibt}P_{\mathrm{inh}}(b)db \label{LT_INH}%
\end{equation}
decays on a time scale $T_{2}^{\ast}\sim$ inverse of the characteristic width
of the static noise distribution $P_{\mathrm{inh}}(b)$. Such decay by
classical ensemble averaging over static noises is called inhomogeneous
dephasing ($T_{2}^{\ast}$ process). For the commonly encountered Gaussian
distribution
\begin{equation}
P_{\mathrm{inh}}(b)=\frac{1}{\sqrt{2\pi}b_{\mathrm{rms}}}e^{-b^{2}%
/(2b_{\mathrm{rms}}^{2})}, \label{PENS_GAU}%
\end{equation}
the spin coherence shows the Gaussian decay:
\begin{subequations}
\label{L_T2STAR}%
\begin{align}
L_{\mathrm{inh}}(t)  &  =e^{-(t/T_{2}^{\ast})^{2}},\\
T_{2}^{\ast}  &  =\frac{\sqrt{2}}{b_{\mathrm{rms}}}.
\end{align}
As will be discussed in Sec. \ref{SEC_DD_FILTER}, the $T_{2}^{\ast}$ process
can be completely removed by spin echo techniques.

\subsection{Summary}

Including the unitary evolution under the external field [Eq. (\ref{H0})] and
a fixed local field $b$, as well as the $T_{1}$ and $T_{\varphi}$ processes
caused by rapidly fluctuating noises that lose memory much faster than
central spin decoherence [Eqs. (\ref{DRHO_T1}) and (\ref{DRHO_TPHI})], the
density matrix of the central spin obeys the Lindblad master equation
\end{subequations}
\[
\dot{\rho}_{b}(t)=-i[\hat{H}_{0}+b\hat{S}_{z},\hat{\rho}_{b}(t)]-%
\begin{bmatrix}
\frac{\lbrack\hat{\rho}_{b}(t)]_{\uparrow\uparrow}-\rho_{\uparrow\uparrow
}^{\mathrm{eq}}}{T_{1}} & \frac{[\hat{\rho}_{b}(t)]_{\uparrow\downarrow}%
}{T_{2}}\\
\frac{\lbrack\hat{\rho}_{b}(t)]_{\downarrow\uparrow}}{T_{2}} & \frac
{[\hat{\rho}_{b}(t)]_{\downarrow\downarrow}-\rho_{\downarrow\downarrow
}^{\mathrm{eq}}}{T_{1}}%
\end{bmatrix}
,
\]
where $T_{2}\equiv\lbrack1/(2T_{1})+1/T_{\varphi}]^{-1}$ ($\leq2T_{1}$) is the
spin dephasing time. In the presence of inhomogeneous dephasing, the density
matrix $\hat{\rho}(t)$ is obtained by averaging $\hat{\rho}_{b}(t)$ over the
distribution of $b$. Note that although spin relaxation imposes an upper limit
on the spin dephasing time via $T_{2}\leq2T_{1}$, in typical cases of central
spin decoherence $T_{2}$ is much shorter than $T_{1}$ and is limited by pure
dephasing ($T_{\varphi}$ and $T_{2}^{\ast}$ processes).

\section{Semiclassical noise theory for spin decoherence}

\label{SEC_SEMICLASSICAL}

A simple theoretical treatment of spin decoherence is to describe the
environment as a source of classical magnetic noise $\mathbf{\tilde{b}}(t)$
with zero mean $\langle\mathbf{\tilde{b}}(t)\rangle=0$, so the central spin
Hamiltonian is
\begin{equation}
\hat{H}(t)=\omega_{0}\hat{S}_{z}+\mathbf{\tilde{b}}(t)\cdot\hat{\mathbf{S}}.
\label{HAMIL_SEMI}%
\end{equation}
The time-dependent transverse noises $\tilde{b}_{\pm}(t)\equiv\tilde{b}%
_{x}(t)\pm i\tilde{b}_{y}(t)$ could randomly tilt the precession axis away
from the $z$ axis, flip the central spin between the unperturbed eigenstates
$|\uparrow\rangle$ and $|\downarrow\rangle$, and hence induce spin relaxation.
The longitudinal noise $\tilde{b}_{z}(t)$ randomly modulates the central spin
precession frequency along the $z$ axis and induce pure dephasing. Here we
consider a strong external magnetic field and hence a large unperturbed
precession frequency $\omega_{0}$, so that the noise can be treated as a perturbation.

\subsection{Basic concept of classical noise}

\begin{figure}[ptb]
\includegraphics[width=\columnwidth,clip]{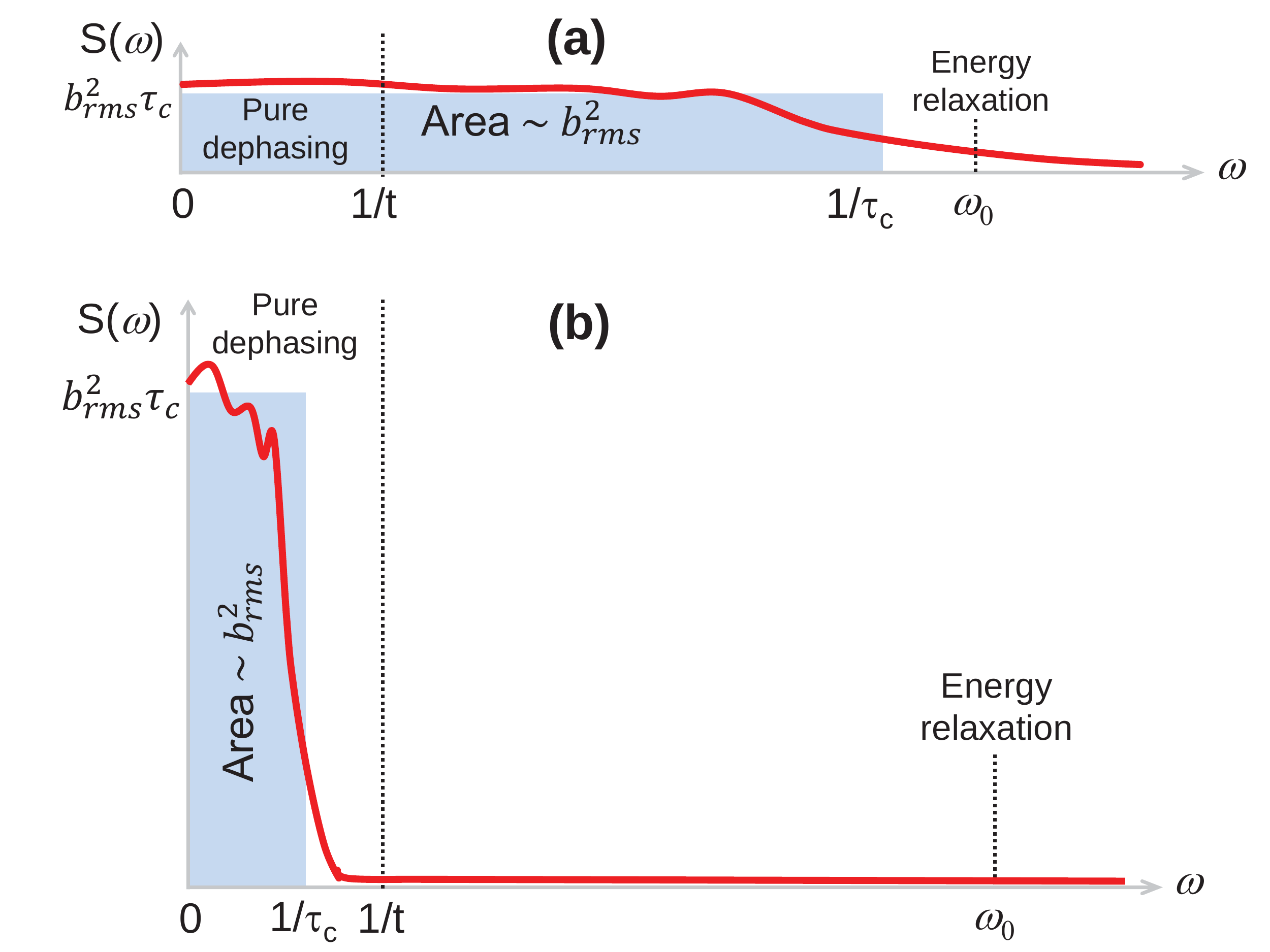}
\caption{Schematic of the noise spectra for (a) a noise that
fluctuates on a time scale $\tau_{c}\ll t$ and (b) a noise that
fluctuates on a time scale $\tau_{c}\gtrsim t$, where $t$ is the evolution
time.}%
\label{G_NOISE_SPECTRUM}%
\end{figure}

We take a real, scalar noise $\tilde{b}(t)$ with zero mean $\langle\tilde
{b}(t)\rangle=0$ to explain some basic concepts of classical noises. A
classical noise is specified by the probability distribution for each
realization of the noise, e.g., the probability distribution $P(b_{0}%
,b_{1},\cdots)$ for the noise $\tilde{b}_{n}\equiv\tilde{b}(t_{n})$ at all the
time points $t_{n}\equiv n\Delta t$. Below we
introduce two important characteristics of noises: statistics and
auto-correlations (or equivalently spectra). We will particularly focus on Gaussian noises, which are the simplest and yet a commonly encountered type of noise statistics. Among various noise spectra, we highlight two simple cases, namely, static noises and rapidly fluctuating noises that lose memory much faster than central spin decoherence.

\subsubsection{Statistics}

According to the form of $P(b_{0},b_{1},\cdots)$, noises are often classified
as Gaussian or non-Gaussian. Gaussian noises are one of the simplest and most
widely encountered noises. For a Gaussian noise, the random variables
$\tilde{b}_{0},\tilde{b}_{1},\cdots$ obey the multivariate normal
distribution
\begin{equation}
P(b_{0},b_{1},\cdots)\propto e^{-(1/2)\sum_{ij}b_{i}(\mathbf{C}^{-1}%
)_{ij}b_{j}}, \label{PB_DISCRETE}%
\end{equation}
where $\mathbf{C}^{-1}$ is a positive-definite symmetric matrix. In the continuous form, the Gaussian distribution as a functional of the noise
$\tilde{b}(t)$ has the form $P[b(t)]\propto e^{(-1/2)\int dt_{1}\int
dt_{2}b(t_{1})C^{-1}(t_{1},t_{2})b(t_{2})}$, where $C^{-1}(t_{1},t_{2})$ is a
positive-definite symmetric matrix. As a key property, an arbitrary linear
combination $\tilde{\varphi}\equiv\sum_{n}c_{n}\tilde{b}_{n}$ of Gaussian
random variables is still Gaussian, i.e., still obeys normal distribution.
Averaging over Gaussian noises can be obtained explicitly, e.g.,
\begin{equation}
\langle e^{i\tilde{\varphi}}\rangle=e^{-\langle\tilde{\varphi}^{2}\rangle/2},
\label{EXP_GAU}%
\end{equation}
which can be readily verified by assuming that $\tilde{\varphi}$ obeys Gaussian
distribution $P(\varphi)\equiv e^{-\varphi^{2}/(2\sigma^{2})}/(\sqrt{2\pi
}\sigma)$. As suggested by Eq. (\ref{PB_DISCRETE}), the distribution and hence
all moments of the Gaussian noise are completely determined by the matrix
$\mathbf{C}$:
\begin{subequations}
\label{WICK_GAU}%
\begin{align}
\langle\tilde{b}_{i}\tilde{b}_{j}\rangle &  =(\mathbf{C})_{i,j}%
,\label{WICK_GAU_1}\\
\langle\tilde{b}_{i}\tilde{b}_{j}\tilde{b}_{k}\tilde{b}_{l}\rangle &
=(\mathbf{C})_{i,j}(\mathbf{C})_{k,l}+(\mathbf{C})_{i,k}(\mathbf{C}%
)_{j,l}+(\mathbf{C})_{i,l}(\mathbf{C})_{j,k},\\
&  \cdots\nonumber\\
\langle\tilde{b}_{i_{1}}\tilde{b}_{i_{2}}\cdots\tilde{b}_{i_{2M}}\rangle &
=\sum_{\mathrm{P}}(\mathbf{C})_{i_{p_{1}}i_{p_{2}}}(\mathbf{C})_{i_{p_{3}%
}i_{p_{4}}}\cdots(\mathbf{C})_{i_{p_{2M-1}}i_{p_{2M}}},
\end{align}
where $\sum_{\mathrm{P}}$ runs over all possible pairings of $\{i_{1}%
,\cdots,i_{2M}\}$. Equation (\ref{WICK_GAU}) is the Wick's theorem for
Gaussian noises, and Eq. (\ref{WICK_GAU_1}) shows that the matrix $\mathbf{C}$
in Eq. (\ref{PB_DISCRETE}) is the covariance matrix of the Gaussian noise.

\subsubsection{Noise auto-correlations}

\label{SEC_NOISE_AUTOCORRELATION}

A classical noise is usually characterized by its auto-correlation%
\end{subequations}
\begin{equation}
C(\tau)=\langle\tilde{b}(\tau)\tilde{b}(0)\rangle, \label{CTAO}%
\end{equation}
or equivalently the noise spectrum (the power distribution)
\begin{equation}
S(\omega)\equiv\int e^{i\omega\tau}C(\tau)d\tau, \label{SOMEGA}%
\end{equation}
both of which are even functions. The auto-correlation $C(\tau)$ is usually
maximal at $\tau=0$ and decays with increasing $|\tau|$. For example, the electron spin bath is usually modelled
by the Ornstein-Uhlenbeck noise
\cite{DobrovitskiPRL2009,WitzelPRB2012,WitzelPRB2014,LangeScience2010}, which
is Gaussian and has the auto-correlation
\begin{equation}
C(\tau)=b_{\mathrm{rms}}^{2}e^{-|\tau|/\tau_{c}} \label{OU_CORRELATION}%
\end{equation}
and the noise spectrum
\begin{equation}
S(\omega)=2\pi b_{\mathrm{rms}}^{2}\delta^{(1/\tau_{c})}(\omega),
\label{OU_SPECTRUM}%
\end{equation}
where $\delta^{(\gamma)}(\Delta)\equiv(\gamma/\pi)/(\Delta^{2}+\gamma^{2})$ is
the Lorentzian shape function.

The auto-correlation or noise spectrum has three important properties:
auto-correlation (or memory) time $\tau_{c}$, the behavior of high-frequency
cutoff, and the noise power $b_{\mathrm{rms}}^{2}\equiv\langle\tilde{b}%
^{2}(0)\rangle=\langle\tilde{b}^{2}(t)\rangle$. The \textit{auto-correlation
time} $\tau_{c}$, which quantifies how fast the noise fluctuates, is the
characteristic time for the auto-correlation to decay. Equivalently,
$1/\tau_{c}$ is the characteristic cutoff frequency above which the noise
spectrum decays significantly [see Eqs. (\ref{OU_CORRELATION}) and
(\ref{OU_SPECTRUM}) for the Ornstein-Uhlenbeck noise]. If $\tau_{c}$ is large
compared with the achievable time scale of control over the central spin and
the high-frequency tail of the spectrum decays faster than power-law decay,
then the noise is said to have a hard high-frequency cutoff. Otherwise, the
noise has a soft cutoff. For example, the noise spectrum of the Debye phonon
bath \cite{UhrigPRL2007}, $S(\omega)=2\alpha\omega\Theta(\omega_{D}-\omega)$
with $\Theta(\omega)$ the Heaviside step function, has a hard cut-off, while
that of the Ornstein-Uhlenbeck noise has a soft cutoff as it decays as
$1/\omega^{2}$ at high frequency. The \textit{noise power} is equal to the
area of the noise spectrum:
\[
b_{\mathrm{rms}}^{2}=\int_{-\infty}^{\infty}S(\omega)\frac{d\omega}{2\pi}%
=\int_{0}^{\infty}S(\omega)\frac{d\omega}{\pi}.
\]
For a fixed noise power, rapidly fluctuating noise has a low and
broad spectrum [Fig. \ref{G_NOISE_SPECTRUM}(a)], while slowly fluctuating
noise has a high and narrow spectrum [Fig.
\ref{G_NOISE_SPECTRUM}(b)]. As will be discussed in Sec. \ref{SEC_NOISE_T2},
the broad noise spectrum underlies the motional narrowing phenomenon in magnetic
resonance spectroscopy \cite{AndersonJPSJ1954,AbragamBook1961}.

\subsubsection{Markovian and non-Markovian noises and stochastic processes}
Considering that there is considerable inconsistency in the terminology of Markovian/non-Markovian stochastic processes, noises, and decoherence, here we would like to make clear our usage of terminology, yet without the intention of unifying the usage in the vast literature.
We note that it is useful to distinguish the {\em noise} and the {\em stochastic process} (such as phonon scattering, atom-atom collisions,
and nuclear spin flip-flips) that causes the noise.

A classical noise as the collection of the random variables $\tilde{b}_{n}\equiv\tilde{b}(t_{n})$ at all the
time points $t_{n}\equiv n\Delta t$ is characterized by
the probability distribution $P(b_{0})$ of $\tilde{b}_{0}$ and the probability
distribution $P(b_{n}|b_{0},\cdots,b_{n-1})$ of $\tilde{b}_{n}$ conditioned on
$\tilde{b}_{k}$ being $b_{k}$ ($k=0,1,\cdots,n-1)$. The noise is caused by certain microscopic stochastic processes.
A {\em stochastic process} is called Markovian or memoryless if the distribution of $\tilde{b}_{n}$ depends on $\tilde{b}_{n-1}$
only, i.e., $P(b_{n}|b_{0},\cdots,b_{n-1})=P(b_{n}|b_{n-1})$, so that the
probability distribution of the noise can be written as
\begin{equation}
P(b_{0},b_{1},b_{2},\cdots)=P(b_{0})P(b_{1}|b_{0})P(b_{2}|b_{1})\cdots.
\label{PROB_ALL}%
\end{equation}
Physically, this occurs when the stochastic process (such as a phonon scattering, an atom-atom collision, or a nuclear spin flip-flop) takes a time much shorter than the timescale under consideration. For example, the atom-atom collision is Markovian under the impact approximation, and a phonon scattering is Markovian for a timescale much greater than $\sim$ 1 picosecond. On the other hand, a noise, being caused by either a Markovian or non-Markovian stochastic process, is labeled as Markovian or memoryless when its auto-correlation time $\tau_{c}$ is much shorter than the central spin
decoherence time. In general a Markovian or non-Markovian noise could be produced by either a non-Markovian or Markovian stochastic process.

For example, the Ornstein-Uhlenbeck noise [Eqs. (\ref{OU_CORRELATION}) and (\ref{OU_SPECTRUM})] is caused by the Ornstein-Uhlenbeck process, which is characterized by a Gaussian distribution $P(b_{0})=e^{-b_{0}^{2}%
/(2b_{\mathrm{rms}}^{2})}/(\sqrt{2\pi}b_{\mathrm{rms}})$ for the initial value
$\tilde{b}_{0}$ and a Gaussian conditional distribution for $\tilde{b}_{n}$ \cite{KlauderPR1962}:
\begin{equation}
P(b_{n}|b_{n-1},\cdots,b_{0})=P(b_{n}|b_{n-1})=\frac{e^{-(b_{n}-e^{-\Delta
t/\tau_{c}}b_{n-1})^{2}/(2\sigma^{2})}}{\sqrt{2\pi}\sigma}, \label{MAR_GAU}%
\end{equation}
where $\sigma=b_{\mathrm{rms}}\sqrt{1-e^{-2\Delta t/\tau_{c}}}$. Obviously, the Ornstein-Uhlenbeck \textit{process} is Markovian since the distribution of $\tilde{b}_{n}$ only depends on ${b}_{n-1}$. However, the Ornstein-Uhlenbeck \textit{noise} has the auto-correlation $\langle\tilde{b}_{n}\tilde{b}_{m}\rangle=b_{\mathrm{rms}}^{2}e^{-|t_{n}-t_{m}|/\tau_{c}}$ [see Eq. (\ref{OU_CORRELATION}) for its continuous form], so it could be either Markovian or non-Markovian depending on whether or not its auto-correlation time $\tau_{c}$ is much larger than the central spin decoherence time. In addition, the Ornstein-Uhlenbeck noise is also Gaussian since its distribution function $P(b_{0},b_{1},\cdots)$ can be put into the form of Eq. (\ref{PB_DISCRETE}).

Under the classification based on $\tau_{c}$, two kinds of noises are
relatively simple: quasistatic noise with $\tau_{c}\gg$ duration of each
measurement cycle ($\sim$ central spin decoherence time), and Markovian noise
with $\tau_{c}\ll$ duration of each measurement cycle. The static noise
$\tilde{b}(t)=\tilde{b}$ is completely specified by its static distribution
$P_{\mathrm{inh}}(b)$, so inhomogeneous dephasing caused by static noise can
be easily treated (see Sec. \ref{SEC_T2STAR}). The Markovian noise gives
memoryless random jumps of the central spin, as described intuitively in Sec.
\ref{SEC_T1} (for $T_{1}$ process) and Sec. \ref{SEC_TPHI} (for $T_{\varphi}$
process) in terms of two phenomenological jump rates $\gamma$ and
$\gamma_{\varphi}$.

\subsection{Spin relaxation by transverse noises}

\label{SEC_NOISE_T1}

For the sake of simplicity, let us assume $\tilde{b}_{z}(t)=0$. The transverse
noise induced central spin flip can be understood in a simple physical picture
first proposed by Bloembergen, Purcell, and Pound
\cite{BloembergenPR1948,AbragamBook1961}:\ the fourier spectrum $\tilde
{b}_{\pm}(\omega)$ of the transverse noises $\tilde{b}_{\pm}(t)$ may have
nonzero components near the unperturbed spin precession frequency $\omega_{0}$
and these components would induce resonant transitions between the two
unperturbed eigenstates $|\uparrow\rangle$ and $|\downarrow\rangle$ at a rate
proportional to the noise spectrum $S(\omega)\equiv\int\langle\tilde{b}%
_{-}(\tau)\tilde{b}_{+}(0)\rangle e^{i\omega\tau}d\tau\propto\langle|\tilde
{b}_{\pm}(\omega)|^{2}\rangle$ at frequency $\omega_{0}$.

Usually the noise must fluctuate rapidly ($\tau_{c}\lesssim1/\omega_{0}$) in
order for its spectrum to have a significant high frequency component at
$\omega_{0}$, so usually $\tau_{c}\ll$ central spin relaxation time ($T_{1}$),
i.e., the noise is Markovian. When $\langle\tilde{b}_{-}(t)\tilde{b}%
_{+}(t^{\prime})\rangle$ is the only nonvanishing noise auto-correlation, the
Born-Markovian approximation \cite{AbragamBook1961} gives the intuitive result
[Eq. (\ref{DRHO_T1})] for the environment induced central spin evolution, with
$\beta=0$ (i.e., the classical noise is equivalent to an environment at
infinite temperature) and an explicit expression for the central spin jump
rate:
\[
\gamma=\frac{1}{2T_{1}}=\frac{S(\omega_{0})}{4},
\]
which is the noise spectrum at the central spin transition frequency
$\omega_{0}$ (as spin relaxation involves an energy transfer $\omega_{0}$),
thus a rapidly fluctuating Markovian noise with $\tau_{c}\lesssim1/\omega_{0}$
contributes significantly to spin relaxation [Fig. \ref{G_NOISE_SPECTRUM}(a)],
while non-Markovian noises contribute negligibly [Fig. \ref{G_NOISE_SPECTRUM}(b)].

\subsection{Pure dephasing by longitudinal noises}

\label{SEC_NOISE_T2}

Here we assume $\tilde{b}_{x}(t)=\tilde{b}_{y}(t)=0$ and write $\tilde{b}%
_{z}(t)$ as $\tilde{b}(t)$ for brevity. In the interaction picture with
respect to $\hat{H}_{0}$, the Hamiltonian
\begin{equation}
\hat{H}(t)=\hat{S}_{z}\tilde{b}(t), \label{HAMIL_CLASSICAL}%
\end{equation}
describes the random jumps of the central spin transition frequency or
equivalently diffusion of the resonance line (similar to Brownian motion).
Therefore this model has been known as \textit{random frequency modulation} or
\textit{spectral diffusion} in the context of magnetic resonance spectroscopy
following the pineering work of
Anderson~\cite{AndersonRMP1953,AndersonJPSJ1954,KlauderPR1962} and
Kubo~\cite{KuboJPSJ1954}. In the context of quantum computing, this model was
elaborated by de Sousa and Das Sarma
\cite{SousaPRB2003a,SousaPRB2003,SousaPRB2005} to explain the spin echo
experiments for donor electron spins in silicon
\cite{ChibaJPSJ1972,TyryshkinPRB2003,AbePRB2004,FerrettiPRB2005,TyryshkinJPC2006}%
. The theory gives reasonable order-of-magnitude agreement (within a factor of
3) for the dephasing time, but fails to explain the $e^{-\tau^{2}}$ decay of
the echo envelope \cite{SousTAP2009}.

For a general noise, a random relative phase
\begin{equation}
\tilde{\varphi}(t)\equiv\int_{0}^{t}\tilde{b}(t^{\prime})dt^{\prime}
\label{PHI_CLASSICAL}%
\end{equation}
is accumulated between the unperturbed eigenstates $|\uparrow\rangle$ and
$|\downarrow\rangle$, leading to the decay of the off-diagonal coherence
\begin{equation}
L(t)=\langle e^{-i\tilde{\varphi}(t)}\rangle. \label{L_CLASSICAL}%
\end{equation}
In contrast to spin relaxation caused by the high frequency part (near
$\omega_{0})$ of the noise, the pure dephasing is dominated by low-frequency
part of the noise (see Fig. \ref{G_NOISE_SPECTRUM}), because high frequency
components $\omega\gg1/t$ are effectively averaged out in Eq.
(\ref{PHI_CLASSICAL}).

Significant dephasing appears when the root-mean-square phase fluctuation
$\sqrt{\langle\tilde{\varphi}^{2}(t)\rangle}$ attains unity, i.e., the
dephasing time $T_{2}$ can be estimated from $\langle\tilde{\varphi}^{2}%
(T_{2})\rangle=1$, where
\begin{equation}
\langle\tilde{\varphi}^{2}(t)\rangle=\int_{0}^{t}dt_{1}\int_{0}^{t}%
dt_{2}\ \langle\tilde{b}(t_{1})\tilde{b}(t_{2})\rangle.
\label{PHI_RMS_CLASSICAL}%
\end{equation}
Here the accumulation of the random phase depends crucially on the ratio
between $\tau_{c}$ and $T_{2}$:

\leftmargini=3mm

\begin{enumerate}
\item Quasi-static noise ($\tau_{c}\gg1/b_{\mathrm{rms}}\sim T_{2}$). Here
$\tilde{\varphi}(t)\approx\tilde{b}t$ and hence the phase fluctuation
$\sqrt{\langle\tilde{\varphi}^{2}(t)\rangle}\approx b_{\mathrm{rms}}t$
increases linearly with time. This gives inhomogeneous dephasing on a time
scale $T_{2}=T_{2}^{\ast}\sim1/b_{\mathrm{rms}}\ll\tau_{c}$, consistent with
the discussion in Sec. \ref{SEC_T2STAR}. In this regime, the dephasing time is
determined only by the noise power and is independent of $\tau_{c}$.

\item Markovian noise ($\tau_{c}\ll1/b_{\mathrm{rms}}\ll T_{2}$). The noise tends to
average out itself during a single measurement cycle, leading to slow,
diffusive increase of the phase fluctuation $\sqrt{\langle\tilde{\varphi}%
^{2}(t)\rangle}\sim(b_{\mathrm{rms}}\tau_{c})\sqrt{t/\tau_{c}}$. This result
can also be obtained from Eq. (\ref{PHI_RMS_CLASSICAL}) by noting that only
$|t_{1}-t_{2}|\lesssim\tau_{c}$ contributes significantly to the integral.
This gives \textquotedblleft true\textquotedblright\ decoherence on a time
scale $T_{2}=T_{\varphi}\sim1/(b_{\mathrm{rms}}^{2}\tau_{c})\gg 1/b_{\mathrm{rms}}\gg\tau_{c}$.
Actually, the use of Born-Markovian approximation recovers the intuitive
result [Eq. (\ref{DRHO_TPHI})] with an explicit expression for the central
spin jump rate:%
\begin{equation}
\gamma_{\varphi}=\frac{1}{2T_{\varphi}}=\frac{S(0)}{4}\sim b_{\mathrm{rms}}%
^{2}\tau_{c}\ll b_{\mathrm{rms}}, \label{GAMMA_PHI}%
\end{equation}
which is the noise spectrum at zero frequency (as pure dephasing involves no
energy transfer). The above discussions show that faster fluctuations of the
noise lead to longer dephasing time or, in terms of the fourier transform of
$L(t)$, narrower magnetic resonance line. This is the \textit{motional
narrowing} phenomenon in magnetic resonance spectroscopy
\cite{AndersonJPSJ1954,AbragamBook1961}, where the random motion of atoms
makes the magnetic noise fluctuate rapidly and hence reduces the width of the
magnetic resonance line of the central spin.
\end{enumerate}

On sufficiently short time scales, any noise with a hard high-frequency cutoff
becomes static and the small random phase can be treated up to the second
order to give Gaussian inhomogeneous dephasing $L_{\mathrm{inh}}%
(t)=e^{-(b_{\mathrm{rms}}t)^{2}/2}$ [cf. Eq. (\ref{L_T2STAR})]. However, the
entire dephasing profile over the time scale $\sim T_{2}$ depends on the
specific statistics and auto-correlation of the noise.

If the noise is Gaussian, then the dephasing can be obtained from Eq.
(\ref{EXP_GAU}) as \cite{AndersonRMP1953}
\[
L(t)=e^{-\langle\tilde{\varphi}^{2}(t)\rangle/2}.
\]
According to the discussions following Eq. (\ref{PHI_RMS_CLASSICAL}), quasi-static
noise $(b_{\mathrm{rms}}\tau_{c}\gg1$) gives Gaussian inhomogeneous dephasing
$L_{\mathrm{inh}}(t)=e^{-(t/T_{2}^{\ast})^{2}}$ on a short time scale
$T_{2}^{\ast}\sim\sqrt{2}/b_{\mathrm{rms}}\ll\tau_{c}$, consistent with Eq.
(\ref{L_T2STAR}). Markovian noise ($b_{\mathrm{rms}}\tau_{c}\ll1$)\ gives
exponential \textquotedblleft true\textquotedblright\ decoherence
$L(t)=e^{-t/T_{\varphi}}$ on a much longer time scale $T_{\varphi}%
\sim1/(b_{\mathrm{rms}}^{2}\tau_{c})\gg\tau_{c}$, consistent with Eq.
(\ref{GAMMA_PHI}). In the intermediate regime, the dephasing profile depends
sensitively on the noise spectrum, e.g., the spectrum of the
Ornstein-Uhlenbeck noise in Eqs. (\ref{OU_SPECTRUM}) gives
\[
L(t)=\exp(-b_{\mathrm{rms}}^{2}\tau_{c}t+b_{\mathrm{rms}}^{2}\tau_{c}%
^{2}(1-e^{-|t|/\tau_{c}})),
\]
which reduces to the exponential decoherence with $T_{\varphi}%
=1/(b_{\mathrm{rms}}^{2}\tau_{c})$ for $t\gg\tau_{c}$ and the Gaussian
inhomogeneous dephasing with $T_{2}^{\ast}=\sqrt{2}/b_{\mathrm{rms}}$ for
$t\ll\tau_{c}$.

\section{Semi-classical noise theory of dynamical decoupling}

\label{SEC_DD_FILTER}

Dynamical decoupling (DD) is a powerful approach to suppressing the central
spin decoherence. The key idea is to dynamically average out the coupling of
the central spin to the environment by frequently flipping the central spin.
The DD approach originated from the Hahn echo in nuclear magnetic resonance
\cite{HahnPR1950} and was later developed for high-precision magnetic
resonance spectroscopy \cite{MehringBook1983,RhimPRL1970,Haeberlen1976}. Then
the idea of DD was introduced in quantum computing
\cite{ViolaPRA1998,BanJMO1998,ZanardiPLA1999,ViolaPRL1999}, which stimulated
numerous studies on applications and extensions to suppressing qubit
decoherence for quantum computing (see Ref. \cite{YangFP2011} for a review).

DD can efficiently suppress decoherence when the DD induced central spin flip
is much faster than the auto-correlation time $\tau_{c}$ of the environmental
noise, so that the lost coherence can be retrieved before it is dissipated
irreversibly in the environment. According to Sec. \ref{SEC_NOISE_T1}, spin
relaxation is usually dominated by Markovian noise with $\tau_{c}%
\lesssim1/\omega_{0}$, while flipping the central spin usually requires a
duration $\gtrsim1/\omega_{0}$, thus DD is inefficient for suppressing spin
relaxation. As discussed in Sec. \ref{SEC_NOISE_T2}, pure dephasing is usually
dominated by non-Markovian noises and especially static noise, so DD is
efficient for combating pure dephasing. Therefore, we only consider pure
dephasing in this section.

In a general $N$-pulse DD scheme, the $N$ instantaneous $\pi$-pulses are
applied successively at $\tau_{1}<\tau_{2}<\cdots<\tau_{N}$ to induce the flip
between $|\uparrow\rangle$ and $|\downarrow\rangle$ and the central spin is
measured at a later time $t_{\mathrm{d}}$. In the Schr\"{o}dinger picture, the
central spin Hamiltonian consists of the external field term $\hat{H}_{0}$
[Eq. (\ref{H0})], the DD control term
\begin{equation}
\hat{H}_{c}(t)=\sum_{n=1}^{N}\pi\delta(t-\tau_{n})\hat{S}_{x}, \label{HCT}%
\end{equation}
and the noise term $\tilde{b}(t)\hat{S}_{z}$. A convenient way is to work in
the interaction\ picture with respect to $\hat{H}_{0}+\hat{H}_{c}(t)$, where
the central spin Hamiltonian is [cf. Eq. (\ref{HAMIL_CLASSICAL})]%
\begin{equation}
\hat{H}(t)=s(t)\tilde{b}(t)\hat{S}_{z} \label{HT_SDD}%
\end{equation}
and $s(t)$ is the DD modulation function:\ it starts from $s(0)=+1$ and
changes its sign every time the central spin is flipped by a $\pi$-pulse,
i.e., each $\pi$-pulse in the DD switches the sign of the environmental noise.
The spin decoherence in the absence of any control is called free-induction
decay (FID), which corresponds to a constant modulation function
$s(t)\equiv+1$.

Intuitively, when the sign switch by DD is more frequent than the fluctuation
of the noise $\tilde{b}(t)$ ($\tau_{c}>$ pulse interval), DD could effectively
speed up the noise fluctuation and suppress dephasing efficiently (reminiscent
of motional narrowing). On the other hand, when the sign switch coincides with
the characteristic fluctuation of a noise, DD could resonantly enhance the
effect of the noise $\tilde{b}(t)$, causing rapid decoherence. Below we
discuss two important cases: static noises and and Gaussian noises.

If the noise is static during each measurement cycle $[0,t_{\mathrm{d}}]$,
then $\tilde{\varphi}(t_{\mathrm{d}})=\tilde{b}\int_{0}^{t_{\mathrm{d}}%
}s(t)dt$ vanishes when $t_{\mathrm{d}}$ satisfies the echo condition:
\begin{equation}
\int_{0}^{t_{\mathrm{d}}}s(t)dt=0. \label{PHASE_FOCUS}%
\end{equation}
This means that a static noise can be completely eliminated at the echo time
$t_{\mathrm{d}}$. The simplest DD scheme is Hahn echo \cite{HahnPR1950}, where
a $\pi$-pulse is applied at $\tau$ followed by a measurement at $t_{\mathrm{d}%
}=2\tau$.

\begin{figure}[ptb]
\includegraphics[width=\columnwidth,clip]{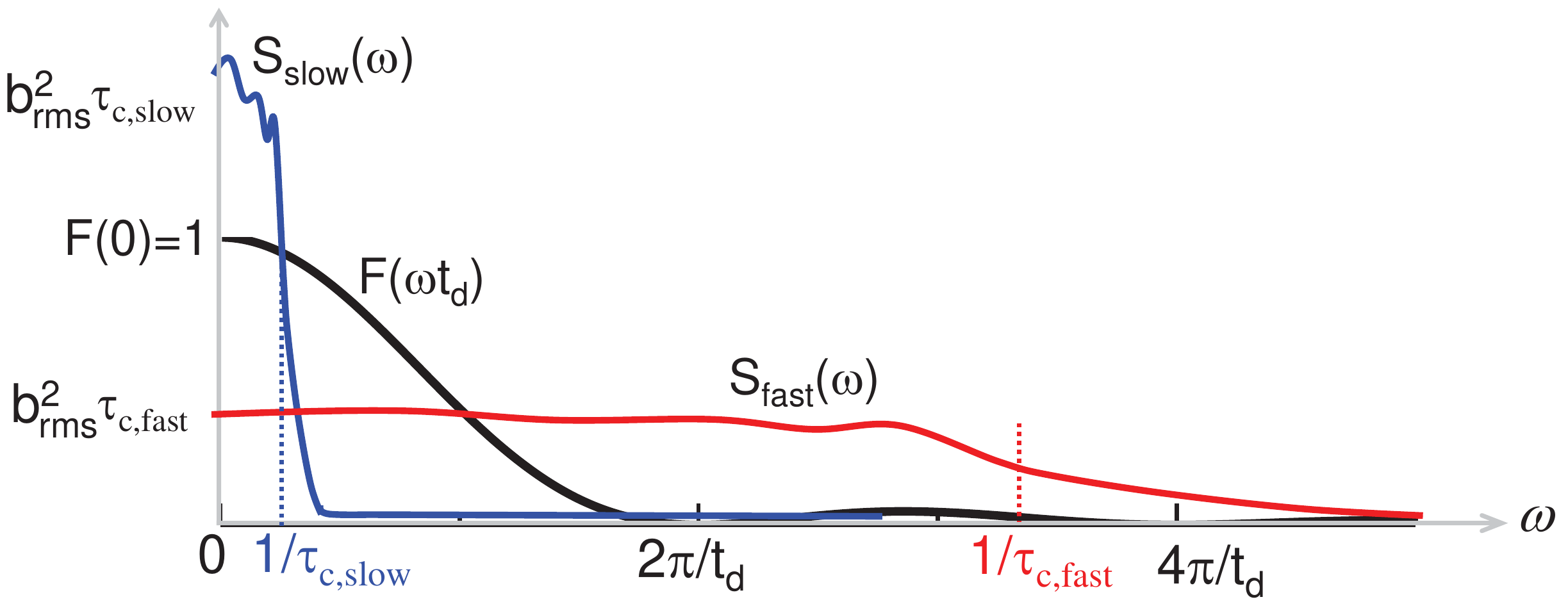}
\caption{Noise
filter for FID and spectra of slowly fluctuating (blue line, $\tau
_{c,\mathrm{slow}}\gg t_{\mathrm{d}}$) and fast fluctuating (red line,
$\tau_{c,\mathrm{fast}}\ll t_{\mathrm{d}}$) noises.}%
\label{G_FILTER_FID}%
\end{figure}

For Gaussian noises, the central spin dephasing is completely determined by
the noise auto-correlation:
\begin{equation}
L(t_{\mathrm{d}})=e^{-\langle\tilde{\varphi}^{2}(t_{\mathrm{d}})\rangle/2},
\label{LTD}%
\end{equation}
where \cite{CywinskiPRB2008}
\begin{equation}
\langle\tilde{\varphi}^{2}(t_{\mathrm{d}})\rangle=t_{\mathrm{d}}^{2}%
\int_{-\infty}^{\infty}S(\omega)F(\omega t_{\mathrm{d}})\frac{d\omega}{2\pi}
\label{PHIDD_RMS_CLASSICAL}%
\end{equation}
is determined by the overlap integral of the noise spectrum $S(\omega
)\equiv\int_{-\infty}^{\infty}\langle\tilde{b}(t)\tilde{b}(0)\rangle
e^{i\omega t}dt=S(-\omega)$ and the dimensionless noise filter
\begin{equation}
F(\omega t_{\mathrm{d}})\equiv\frac{1}{t_{\mathrm{d}}^{2}}\left\vert \int
_{0}^{t_{\mathrm{d}}}s(t)e^{-i\omega t}dt\right\vert ^{2}=F(-\omega
t_{\mathrm{d}}), \label{FDD}%
\end{equation}
which is related to the fourier transform of the DD modulation function $s(t)$
and obeys $F(\omega t_{\mathrm{d}})\leq1$ as well as the normalization
$\int_{-\infty}^{\infty}F(\omega t_{\mathrm{d}})d\omega=2\pi/t_{\mathrm{d}}$.

This noise filter formalism \cite{CywinskiPRB2008} provides a physically
transparent undertanding of dephasing caused by Gaussian noise and its control
by DD in the frequency domain, e.g., coherence protection can be achieved by
designing the noise filter to minimize the overlap integral in Eq.
(\ref{PHIDD_RMS_CLASSICAL}).

\begin{figure}[ptb]
\includegraphics[width=\columnwidth,clip]{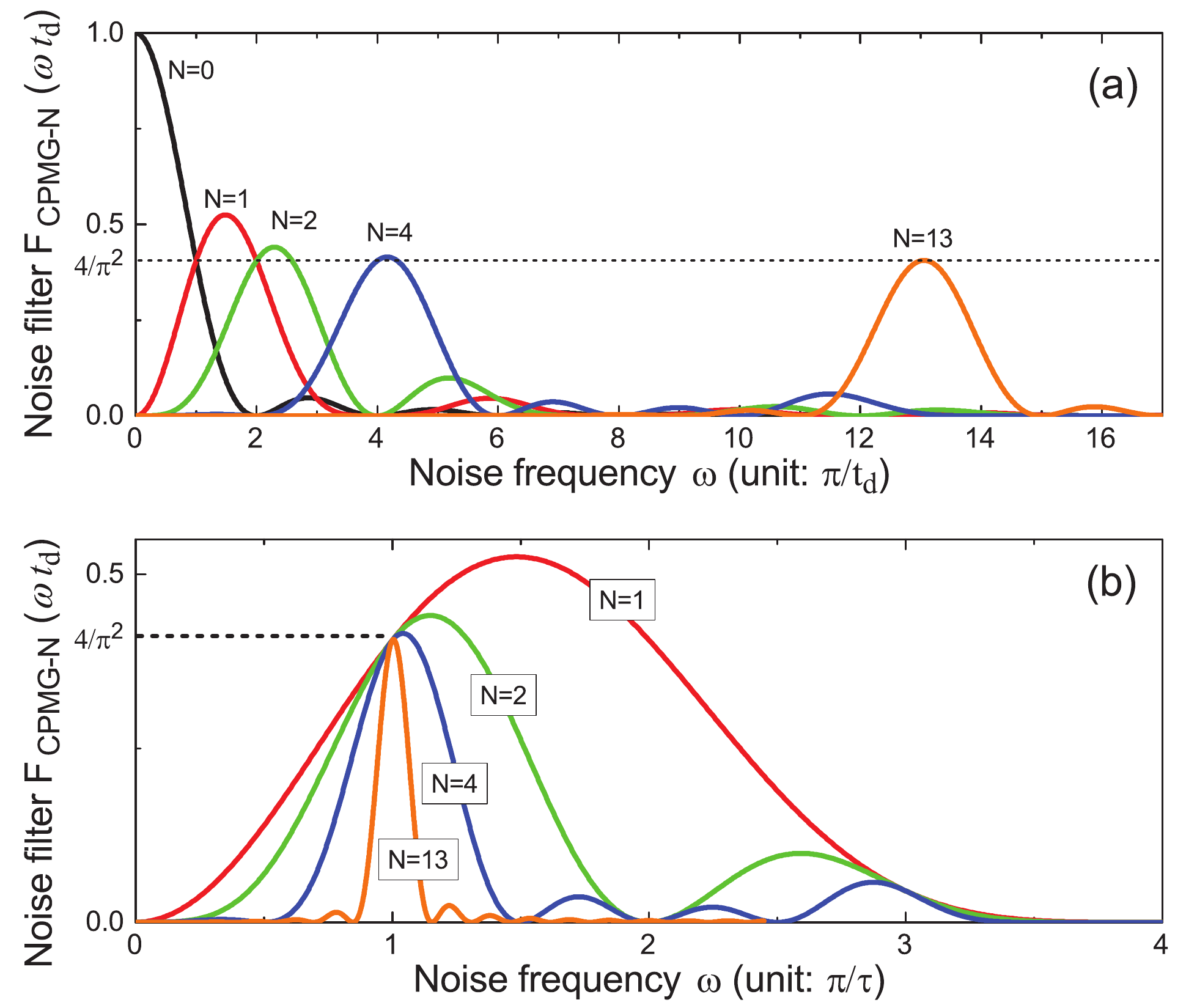}
\caption{(a) Noise
filter functions for FID $(N=0)$ and CPMG-$N$ sequence for fixed total
evolution time $t_{\mathrm{d}}$. (b) CPMG-$N$ filter function for fixed pulse
interval $\tau$.}%
\label{G_DD_FILTER}%
\end{figure}

For FID, the filter
\begin{equation}
F_{\mathrm{FID}}(\omega t_{\mathrm{d}})=\frac{\sin^{2}(\omega t_{\mathrm{d}%
}/2)}{(\omega t_{\mathrm{d}}/2)^{2}}\equiv\operatorname{sinc}^{2}\frac{\omega
t_{\mathrm{d}}}{2} \label{FILTER_FID}%
\end{equation}
passes\ low-frequency noises ($\omega\lesssim\pi/t_{\mathrm{d}}$) but
attenuates high frequency noises ($\omega\gtrsim\pi/t_{\mathrm{d}}$) (black
solid line in Fig. \ref{G_FILTER_FID}), i.e., low frequency noises are most
effective in causing pure dephasing. For quasi-static noise ($\tau_{c}\gg
t_{\mathrm{d}}$), the noise spectrum (blue line in Fig. \ref{G_FILTER_FID}) is
well within the low-pass regime of the filter (black line in Fig.
\ref{G_FILTER_FID}), so all noise power passes, leading to rapid inhomogeneous
dephasing [Eq. (\ref{L_T2STAR})]. For Markovian noise ($\tau_{c}\ll
t_{\mathrm{d}}$), the noise spectrum is broad (red line in Fig.
\ref{G_FILTER_FID}) and remains nearly a constant $S(\omega)\approx\bar{S}\sim
b_{\mathrm{rms}}^{2}\tau_{c}$ within the low-pass regime, so $\langle\tilde{\varphi}^{2}(t_{\mathrm{d}})\rangle\approx\bar{S}t_{\mathrm{d}}$ leads to exponential dephasing on a time
scale $\sim1/\bar{S}\sim1/(b_{\mathrm{rms}}^{2}\tau_{c})$ $\gg$ inhomogeneous
dephasing time.

A particularly interesting DD sequence is the $N$-pulse Carr$-$Purcell$-$%
Meiboom$-$Gill (CPMG-$N$) \cite{CarrPR1954,MeiboomRSI1958} consisting of $N$
instantaneous $\pi$-pulses applied at $\tau_{n}=t_{\mathrm{d}}(n-1/2)/N$
$(n=1,2,\cdots,N)$, respectively. The filter for CPMG-$N$ control is%
\[
F_{\mathrm{CPMG-N}}(\omega t_{\mathrm{d}})=2\frac{\sin^{4}\frac{\omega
t_{\mathrm{d}}}{4N}}{\cos^{2}\frac{\omega t_{\mathrm{d}}}{2N}}\frac{1\mp
\cos(\omega t_{\mathrm{d}})}{(\omega t_{\mathrm{d}}/2)^{2}}%
\]
(upper sign for even $N$ and lower sign for odd $N$), which, for $N\gg 1$, has a primary peak
at $\omega=N\pi/t_{\mathrm{d}}=\pi/\tau$ ($\tau\equiv t_{\mathrm{d}}/N$ is the
pulse interval$)$ and a bandwidth $\sim\pi/t_{\mathrm{d}}$ (see Fig.
\ref{G_DD_FILTER}). Near this peak,
\[
F_{\mathrm{CPMG-N}}(\omega t_{\mathrm{d}})\approx\frac{4}{\pi^{2}%
}e^{-t_{\mathrm{d}}^{2}(\omega-N\pi/t_{\mathrm{d}})^{2}/12}=\frac{4}{\pi^{2}%
}e^{-N^{2}\tau^{2}(\omega-\pi/\tau)^{2}/12}.
\]
As mentioned before, for the DD to be efficient, the pulses must be applied
faster than the noise auto-correlation time ($\tau<\tau_{c}$). For CPMG-$N$,
this is equivalent to that the filter's peak frequency $\pi/\tau>$ noise
cutoff frequency $1/\tau_{c}$. For Markovian noise with $\tau_{c}\ll\tau$, the
noise spectrum is nearly constant over the entire band-pass window of the
filter, so DD has no effect.

\section{Quantum noise versus classical noise}

\label{SEC_QUANTUM_NOISE}

In the semiclassical theory of central spin decoherence, the central spin is
treated as a quantum object, while the spin bath is approximated by a
classical noise. In spin-based solid-state quantum technologies, the nanoscale
spin bath is also a quantum object and requires a quantum description. Within
the characteristic time scale of central spin decoherence, the central spin
and the spin bath can be regarded as a closed quantum system (see Fig.
\ref{G_SPINBATH}). Here we are only interested in the most relevant mechanism
for electron spin decoherence in nuclear spin baths:\ pure dephasing, i.e., we
assume that the central spin transition frequency $\omega_{0}$ is far beyond
the high-frequency cutoff of the bath noise spectrum. In this case, the
central spin and the bath are described by a general pure dephasing
Hamiltonian \cite{YaoPRB2006,LiuNJP2007,YaoPRL2007}%
\begin{equation}
\hat{H}=\hat{H}_{B}+\hat{b}\hat{S}_{z} \label{HAMIL1}%
\end{equation}
in the interaction picture with respect to $\hat{H}_{0}$ [Eq. (\ref{H0})],
where $\hat{H}_{B}$ is the bath Hamiltonian and $\hat{b}$ is the bath noise
operator coupled to the central spin.

Below we will classify the noises from the spin bath into two categories
according to their natures, namely, static thermal noises and dynamical
quantum noises. It should be noted that the noises from the ``rest of
universe'' (Fig. \ref{G_SPINBATH}), which is taken as classical, can be static
or dynamical. Ultimately, all noises have a quantum origin (e.g., the thermal
distribution of a spin bath can be ascribed to entanglement between the bath
and the rest of universe). Here the static thermal noise and the dynamical
quantum noise are differentiated in the sense that the spin bath and the
central spin are regarded as a closed quantum system in the timescale of interest.

\subsection{Static thermal noises}

\label{SEC_THERMAL_NOISE}

The initial state of the bath is the maximally mixed thermal state (relevant for
nuclear spin baths):
\begin{equation}
\hat{\rho}_{B}^{\mathrm{eq}}=\frac{\hat{I}}{\operatorname*{Tr}\hat{I}}%
=\sum_{J}P_{J}|J\rangle\langle J|, \label{RHOB}%
\end{equation}
When $[\hat{b},\hat{H}_{B}]=0$, $\{|J\rangle\}$ can be chosen as the common
eigenstates of $\hat{b}$ and $\hat{H}_{B}$. If the initial state of the bath were a
pure state $|J\rangle$, then it would remain in $|J\rangle$, and during the
measurement cycle the central spin would evolve under a constant noise field
$b_{J}=\langle J|\hat{b}|J\rangle$ from $\hat{\rho}(0)$ to $\hat{\rho}_{J}(t)\equiv e^{-ib_{J}\hat{S}%
_{z}t}\hat{\rho}(0)e^{ib_{J}\hat{S}_{z}t}$ with an oscillating off-diagonal
coherence $L(t)=e^{-ib_{J}t}$, while the coupled system would evolve as%
\[
\hat{\rho}(0)\otimes|J\rangle\langle J|\overset{\mathrm{evolution}%
}{\longrightarrow}\hat{\rho}_{J}(t)\otimes|J\rangle\langle J|.
\]
The ensemble average over the thermal distribution in Eq. (\ref{RHOB}) gives the
evolution%
\[
\hat{\rho}(0)\otimes\sum_{J}P_{J}|J\rangle\langle J|\overset
{\mathrm{evolution}}{\longrightarrow}\sum_{J}\hat{\rho}_{J}(t)\otimes
P_{J}|J\rangle\langle J|
\]
which coincides with the decoherence induced by a static noise with the
distribution $P_{\mathrm{inh}}(b)\equiv\sum_{J}P_{J}\delta(b-b_{J})$ (see Eq.
(\ref{LT_INH}) of Sec. \ref{SEC_T2STAR}). In this sense, the thermal noise
(caused by the thermal distribution of the bath states) amounts to
inhomogeneous dephasing. The static thermal noise usually dominates the FID of
central spin coherence, but it can be completely removed by DD at the echo time.

\subsection{Dynamical quantum noises}

\label{SEC_QUANTUM_DYN}

When $[\hat{H},\hat{b}]\neq0$, the eigenstate of the noise operator $\hat{b}$
is not necessarily the eigenstate of $\hat{H}_{B}$. Thus even if the bath is
initially in an eigenstate of $\hat{b}$, the intrinsic bath Hamiltonian
$\hat{H}_{B}$ would drive the bath into different eigenstates, producing a
dynamical noise on the central spin. This noise is best described in the
interaction picture of the bath, where\ the total Hamiltonian
\begin{equation}
\hat{H}(t)=\hat{b}(t)\hat{S}_{z} \label{H_QUANTUM}%
\end{equation}
with the noise operator in the interaction picture, $\hat{b}(t)\equiv
e^{i\hat{H}_{B}t}\hat{b}e^{-i\hat{H}_{B}t}$, being the quantum analog of the
classical noise $\tilde{b}(t)$. Note that if $[\hat{H},\hat{b}]=0$, then
$\hat{b}(t)$ would have no time depehence. Thus the dynamical nature of the
noise is ascribed to the quantum nature of the bath \footnote{Here the central
spin and the spin bath forms a close system, so the thermal noise from the
spin bath is static and the quantum noise from the spin bath is dynamical.
Noises from other environments that are not explicitly included in our model
[Eq. (\ref{HAMIL1})] can also be dynamical and often treated as classical.}.
The quantum noise $\hat{b}(t)$ at different times, in contrast to the
classical noise, does not commute in general. So the decoherence of the
central spin,
\[
L(t)=\langle(\mathcal{\bar{T}}e^{-(i/2)\int_{0}^{t}\hat{b}(t^{\prime
})dt^{\prime}})(\mathcal{T}e^{-(i/2)\int_{0}^{t}\hat{b}(t^{\prime})dt^{\prime
}})\rangle
\]
involves the time-ordering (anti-time-ordering) superoperator $\mathcal{T}$
($\mathcal{\bar{T}}$). For a large many-body bath, the effect of the dynamical noise
is similar for most initial states $|J\rangle$. Therefore the central spin
coherence can be approximated as
\begin{equation}
L(t)\approx L_{\mathrm{inh}}(t)L_{\mathrm{dyn}}(t) \label{FID_FACTORIZE}%
\end{equation}
up to a global phase factor, i.e., the decoherence can be separated into the
effect of the static thermal noise, i.e., $L_{\mathrm{inh}}(t)$ in Eq.
(\ref{LT_INH}), and that due to the dynamical quantum noise (the
\textquotedblleft true\textquotedblright\ decoherence), i.e.,
\begin{equation}
L_{\mathrm{dyn}}(t)=\langle J|(\mathcal{\bar{T}}e^{-(i/2)\int_{0}^{t}\hat
{b}(t^{\prime})dt^{\prime}})(\mathcal{T}e^{-(i/2)\int_{0}^{t}\hat{b}%
(t^{\prime})dt^{\prime}})|J\rangle. \label{LDYNT}%
\end{equation}
Note that $|L_{\mathrm{dyn}}(t)|$ is similar for most initial
states $|J\rangle$ of a large many-body bath.

In the presence of DD control Hamiltonian $\hat{H}_{c}(t)$ [Eq. (\ref{HCT})],
we can work in the interaction picture with respect to $\hat{H}_{0}+\hat
{H}_{c}(t)+\hat{H}_{B}$, where the total Hamiltonian is%
\begin{equation}
\hat{H}(t)=s(t)\hat{b}(t)\hat{S}_{z}. \label{HDD_QUANTUM}%
\end{equation}
At the echo time, the static thermal noise is completely removed, so the
central spin undergoes \textquotedblleft true\textquotedblright\ decoherence
due to the dynamical quantum noise:%
\begin{align}
L(t_{\mathrm{d}})  &  =\langle(\mathcal{\bar{T}}e^{-(i/2)\int_{0}%
^{t_{\mathrm{d}}}s(t)\hat{b}(t)dt})(\mathcal{T}e^{-(i/2)\int_{0}%
^{t_{\mathrm{d}}}s(t)\hat{b}(t)dt})\rangle\\
&  \approx\langle J|(\mathcal{\bar{T}}e^{-(i/2)\int_{0}^{t_{\mathrm{d}}%
}s(t)\hat{b}(t)dt})(\mathcal{T}e^{-(i/2)\int_{0}^{t_{\mathrm{d}}}s(t)\hat
{b}(t)dt})|J\rangle,\label{EQ39}
\end{align}
where the second line is similar for most initial states $|J\rangle$.

\subsection{Quantum Gaussian noises}

\label{SEC_QUAN_GAUSS} A close analogy to the classical noise model is possible when
the commutator $[\hat{b}(t_{1}),\hat{b}(t_{2})]$ is a c-number, so that
$\mathcal{T}$ and $\mathcal{\bar{T}}$ play no role up to a phase factor. This
happens when the bath state $\hat{\rho}_{B}^{\mathrm{eq}}$ can be mapped to a
non-interacting bosonic state and $\hat{b}(t)$ can be mapped to a bosonic
field operator (i.e., a linear combination of creation and annihilation
operators), so that the quantum noise is Gaussian. In this case, the
off-diagonal coherence assumes exactly the same form as Eq. (\ref{L_CLASSICAL}%
) for classical Gaussian noise:%
\[
L(t_{\mathrm{d}})=\langle e^{-i\hat{\varphi}(t_{\mathrm{d}})}\rangle,
\]
where%
\[
\hat{\varphi}(t_{\mathrm{d}})\equiv\int_{0}^{t_{\mathrm{d}}}s(t)\hat{b}(t)dt
\]
is the quantum analog to the classical random phase $\tilde{\varphi
}(t_{\mathrm{d}})$. Using linked-cluster expansion for non-interacting bosons
(see Sec. \ref{SEC_LCE_RDT}) and assuming $\langle\hat{b}(t)\rangle=0$ (just
for simplicity) gives an \textit{exact} result%
\begin{align}
L(t_{\mathrm{d}})  &  =e^{-\langle\hat{\varphi}^{2}(t_{\mathrm{d}})\rangle/2
},\label{LCE1}\\
\langle\hat{\varphi}^{2}(t_{\mathrm{d}})\rangle &  =\int_{0}^{t_{\mathrm{d}}%
}dt_{1}\int_{0}^{t_{\mathrm{d}}}dt_{2}\ s(t_{1})s(t_{2})\langle\{\hat{b}%
(t_{1}),\hat{b}(t_{2})\}/2\rangle.\nonumber
\end{align}
The above equation has exactly the same form as the classical case [Eq.
(\ref{LTD})].


The quantum Gaussian noise is best illustrated in the spin-boson model
\cite{UhrigPRL2007}, in which the central spin is linearly coupled to a
collection of non-interaction bosonic modes $\{\hat{b}_{m}\}$ in thermal
equilibrium, corresponding to $\hat{H}_{B}=\sum_{m}\omega_{m}\hat{b}%
_{m}^{\dagger}\hat{b}_{m}$ and $\hat{b}=\sum_{m}\lambda_{m}(\hat{b}%
_{m}^{\dagger}+\hat{b}_{m})$. Under DD control, the total Hamiltonian in the
interaction picture assumes the standard form (Eq. (\ref{HDD_QUANTUM})), where
the quantum noise
\[
\hat{b}(t)=\sum_{m}\lambda_{m}(\hat{b}_{m}^{\dagger}e^{i\omega_{m}t}+\hat
{b}_{m}e^{-i\omega_{m}t})
\]
is Gaussian. The quantum noise spectrum as the fourier transform of
$\langle\{\hat{b}(t_{1}),\hat{b}(t_{2})\}/2\rangle$ are readily obtained as%
\[
S(\omega)=2\pi\sum_{m}\lambda_{m}^{2}[\bar{n}(\omega_{m})+1/2][\delta
(\omega+\omega_{m})+\delta(\omega-\omega_{m})],
\]
where $\bar{n}(\omega)=1/(e^{\beta\omega}-1)$ is the Bose-Einstein
distribution. The exact central spin dephasing is obtained by substituting
this spectrum into the noise filter formalism (Eqs. (\ref{LTD}) and
(\ref{PHIDD_RMS_CLASSICAL})).

\subsection{Can quantum baths be simulated by classical noises?}

The key difference between classical noises and quantum noises is that the
former commutes at different times, while the latter does not. This means that
the action of $\hat{b}(t)$ at an earlier time changes its action on the bath
evolution at a later time. By contrast, in the classical model [Eqs.
(\ref{PHI_CLASSICAL}) and (\ref{L_CLASSICAL})], only the integral of the
classical noise matters, i.e., the classical noise at different times do not
influence each other. In the presence of DD control, we need to replace
$\hat{b}(t)$ with $s(t)\hat{b}(t)$. Therefore, the sign switch of $\hat{b}(t)$
due to a DD pulse at an earlier time may change the action of $\hat{b}(t)$ at
a later time, i.e., controlling the central spin may change the quantum noise
itself. This is the so-called quantum back-action from the central spin
\cite{ZhaoPRL2011,HuangNatCommun2011,MaPRB2015}: the evolution of the quantum bath conditioned
on the central spin state (see Sec. \ref{SEC_QUANTUM_PICTURE} for details)
governs the quantum noise.

It is desirable to simulate quantum baths (or equivalently quantum noises)
with classical noises. First, computing central spin decoherence caused by a
quantum bath requires a large amount of numerical simulations of the many-body
dynamics of the bath, while computing the decoherence caused by classical
noises, especially classical Gaussian noise, is much simpler. Second,
controlling the central spin does not change the classical noise, so the noise
filter formalism of DD allows efficient reconstruction of the classical noise
\cite{AlvarezPRL2011,BarGillNatCommun2012,BylanderNatPhys2011,CywinskiPRA2014,MuhonenNatNano2014}%
, which in turn can be used to efficiently design optimal quantum control to
suppress the central spin decoherence. By contrast, controlling the central
spin can actively change the quantum noise itself. On the one hand, this
provides more flexibility in engineering the quantum noise. On the other hand,
this makes it impossible to describe the quantum noise without referring to
the control over the central spin.

The question is under what circumstances can a quantum bath be approximated by
a classical noise, i.e., given a central spin in a quantum bath, is it
possible to find a classical noise (Gaussian or non-Gaussian) that is capable
of faithfully reproducing the decoherence of the central spin under all
classical controls (not necessarily DD)? The answer to this general question
is still absent due to the existence of a diverse range of classical noises
and controls. Here we restrict ourselves to a simpler question: is it possible
to find a Gaussian noise to faithfully reproduce the decoherence of the
central spin under all possible classical controls? According to Sec.
\ref{SEC_QUAN_GAUSS}, this is possible when the quantum noise is Gaussian,
i.e., when the state of the bath can be mapped to a noninteracting bosonic
state and the quantum noise can be mapped to a bosonic field operator (i.e., a
linear combination of creation and annihilation operators) such as the
spin-boson model in Sec. \ref{SEC_QUAN_GAUSS}. Actually, according to Eq.
(\ref{LCE1}), if a quantum noise is Gaussian, it is equivalent to a classical
noise that has the same noise spectrum. Therefore, the question of
approximating a quantum bath as a classical Gaussian noise is equivalent to
the question about Gaussian nature of the quantum bath.

\subsubsection{One-spin bath}

\label{SEC_GENERAL_CONCLUSION}

To illustrate the condition for the Gaussian noise approximation to be valid,
let us first consider the simplest spin ``bath'', namely, a bath that has only
one spin-1/2 $\mathbf{\hat{I}}_{m}$. Without loss of generality we assume the
bath Hamiltonian $\hat{H}_{B}=\omega_{m}\hat{I}_{m}^{z}$ and the noise
operator as $\hat{b}=2\lambda_{m}\hat{I}_{m}^{x}$ (therefore the bath causes a
dynamical quantum noise as that in Sec. \ref{SEC_QUANTUM_DYN}). The initial
state of the bath is taken as the spin-down eigenstate of its intrinsic
Hamiltonian. Under either the \textit{short-time condition} $|\lambda
_{m}|t_{\mathrm{d}}\ll1$ or \textit{off-resonant condition} $|\lambda_{m}%
|\ll|\omega_{m}|$, the coupling to the central spin only weakly perturbs the
bath, so we can map the initial state of the bath into the vacuum state
$|0\rangle_{m}$ of a Holstein-Primakoff boson mode $\{\hat{b}_{m},\hat{b}%
_{m}^{\dagger}\}$:
\begin{align}
&  \hat{I}_{m}^{-}=\left(\sqrt{1-\hat{b}_{m}^{\dagger}\hat{b}_{m}}\right)\hat{b}_{m}%
\approx\hat{b}_{m},\\
&  \hat{I}_{m}^{z}=\hat{b}_{m}^{\dagger}\hat{b}_{m}-1/2.
\end{align}
Then we have $\hat{H}_{B}=\omega_{m}\hat{b}_{m}^{\dagger}\hat{b}_{m}%
-\omega_{m}/2$ and $\hat{b}\approx\lambda_{m}(\hat{b}_{m}+\hat{b}_{m}%
^{\dagger})$ and recover the single-mode version of the spin-boson model,
which has been discussed in Sec. \ref{SEC_QUAN_GAUSS}. Substituting the
quantum noise spectrum $S(\omega)=\pi\lambda_{m}^{2}[\delta(\omega+\omega
_{m})+\delta(\omega-\omega_{m})]$ into the noise filter formalism immediately
gives the central spin decoherence under Gaussian noise approximation:
\[
L_{\mathrm{Gau}}(t_{\mathrm{d}})=e^{-\lambda_{m}^{2}F(\omega_{m}t_{\mathrm{d}%
})t_{\mathrm{d}}^{2}/2},
\]
where $F(z)$ is the noise filter determined by the DD sequence. Under either
the short-time condition $|\lambda_{m}|t_{\mathrm{d}}\ll1$ or off-resonant
condition $|\lambda_{m}|\ll|\omega_{m}|$, the central spin decoherence caused
by this bath spin is small and the Gaussian approximation results indeed agree
well with the exact results, e.g., the FID
\begin{align*}
&  L_{\mathrm{Gau}}(t)=e^{-(2\lambda_{m}^{2}/\omega_{m}^{2})\sin^{2}%
(\omega_{m}t/2)},\\
&  L(t)=1-\frac{2\lambda_{m}^{2}}{\lambda_{m}^{2}+\omega_{m}^{2}}\sin^{2}%
\frac{\sqrt{\lambda_{m}^{2}+\omega_{m}^{2}}t}{2},
\end{align*}
and Hahn echo at $t_{\mathrm{d}}=2\tau$:
\begin{align*}
&  L_{\mathrm{Gau}}(2\tau)=e^{-(8\lambda_{m}^{2}/\omega_{m}^{2})\sin
^{4}(\omega_{m}\tau/2)},\\
&  L(2\tau)=1-\frac{8\lambda_{m}^{2}\omega_{m}^{2}}{(\lambda_{m}^{2}%
+\omega_{m}^{2})^{2}}\sin^{4}\frac{\sqrt{\lambda_{m}^{2}+\omega_{m}^{2}}\tau
}{2}.
\end{align*}

If the bath consists of many independent spin-1/2's, then we can map the
initial state of the $m$th bath spin into the vacuum state of the $m$th
Holstein-Primakoff boson mode and obtain the many-mode spin-boson model
discussed in Sec. \ref{SEC_QUAN_GAUSS}.

\subsubsection{Many-body bath}

A spin bath that has many-body interactions can be in general written as
$\hat{H}_{B}=\sum_{m}\varepsilon_{m}|m\rangle\langle m|$ and its initial state
can be taken as an eigenstate $|k\rangle$. Generally, the noise operator
$\hat{b}$ could induce the excitations $|k\rangle\rightarrow|m\rangle$ $(m\neq
k)$ with amplitudes $\lambda_{mk}\equiv\langle m|\hat{b}|k\rangle$ and energy
costs $\omega_{mk}\equiv\varepsilon_{m}-\varepsilon_{k}$. When all excitations
are off-resonant ($|\lambda_{mk}|\ll|\omega_{mk}|)$ or when the time is short
$|\lambda_{mk}|t_{\mathrm{d}}\ll1$, we can approximate the excitation by a
boson mode $\hat{b}_{mk}$ to obtain a spin-boson model, where $\hat{b}%
\approx\sum_{m(\neq k)}(\lambda_{mk}\hat{b}_{mk}^{\dagger}+h.c.)$ and $\hat
{H}_{B}\approx\varepsilon_{k}+\sum_{m(\neq k)}\omega_{mk}\hat{b}_{mk}%
^{\dagger}\hat{b}_{mk}$.

\subsubsection{Electronic and nuclear spin baths}

For a central electron spin in an electron spin bath, the central spin and the
bath spins are alike and are typically coupled together through magnetic
dipolar interactions. Thus the central spin decoherence caused by many bath
spins is usually much faster than the bath spin evolution caused by a single
central spin, i.e., within the time scale of the central spin decoherence, the short-time condition is satisfied and the quantum
noise from the electron spin bath can be approximated by classical Gaussian
noise. This has been confirmed by many theoretical and experimental studies
\cite{HansonScience2008,WitzelPRB2012,DobrovitskiPRL2009,LangeScience2010,WangPRB2013}%
, where the noise spectrum obtained by fitting the central spin decoherence
under different DD controls agrees with a widely used classical Gaussian
noise: the Ornstein-Uhlenbeck noise [Eqs. (\ref{OU_CORRELATION}) and
(\ref{OU_SPECTRUM})]. Witzel \textit{et al. }\cite{WitzelPRB2014} further
demonstrates that the spectrum of the quantum noise directly calculated from
the quantum many-body theory (see Sec. \ref{SEC_CCE_CORRELATION}) agrees
reasonably with the Ornstein-Uhlenbeck noise and can well describe the central
spin decoherence under various DD control, unless a few bath spins are
strongly coupled to the central spin. In that case, the quantum noise is
dominated by a few strongly coupled bath spins and cannot be approximated as
classical Gaussian noise.

For a central electron spin in a nuclear spin bath, the hyperfine interaction
(HFI) between the bath spin and the central spin is much stronger than the
magnetic dipolar interaction between nuclear spins, but could be weaker than
the Zeeman splitting of individual nuclear spins under a strong magnetic field
(see Sec. \ref{SEC_SYSTEM} for various interactions in paradigmatic physical
systems). In other words, the off-resonant condition could be satisfied for
the evolution of individual nuclear spins, but does not for the evolution of
nuclear spin clusters. Two situations have been found where the nuclear spin
bath can be approximated by classical Gaussian noise:

\leftmargini=3mm

\begin{figure}[ptb]
\includegraphics[width=\columnwidth]{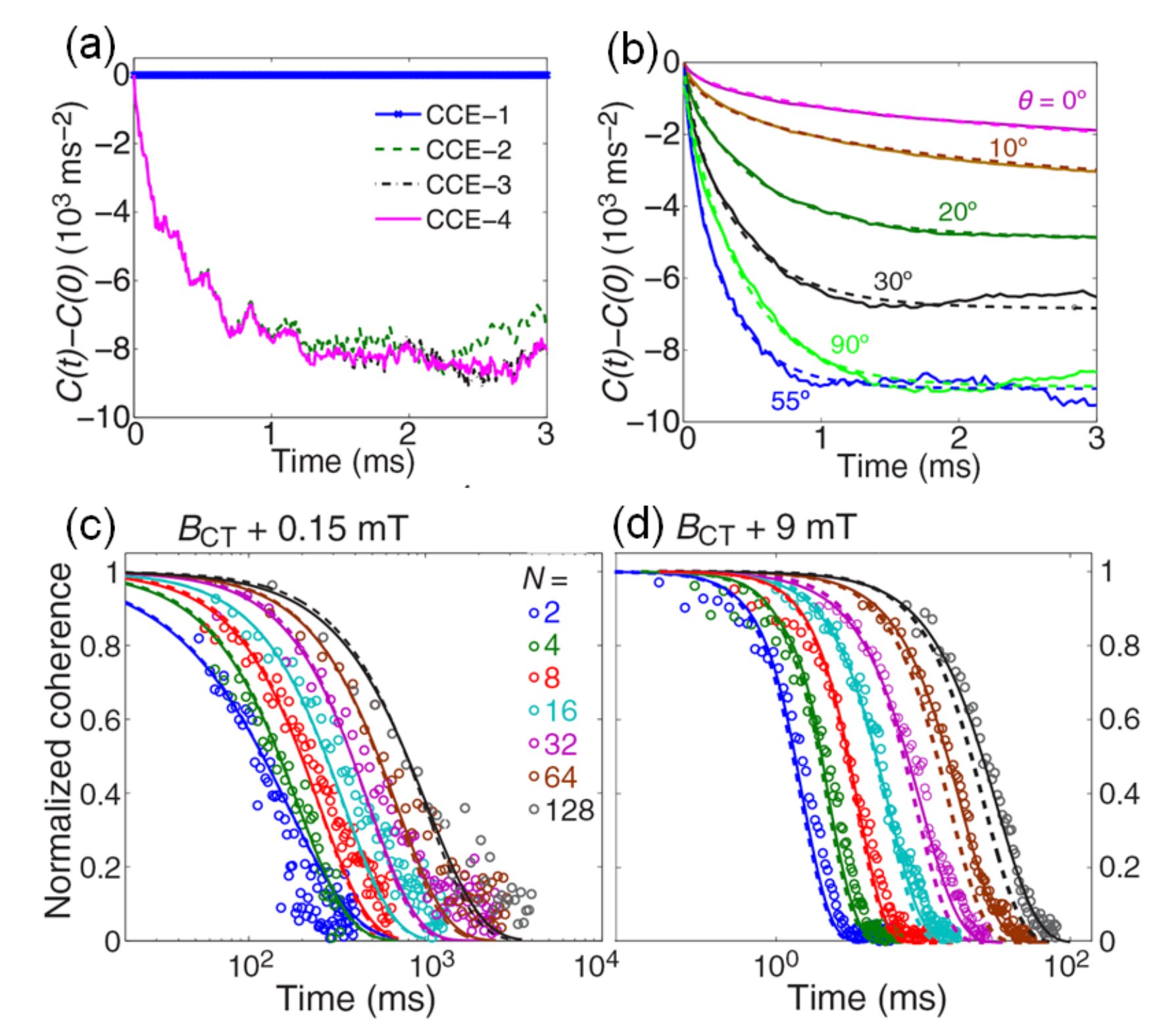}
 \caption{ Experimental
test of Gaussian noise model in $^{\mathrm{nat}}$Si:Bi system. (a) Relative
auto-correlation function $C(t)-C(0)$ of the $^{29}$Si nuclear spin bath at
the ``clock" transtion of bisumth donors in silicon ($B_{\mathrm{CT}}=79.9$
mT) calculated by the CCE method (CCE-$M$ denotes the $M$th-order CCE
truncation by keeping cluster correlations up to a certain size $M$, see Sec.
\ref{SEC_CCE_CORRELATION} for details). Here a specific nuclear spin
configuration is chosen with the external magnetic field $\mathbf{B}%
\parallel[110]$. (b) $C(t)-C(0)$ (solid lines) at the ``clock" transtion for
several magnetic field orientations in the $[001]-[110]$ plane with
$\theta=0^{\circ}$ corresponding to [001]. Results are obtained by averaging
over 50 different nuclear spin configurations. Dashed lines are fits of the
form $\Delta^{2}\{\text{exp}[-(|t|/\tau)^{n}]-1\}$. (c,d) Comparisons of
electron spin decoherence obtained by the quantum model (solid lines), the
semiclassical model (dashed lines), and the experimental measurement (circles)
for the magnetic fields near the ``clock" transtion. Here $N$=2, 4, 8, 16, 32,
64, 128 corresponds to the DD control CPMG-2, XY-4, XY-8, XY-16,
(XY-16)$\times$2, (XY-16)$\times$4, (XY-16)$\times$8. In theoretical
calculations, CPMG-$N$ is equivalent to XY-DD. Extracted from Refs.
\cite{MaPRB2015}. }%
\label{GaussTest}%
\end{figure}

\begin{enumerate}
\item Anisotropic HFI [Eq. (\ref{DHFI})] and intermediate magnetic field. Here
the magnetic field is not too strong such that central spin decoherence is
dominated by the noise from individual nuclear spins instead of nuclear spin
pairs, and not too weak such that the nuclear spin Zeeman splitting $\gg$ HFI
(off-resonant condition satisfied). Tuning the magnetic field allows the
cross-over between Gaussian and non-Gaussian behaviors, as observed
experimentally for the $^{13}$C nuclear spin bath in NV center
\cite{ReinhardPRL2012,LiuSciRep2012}.

\item Decoherence of electron-nuclear hybrid spin-1/2 near the so-called
\textquotedblleft clock\textquotedblright\ transitions of Bi donor in silicon
\cite{MaPRB2015}. Near the \textquotedblleft clock\textquotedblright%
\ transition, electron-nuclear hybridization dramatically suppresses the HFI
between the hybrid spin-1/2 and the $^{29}$Si nuclear spin bath. This leads to
two effect. First, it prolongs the coherence time by two orders of magnitudes
(from $\sim0.8$ ms to $\sim90$ ms) \cite{WolfowiczNatNano2013}. Second, when
the suppressed HFI becomes weaker than the intrinsic $^{29}$Si bath dynamics
(off-resonant condition satisfied), the bath can be well approximated by
classical Gaussian noise [with the bath auto-correlation funciton shown in
Figs. \ref{GaussTest} (a) and (b)], as confirmed by the excellent agreement
between the semi-classical model with a Gaussian noise, the exact results from
the quantum many-body theory, and experimental measurements \cite{MaPRB2015},
as shown in Figs. \ref{GaussTest} (c) and (d). Away from the \textquotedblleft
clock\textquotedblright\ transitions, the HFI becomes larger and the Gaussian
noise model is no longer valid.
\end{enumerate}

\subsubsection{Test of Gaussian noise model in real systems}

\begin{figure}[ptb]
\includegraphics[width=\columnwidth]{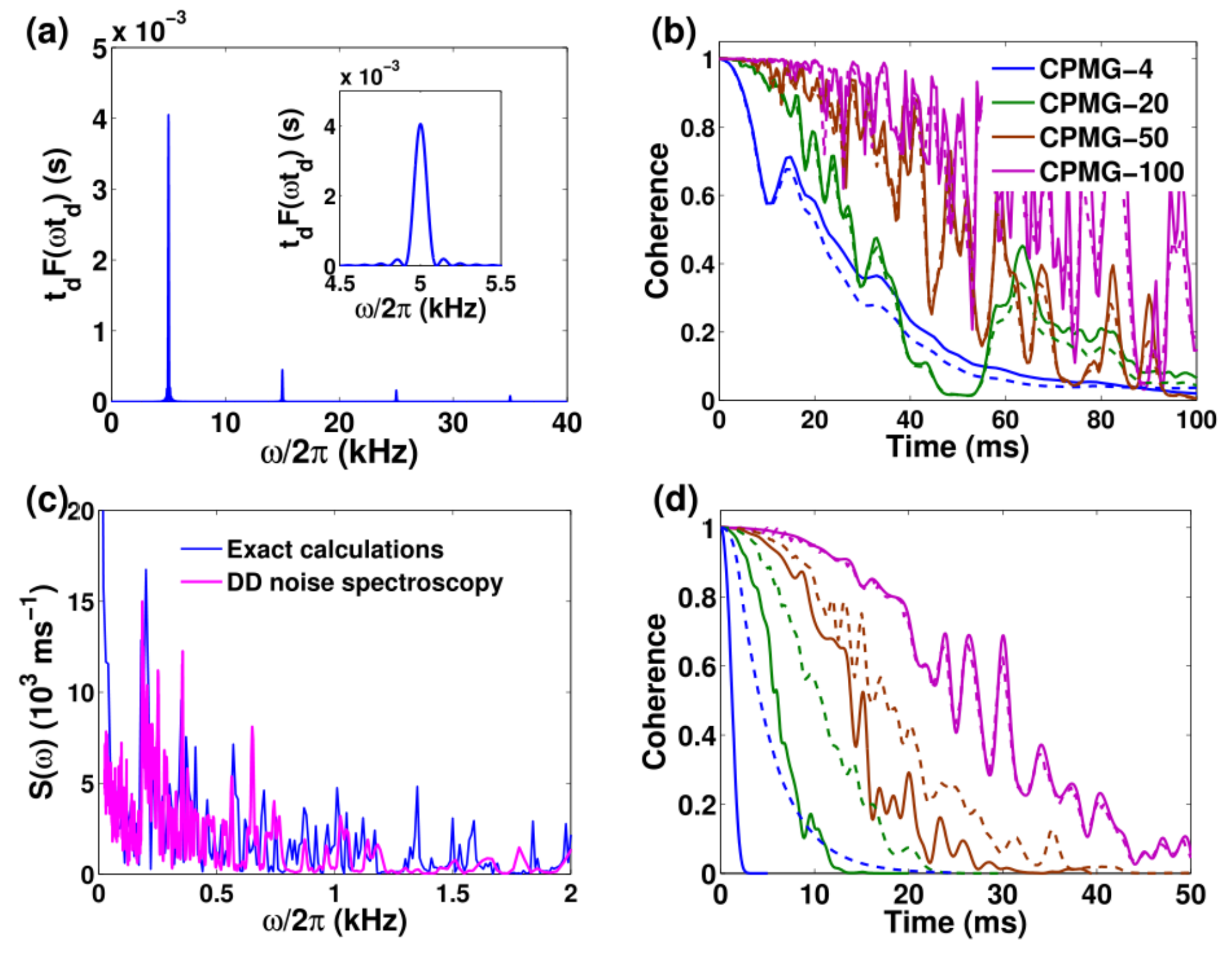}
\caption{ (a)
Filter function $t_{\mathrm{d}} F(\omega t_{\mathrm{d}})$ for CPMG-100 noise
spectroscopy with $t_{\mathrm{d}}=10$~ms. (b) Calculated Bi donor electron spin decoherence
under an exact quantum model (solid lines) and semiclassical model obtained
from noise spectroscopy of the CPMG-100 DD (dashed lines), evaluated close to
the ``clock" transition ($B_{\text{CT}}+10$ G). (c) Comparison of the noise
spectrum from the CPMG-100 spectral decomposition in (b) to the exact one from
CCE calculations. (d) similar to (b) but for the magentic field far from the
``clock" transition ($B_{\text{CT}}+1000$ G). Extracted from Refs.
\cite{MaPRB2015}. }%
\label{NoiseSpectroscopy}%
\end{figure}

The DD noise spectroscopy method based on the Gaussian noise model has been
widely used to characterize the baths
\cite{AlvarezPRL2011,BylanderNatPhys2011,BarGillNatCommun2012}. The main idea
is to use a specific DD control sequence (such as CPMG-$N$ with large $N$)
with the filter function approximated as a Dirac delta function at $\omega
_{0}=\pm\pi{N}/t_{\mathrm{d}}$ [see Fig.~\ref{NoiseSpectroscopy}(a)],
\[
t_{\mathrm{d}}F(\omega t_{\mathrm{d}})\approx\pi[\delta(\omega-\omega_{0}%
)+\delta(\omega+\omega_{0})],
\]
Then following Eqs.~(\ref{LTD}) and ~(\ref{PHIDD_RMS_CLASSICAL}), the bath noise spectrum can be determined as
\[
S(\pm\omega_{0})=-2\mathrm{ln}[L(t_{\mathrm{d}})]/t_{\mathrm{d}}.
\]
However, this method can reproduce a meaningful bath noise spectrum
only if the the bath can be descibed by a semiclassical Gaussian noise model.
For example, in the $^{\text{nat}}$Si:Bi system, we use the DD noise
spectroscopy method to determine the effective noise spectra corresponding to
the CPMG-100 case, and then use the derived noise spectra to calculate the
spin decoherence under other DD control sequences \cite{MaPRB2015}. Close to
the ``clock" transition, the nuclear spin bath produces approximately a
Gaussian noise, then the DD noise spectroscopy method can not only reproduce
the spin decoherence curves for other DD control [see
Fig.~\ref{NoiseSpectroscopy}(b)], but also well reproduce the exact noise
spectrum obtained from exact quantum calculations
[Fig.~\ref{NoiseSpectroscopy}(c)]. However, far away from the ``clock"
transition, the Gaussian noise appromiation is not valid any more, so we find
increasing discrepancies between the exact decoherence model and the
semiclassical model using the DD noise spectroscopy method as the pulse number
of CPMG-$N$ deviates from 100 [Fig.~\ref{NoiseSpectroscopy}(d)].

\section{Quantum picture of central spin decoherence}

\label{SEC_QUANTUM_PICTURE}

Up to now, we have given two different interpretations of central spin
decoherence. First, random modulation of the central spin's transition
frequency by classical noises (Sec. \ref{SEC_SEMICLASSICAL}) or quantum noises
(Sec. \ref{SEC_QUANTUM_NOISE}). Second, random state collapes of the central
spin due to measurement by the environment (Sec. \ref{SEC_CONCEPT}), but the
environment is not explicitly treated there. In this section, we give a full
quantum picture \cite{WitzelPRB2006,YaoPRB2006,YangPRB2008a} that
substantiates the previous intuitive measurement interpretation of central
spin decoherence.

The starting point is the general pure-dephasing Hamiltonian in Eq.
(\ref{HAMIL1}) for the closed quantum system consisting of the central spin
and the bath \cite{YaoPRB2006,LiuNJP2007,YaoPRL2007}:
\begin{equation}
\hat{H}=\hat{S}_{z}\hat{b}+\hat{H}_{B}=\hat{H}_{+}|\uparrow\rangle
\langle\uparrow|+\hat{H}_{-}|\downarrow\rangle\langle\downarrow|,
\label{HAMIL}%
\end{equation}
where $\hat{H}_{\pm}\equiv\hat{H}_{B}\pm\hat{b}/2$ are the bath Hamiltonians
depending on the central spin states being $|\uparrow\rangle$ or
$|\downarrow\rangle$. The initial state of the bath is the maximally mixed thermal
state [Eq. (\ref{RHOB})]. However, central spin decoherence under DD control
is usually insensitive to the initial state of the bath (as discussed in Sec.
\ref{SEC_QUANTUM_DYN}). This allows us to take a pure state $|J\rangle$
sampled from the thermal ensemble (see Eq. (\ref{RHOB})) as the initial state of the bath
to provide a transparent quantum picture of decoherence
\cite{YaoPRB2006,LiuNJP2007,YaoPRL2007}. Note that a pure initial state of the bath
can in principle be prepared via special methods such as quantum measurements of
the bath \cite{GiedkePRA2006,KlauserPRB2006,StepanenkoPRL2006,CappellaroPRA2012,ShulmanNatCommun2014}
and nonlinear feedback \cite{GreilichScience2007,XuNature2009,SunPRL2012,LattaNatPhys2009,BluhmPRL2010,ToganNature2011,YangPRB2012,YangPRB2013}.

\subsection{Decoherence as a result of measurement by environment}

\begin{figure}[ptb]
\includegraphics[width=\columnwidth]{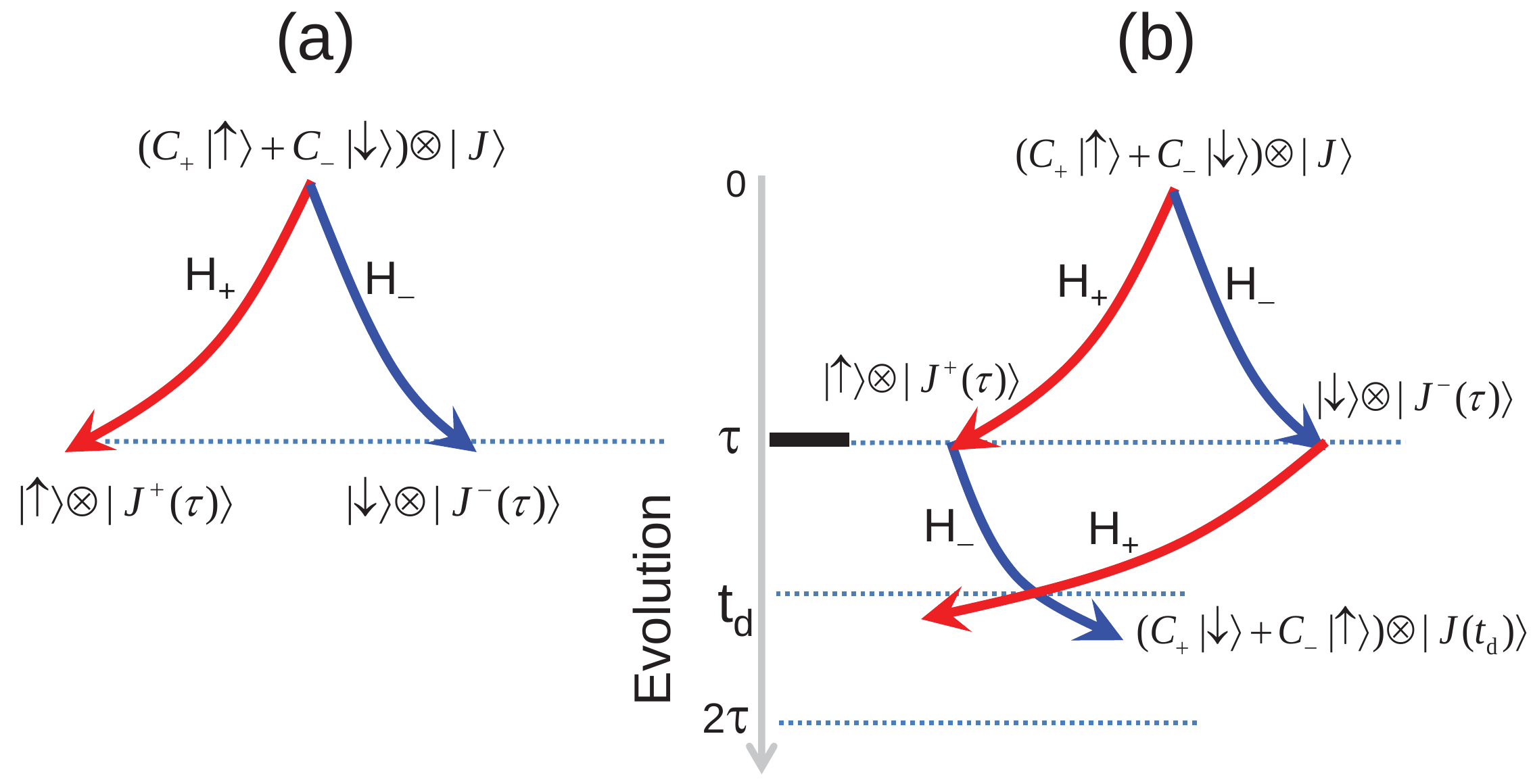}
\caption{Schematic
illustration of the bifurcated bath evolution pathways dependent on the
central spin states: (a) for FID; (b) for the central spin being flipped by a
$\pi$-pulse at an intermediate time. Here the initial state of the spin bath
is assumed to be a pure state $|J\rangle$. }%
\label{G_PATHWAY}%
\end{figure}

Now the initial state of the whole system is the product of the central spin
state $|\psi\rangle=C_{+}|\uparrow\rangle+C_{-}|\downarrow\rangle$ and the
pure bath state $|J\rangle$. The bath undergoes bifurcated evolution
$|J\rangle\rightarrow|J_{\pm}(t)\rangle\equiv e^{-i\hat{H}_{\pm}t}|J\rangle$
[Fig. \ref{G_PATHWAY}(a)], and the coupled system evolves into an entangled
state
\begin{equation}
|\Psi(t)\rangle\equiv C_{+}|\uparrow\rangle\otimes|J_{+}(t)\rangle
+C_{-}|\downarrow\rangle\otimes|J_{-}(t)\rangle. \label{PHI_JT1}%
\end{equation}
During this process, the population of the unperturbed central spin
eigenstates $|\uparrow\rangle$ and $|\downarrow\rangle$ remains unchanged, but
the off-diagonal coherence
\begin{equation}
L(t)=\langle J_{-}(t)|J_{+}(t)\rangle=\langle J|e^{i\hat{H}_{-}t}e^{-i\hat
{H}_{+}t}|J\rangle\label{LT_DEF}%
\end{equation}
generally decays due to the bifurcated bath evolution
\cite{YaoPRB2006,LiuNJP2007,YaoPRL2007}. From the viewpoint of quantum
measurement \cite{BrunePRL1996,ZurekRMP2003}, the central spin state
$|\uparrow\rangle$ ($|\downarrow\rangle$) is recorded in the bath pathway
$|J_{+}(t)\rangle$ ($|J_{-}(t)\rangle$). The off-diagonal coherence between
$|\uparrow\rangle$ and $|\downarrow\rangle$ is the overlap between these two
pathways of the bath. Below we discuss two specific cases.

\leftmargini=3mm

\begin{enumerate}
\item $[\hat{b},\hat{H}_{B}]=0$. The bath initial state $|J\rangle$ can be
chosen as a common eigenstates of $\hat{H}_{B}$ and $\hat{b}$, with
eigenvalues $\varepsilon_{J}$ and $b_{J}$, respectively. Then the two pathways
$|J_{\pm}(t)\rangle=e^{-i(\varepsilon_{J}\pm b_{J}/2)t}|J\rangle$ are
identical up to a phase factor and completely indistinguishable. There is no
quantum entanglement between the central spin and the bath, and the central
spin coherence $L(t)=e^{-ib_{J}t}$ does not decay, but just acquires a phase
due to the static noise field $b_{J}$, consistent with the discussions in Sec.
\ref{SEC_THERMAL_NOISE}.

\item $[\hat{b},\hat{H}_{B}]\neq0$. The initial state $|J\rangle$, if taken as
an eigenstate of $\hat{H}_{B}$, in general is not an eigenstate of $\hat{b}$,
so it undergoes bifurcated evolution into different pathways $|J_{\pm
}(t)\rangle$. Correspondingly, the central spin coherence $L(t)\approx\langle
J_{-}(t)|J_{+}(t)\rangle$ decays due to the bifurcated bath evolution and
hence quantum entanglement between the central spin and the bath. Using
$e^{-i\hat{H}_{+}t}=e^{-i\hat{H}_{B}t}\mathcal{T}e^{-(i/2)\int_{0}^{t}\hat
{b}(t^{\prime})dt^{\prime}}$ and $e^{i\hat{H}_{-}t}=\mathcal{\bar{T}}%
e^{-(i/2)\int_{0}^{t}\hat{b}(t^{\prime})dt^{\prime}}e^{i\hat{H}_{B}t}$, we
immediately see that $\langle J_{-}(t)|J_{+}(t)\rangle$ is just the
\textit{\textquotedblleft true\textquotedblright\ decoherence}
$L_{\mathrm{dyn}}(t)$ caused by the dynamical quantum noise [Eq.
(\ref{LDYNT})], which has been discussed in Sec. \ref{SEC_QUANTUM_DYN}. When
the two pathways of the bath become orthogonal and hence completely
distinguishable at a certain time, the central spin is perfectly measured by
the bath and its off-diagonal coherence vanishes completely.
\end{enumerate}

Finally, we note that upon decomposing the bath states into the unnormalized
common part and the unnormalized difference part as $|J_{\pm}(t)\rangle
\equiv|\tilde{J}_{\mathrm{nc}}(t)\rangle\pm|\tilde{J}_{\mathrm{c}}(t)\rangle$,
the entangled state can be rewritten as
\begin{equation}
|\Psi(t)\rangle=|\psi\rangle\otimes|\tilde{J}_{\mathrm{nc}}(t)\rangle
+\hat{\sigma}_{z}|\psi\rangle\otimes|\tilde{J}_{\mathrm{c}}(t)\rangle,
\label{PHI_JT2}%
\end{equation}
i.e., the central spin state $|\psi\rangle$ and the phase-flipped state
$\sigma_{z}|\psi\rangle$ are recorded in the unnormalized bath states
$|\tilde{J}_{\mathrm{nc}}(t)\rangle$ and $|\tilde{J}_{\mathrm{c}}(t)\rangle$,
respectively. If $|\tilde{J}_{\mathrm{c}}(t)\rangle$ is orthogonal to
$|\tilde{J}_{\mathrm{nc}}(t)\rangle$, then the central spin density matrix
would be an incoherent mixture of $\sqrt{\langle\tilde{J}_{\mathrm{nc}%
}(t)|\tilde{J}_{\mathrm{nc}}(t)\rangle}|\psi\rangle\equiv\hat{M}_{\mathrm{nc}%
}|\psi\rangle$ and $\sqrt{\langle\tilde{J}_{\mathrm{c}}(t)|\tilde
{J}_{\mathrm{c}}(t)\rangle}\hat{\sigma}_{z}|\psi\rangle\equiv\hat
{M}_{\mathrm{c}}|\psi\rangle$, which recovers our intuitive discussion for
Markovian environmental noise in Sec. \ref{SEC_TPHI}.

\subsection{Coherence recovery by dynamical decoupling}

To recover the central spin coherence lost into the bath, it is necessary to
erase the measurement information by making the two bath pathways $|J_{\pm
}(t)\rangle$ identical up to a phase factor. For this purpose, the simplest
approach is Hahn echo \cite{HahnPR1950}, in which a $\pi$-pulse is applied to
the central spin at time $\tau$ to exchange the evolution direction of the two
pathways [see Fig. \ref{G_PATHWAY}(b)]. At $t>\tau$, the two pathways are
$|J_{\pm}(t)\rangle=e^{-i\hat{H}_{\mp}(t-\tau)}e^{-i\hat{H}_{\pm}\tau
}|J\rangle$ and the coupled system evolves into $|\Psi(t)\rangle
=C_{+}|\downarrow\rangle\otimes|J_{+}(t)\rangle+C_{-}|\uparrow\rangle
\otimes|J_{-}(t)\rangle$. The intersection of the two pathways $|J_{+}%
(t_{\mathrm{d}})\rangle\approx e^{i\varphi}|J_{-}(t_{\mathrm{d}})\rangle$ at a
certain time $t_{\mathrm{d}}$ would erase the measurement information and
restores the central spin coherence, as shown in Fig. \ref{G_PATHWAY}(b)
\cite{YaoPRB2006,LiuNJP2007,YaoPRL2007}.

Under a general DD characterized by the DD modulation function $s(t)$, by
working in the interaction picture defined by the control Hamiltonian (Eq.
(\ref{HCT})), the total Hamiltonian becomes $\hat{H}_{B}+s(t)\hat{b}\hat{S}%
_{z}$. The two bath pathways start from $|J\rangle$ and bifurcate into
$|J_{\pm}(t_{\mathrm{d}})\rangle=\hat{U}_{\pm}(t_{\mathrm{d}})|J\rangle$,
where $\hat{U}_{\pm}(t_{\mathrm{d}})\equiv\mathcal{T}e^{-i\int_{0}%
^{t_{\mathrm{d}}}[\hat{H}_{B}\pm s(t)\hat{b}/2]dt}$ are the bifurcated bath
evolution operators. Then the central spin coherence is
\begin{equation}
L(t_{\mathrm{d}})=\langle J|\hat{U}_{-}^{\dagger}(t_{\mathrm{d}})\hat{U}%
_{+}(t_{\mathrm{d}})|J\rangle. \label{LDD_DEF}%
\end{equation}
For example, the FID $L(t)=\langle J|e^{i\hat{H}_{-}\tau}e^{-i\hat{H}_{+}\tau
}|J\rangle$ and Hahn echo $L(t_{\mathrm{d}}=2\tau)=\langle J|e^{i\hat{H}%
_{-}\tau}e^{i\hat{H}_{+}\tau}e^{-i\hat{H}_{-}\tau}e^{-i\hat{H}_{+}\tau
}|J\rangle$. For $[\hat{b},\hat{H}_{B}]=0$, as long as $t_{\mathrm{d}}$
satisfy the echo condition (Eq. (\ref{PHASE_FOCUS})), we have $|J_{\pm
}(t_{\mathrm{d}})\rangle=e^{-i\varepsilon_{J}t_{\mathrm{d}}}|J\rangle$ and
hence $L(t_{\mathrm{d}})=1$, i.e., the phase due to the static noise field
$b_{J}$ is completely refocused.

\subsection{Ensemble average}

\label{SEC_THERMAL_QUANTUM}

In practice, we should use the thermal state $\hat{\rho}_{B}^{\mathrm{eq}}$
[Eq. (\ref{RHOB})] as the initial state of the bath, then we recover the
results in Sec. \ref{SEC_QUANTUM_DYN}. The FID as given by Eq.
(\ref{FID_FACTORIZE}) is the product of inhomogenous dephasing due to the
thermal noise [Eq. (\ref{LT_INH})] and \textit{\textquotedblleft
true\textquotedblright\ decoherence} due to the quantum noise (Eq.
(\ref{LDYNT}) or Eq. (\ref{LT_DEF})) that is almost independent of $|J\rangle$
\cite{YaoPRB2006,ZhaoPRB2012,MaPRB2015}. Under DD, the former is removed, so
$L(t_{\mathrm{d}})=\langle J_{-}(t_{\mathrm{d}})|J_{+}(t_{\mathrm{d}})\rangle$
is \textquotedblleft true\textquotedblright\ decoherence due to the quantum
noise [Eq. (\ref{EQ39}) or (\ref{LDD_DEF})].

The discussions above for a central spin-1/2 can be easily generalized to a
general multi-level system with eigenstates $\{|n\rangle\}$ and the
pure-dephasing Hamiltonian $\sum_{m}\hat{H}_{m}|m\rangle\langle m|$
\cite{ZhaoPRL2011,MaPRB2015}. The off-diagonal coherence $L_{n,m}%
(t)\equiv\langle n|\hat{\rho}(t)|m\rangle$ for a given quantum transition
$|m\rangle\leftrightarrow|n\rangle$ can be mapped to that of a central
spin-1/2 once the states $\{|m\rangle,|n\rangle\}$ are identified as
$\{|\uparrow\rangle,|\downarrow\rangle\}$ with $\hat{H}_{B}\equiv(\hat{H}%
_{m}+\hat{H}_{n})/2$ and $\hat{b}\equiv\hat{H}_{m}-\hat{H}_{n}$.

\section{Physical systems}

\label{SEC_SYSTEM}

Electron spins localized in solid-state nanostructures are promising
candidates of qubits for quantum information processing and quantum sensing.
These \textquotedblleft artificial atoms\textquotedblright\ occur when the
impurities or defects in semiconductor nanostructures produce localized
potentials to confine one or a few electrons or \textit{holes} (i.e., an empty
electron state in the valence band of semiconductors), analagous to electrons
bound to atomic nuclei. For such electron spins, the most relevant
environments are the phonon bath and the nuclear spins of the host lattices.
The electron spins couple indirectly to the phonon baths via spin-orbit
coupling, and couples directly to the nuclear spin baths via the hyperfine
interaction (HFI). In this section, we first introduce these interactions and
then review central spin decoherence due to these interactions in typical
semiconductor nanostructures.

\subsection{Phonon and spin baths}

\subsubsection{Phonon scattering via spin-orbit coupling}

Electric fields are not directly coupled to the electron spin $\hat
{\mathbf{S}}$. Indirect coupling occurs due to the relativistic correction
\[
\hat{H}_{\mathrm{so}}=\frac{1}{2m_{0}^{2}c^{2}}(\nabla V(\hat{\mathbf{r}%
})\times\hat{\mathbf{p}})\cdot\hat{\mathbf{S}}%
\]
to the non-relativistic Hamiltonian for the electron moving in a potential
$V(\mathbf{r})$. Due to this spin-orbit coupling term, the electron spin
eigenstates become mixtures of spin and orbital states, thus fluctuating
electric fields can induce transitions between these eigenstates (i.e., spin
relaxation) \cite{KhaetskiiPRB2000,KhaetskiiPRB2001,WoodsPRB2002} and randomly
modulate the transition frequency (i.e., pure dephasing)
\cite{GolovachPRL2004,SemenovPRL2004}. In carefully designed systems (where
the charge fluctuations are suppressed), the most relevant source of
electrical noises is the lattice vibration (i.e., the phonon bath). The phonon
energy spectrum ranges over a few tens of meV, much larger than the electron
spin transition energy ($\sim\mathrm{\mu eV})$, so the phonon noise is
Markovian and usually limits the electron spin $T_{1}$ and, at high
temperatures, also limits the electron spin $T_{2}$ (see Sec.
\ref{SEC_CONCEPT}). At low temperatures and in light-element materials where
spin-orbit coupling is weak, phonon scattering is suppressed and the
experimentally measured electron spin $T_{1}$ is very long, ranging from tens
of microseconds up to seconds (see \cite{HansonRMP2007} for a review). At low
temperautre, the phonon-limited electron spin $T_{2}$ is estimated as
$T_{2}\approx2T_{1}$ \cite{GolovachPRL2004}, but the experimentally measured
$T_{2}$ is much shorter as limited by the hyperfine interaction with the
nuclear spin bath.

\subsubsection{Hyperfine interaction}

\label{SEC_HFI}%

\begin{table}[tbp] \centering
\begin{tabular}
[c]{lccccc}\hline\hline
& $\ \ $\ $^{75}$As \ \  & $\ ^{113}$In \  & $^{115}$In & \ $^{69}$Ga \  &
\ $^{71}$Ga \ \\\hline
Spin moment $I_{\alpha}$ & $3/2$ & $9/2$ & $9/2$ & $3/2$ & $3/2$\\
Abundance $f_{\alpha}$ & $100\%$ & $4.28\%$ & $95.72\%$ & $60.1\%$ &
$39.9\%$\\
$\gamma_{\alpha}$ (10$^{-3}$ rad ns$^{-1}$T$^{-1}$) & $45.8$ & $58.5$ & $58.6$
& $64.3$ & $81.8$\\
$A_{\alpha}$ (rad ns$^{-1})$ & $69.8$ & $85.1$ & $85.3$ & $56$ &
$73$\\
$Q_{\alpha}$ (10$^{-31}$ m$^{2}$) & $314$ & $759$ & $770$ & $171$ &
$107$\\\hline\hline
\end{tabular}
\caption{Spin moment, natural abundance, gyromagnetic ratio, HFI constant, and quadrupole moment of
some isotopes that appear in III-V semiconductor quantum dots (QDs), with the reduced Planck constant $\hbar=1$ and hence $1\ \mathrm{\mu eV}\approx1.52$ ns$^{-1}$. The quadrupole moments are from Ref. \cite{PyykkoMP2008}. Other data for In and As are from Ref. \cite{LiuNJP2007}, while those for Ga are from Refs. \cite{PagetPRB1977} and \cite{CoishPSSB2009}.}\label{TableI}%
\end{table}%

For a nuclear spin $\hat{\mathbf{I}}_{n\alpha}$ of species $\alpha$ located at
$\mathbf{R}_{n\alpha}$, its magnetic moment $\gamma_{\alpha}\hat{\mathbf{I}%
}_{n\alpha}$ produces a vector potential $\mathbf{A}_{n\alpha}=(\mu_{0}%
/4\pi)(\gamma_{\alpha}\hat{\mathbf{I}}_{n\alpha}\times\boldsymbol{\rho
}_{n\alpha})/\rho_{n\alpha}^{3}$ at the location $\mathbf{r}$ of the electron
with $\boldsymbol{\rho}_{n\alpha}\equiv\mathbf{r}-\mathbf{R}_{n\alpha}$. The
total vector potential $\mathbf{A}\equiv\sum_{n\alpha}\mathbf{A}_{n\alpha}$
due to all the nuclei gives rise to the electron-nuclear magnetic coupling
$\gamma_{e}(\hat{\mathbf{p}}\cdot\mathbf{A}+\mathbf{A}\cdot\hat{\mathbf{p}%
})/2+\gamma_{e}\hat{\mathbf{S}}\cdot(\nabla\times\mathbf{A})$
\cite{AbragamBook1961}, which is the sum of the contact HFI
\[
\hat{H}_{\mathrm{c}}=\frac{2\mu_{0}}{3}\sum_{n\alpha}\gamma_{e}\gamma_{\alpha
}\hat{\mathbf{I}}_{n\alpha}\cdot\hat{\mathbf{S}}\delta(\boldsymbol{\rho
}_{n\alpha}),
\]
the dipolar HFI%
\[
\hat{H}_{\mathrm{d}}=\frac{\mu_{0}}{4\pi}\sum_{n\alpha}\gamma_{e}%
\gamma_{\alpha}\left(  3\frac{(\hat{\mathbf{S}}\cdot\boldsymbol{\rho}%
_{n\alpha})(\hat{\mathbf{I}}_{n\alpha}\cdot\boldsymbol{\rho}_{n\alpha})}%
{\rho_{n\alpha}^{5}}-\frac{\hat{\mathbf{S}}\cdot\hat{\mathbf{I}}_{n\alpha}%
}{\rho_{n\alpha}^{3}}\right)  ,
\]
and the nuclear-orbital interaction%
\[
\hat{H}_{\mathrm{orb}}=\frac{\mu_{0}}{4\pi}\sum_{n\alpha}\gamma_{e}%
\gamma_{\alpha}\frac{\hat{\mathbf{L}}_{n\alpha}\cdot\hat{\mathbf{I}}_{n\alpha
}}{\rho_{n\alpha}^{3}},
\]
where $\gamma_{e}\approx 1.76\times10^{11}$ rad/(s T) is the electron gyromagnetic ratio (positive) for free electrons and $\hat{\mathbf{L}%
}_{n\alpha}\equiv\boldsymbol{\rho}_{n\alpha}\times\hat{\mathbf{p}}$ is the
electron orbital angular momentum around the nucleus. The magnetic interaction
involves the coupling of the electron orbital and electron spin to the nuclear
spin. At low temperature, the localized electron in a nanostructure stays in
its ground orbital $\psi(\mathbf{r})$, so the magnetic interaction should be
averaged over $\psi(\mathbf{r})$ to yield the effective spin-spin interaction.

The spin-spin contact HFI
\begin{equation}
\bar{H}_{\mathrm{c}}=\langle\psi|\hat{H}_{\mathrm{c}}|\psi\rangle
=\sum_{n\alpha}a_{n\alpha}\hat{\mathbf{S}}\cdot\hat{\mathbf{I}}_{n\alpha},
\label{CHFI}%
\end{equation}
where the HFI coefficient $a_{n\alpha}=(2\mu_{0}/3)\gamma_{e}\gamma_{\alpha
}|\psi(\mathbf{R}_{n\alpha})|^{2}$ is determined by the electron density at
the site of the nucleus. The contact HFI is strong for electrons in the
conduction band of III-V semiconductors (mostly $s$-orbital) and silicon
(hybridization of $s$, $p$, and $d$ orbitals), but vanishes in graphene,
carbon nanotubes, and the valence band of III-V semiconductors since their
primary component -- the $p$-orbital -- vanishes at the site of the nucleus
\cite{CoishPSSB2009}. For III-V semiconductors with a non-degenerate
$s$-orbital conduction band minumum at the $\Gamma$ point, the ground orbital
can be written as $\psi(\mathbf{r})$ $=\sqrt{\Omega}F(\mathbf{r}%
)u_{c}(\mathbf{r})$, where $\Omega$ is the unit cell volume, $F(\mathbf{r})$
is the slowly-varying envelope function normalized as $\int|F(\mathbf{r}%
)|^{2}d\mathbf{r}=1$, and $u_{c}(\mathbf{r})$ is the $s$-orbital band-edge
Bloch function that is conveniently normalized as $\int_{\Omega}%
|u_{c}(\mathbf{r})|^{2}d\mathbf{r}=1$, such that $d_{\alpha}\equiv
|u_{c}(\mathbf{R}_{n\alpha})|^{2}$ is the electron density on the nucleus of
species $\alpha$ \cite{PagetPRB1977}. So the HFI coefficient becomes
$a_{n\alpha}=A_{\alpha}\Omega|F(\mathbf{R}_{n\alpha})|^{2}$, where%
\begin{equation}
A_{\alpha}=\frac{2\mu_{0}}{3}\gamma_{e}\gamma_{\alpha}d_{\alpha}
\label{HFI_CONST}%
\end{equation}
is the HFI constant that only depends on the species of the nuclear spin
(through $\gamma_{\alpha}$) and the semiconductor material (through
$d_{\alpha}$). The numerical values of $\gamma_{\alpha}$ and $A_{\alpha}$ for
some relevant isotopes in III-V semiconductor quantum dots (QDs) are listed in
Table \ref{TableI}. For silicon, there are six equivalent conduction band
minima at $\mathbf{k}_{\lambda}=\pm k_{0}\mathbf{e}_{x}$, $\pm k_{0}%
\mathbf{e}_{y}$, $\pm k_{0}\mathbf{e}_{z}$, where $k_{0}\equiv0.85(2\pi
/a_{\operatorname{Si}})$ and $a_{\operatorname{Si}}=5.43$ $%
\operatorname{\text{\AA}}%
$ is the lattice constant of silicon. Thus the ground orbital of a
hydrogen-like donor in silicon is $\psi(\mathbf{r})=(1/\sqrt{6})\sum_{\lambda
}F_{\lambda}(\mathbf{r})u_{\lambda}(\mathbf{r})e^{i\mathbf{k}_{\lambda}%
\cdot\mathbf{r}}$, where $u_{\lambda}(\mathbf{r})e^{i\mathbf{k}_{\lambda}%
\cdot\mathbf{r}}$ is the Bloch function at the $\lambda$th minimum consisting
of $s$, $p$, and $d$ orbitals, with the normalization $\int_{\Omega}|u_{\lambda}(\mathbf{r})|^{2}d\mathbf{r}=\Omega$. The hydrogen-like envelope function associated
with $\pm k_{0}\mathbf{e}_{x}$ is \cite{FeherPR1959a,SousaPRB2003}\
\[
F_{x}(\mathbf{r})=\frac{1}{\sqrt{\pi(na)^{2}nb}}e^{-\sqrt{x^{2}/(nb)^{2}%
+(y^{2}+z^{2})/(na)^{2}}},
\]
with similar expressions for $F_{y}(\mathbf{r})$ and $F_{z}(\mathbf{r})$ by
appropriate permutations of $x,y,z$. Here $a=25.09$ $%
\operatorname{\text{\AA}}%
$ and $b=14.43$ $%
\operatorname{\text{\AA}}%
$ are characteristic lengths for hydrogenic impurities in silicon, $n=0.81$
($0.64$) for phosphorus (bismuth) donors
\cite{SousaPRB2003,RichardPRB2004,ZwanenburgRMP2013}. The donor electron
density at the silicon lattice site $\mathbf{R}_{n}$ is given by
\cite{FeherPR1959a} $|\psi(\mathbf{R}_{n})|^{2}=(2d_{\operatorname{Si}%
}/3)[\sum_{\alpha=x,y,z}F_{\alpha}(\mathbf{R}_{n})\cos(k_{0}\mathbf{R}_{n}%
\cdot\mathbf{e}_{\alpha})]^{2}$, where $d_{\operatorname{Si}}\equiv
|u_{\lambda}(\mathbf{R}_{n})|^{2}\approx186$ is the electron density on the
silicon site in silicon crystal \cite{ShulmanPR1956,SousaPRB2003}.

The spin-spin dipolar HFI
\begin{equation}
\bar{H}_{\mathrm{d}}=\langle\psi|\hat{H}_{\mathrm{d}}|\psi\rangle
=\sum_{n\alpha}\hat{\mathbf{S}}\cdot\mathbf{A}_{n\alpha}\cdot\hat{\mathbf{I}%
}_{n\alpha}, \label{DHFI}%
\end{equation}
where the dipolar HFI tensor
\[
\lbrack\mathbf{A}_{n\alpha}]_{i,j}\equiv\frac{\mu_{0}}{4\pi}\gamma_{e}%
\gamma_{\alpha}\int\frac{|\psi(\mathbf{r})|^{2}}{\rho_{n\alpha}^{3}}\left(
\frac{3\rho_{n\alpha}^{i}\rho_{n\alpha}^{j}}{\rho_{n\alpha}^{2}}-\delta
_{ij}\right)  d\mathbf{r}%
\]
with $i,j=x,y,z$. The dipolar HFI and the nuclear-orbital interaction $\bar
{H}_{\mathrm{orb}}\equiv\langle\psi|\hat{H}_{\mathrm{orb}}|\psi\rangle$ are
negligible for the $s$-orbital conduction band of III-V semiconductors. They
become appreciable for donors in silicon (due to significant $p$- and
$d$-orbital components in the band-edge Bloch functions) and even dominates
for electrons in graphene, carbon nanotubes, and the valence band of III-V
semiconductors
\cite{CoishPSSB2009,FischerPRB2008,TestelinPRB2009,ChekhovichPRL2011,ChekhovichNatPhys2013}%
, where the atomic $p$-orbital is the primary component of the Bloch functions
and hence the contact HFI vanishes. If $\psi(\mathbf{r})$ is localized in the
vicinity of $\mathbf{\bar{r}}$ and far from the nucleus, then the dipolar HFI
$\bar{H}_{\mathrm{d}}=\langle\psi|\hat{H}_{\mathrm{d}}|\psi\rangle=\hat
{H}_{\mathrm{d}}|_{\mathbf{r}\rightarrow\mathbf{\bar{r}}}$ reduces to the
magnetic dipolar interaction between two point-like magnetic moments; while if
$\psi(\mathbf{r})$ overlaps the nucleus, then $\bar{H}_{\mathrm{d}}$ is
dominated by the interaction of the nuclear spin with the on-site electron
spin density \cite{TestelinPRB2009}. Recently the manipulation and decoherence
of valence band electrons (i.e., \textit{holes}) in QDs is under active study
(see Ref. \cite{ChesiEPJP2014} for a review).

\subsubsection{Intrinsic nuclear spin interactions}

\label{SEC_NN_INTERACTION}

The interaction between nuclear spins have been well studied in NMR
experiments and in theories (for a review, see Ref. \cite{SlichterBook1990}). The
direct magnetic dipolar interaction has the dipolar form%
\begin{equation}
\hat{H}_{\mathrm{NN}}^{d}=\frac{1}{2}\sum_{n\alpha\neq m\beta}\frac{\mu_{0}%
}{4\pi}\gamma_{\alpha}\gamma_{\beta}\left(  \frac{\hat{\mathbf{I}}_{n\alpha
}\cdot\hat{\mathbf{I}}_{m\beta}}{R^{3}}-\frac{3(\hat{\mathbf{I}}_{n\alpha
}\cdot\mathbf{R})(\hat{\mathbf{I}}_{m\beta}\cdot\mathbf{R})}{R^{5}}\right)  ,
\label{HDD}%
\end{equation}
where $\mathbf{R}\equiv\mathbf{R}_{n\alpha}-\mathbf{R}_{m\beta}$ is the
relative displacement between the locations $\mathbf{R}_{n\alpha}$ and
$\mathbf{R}_{m\beta}$ of the two nuclei. The indirect nuclear interaction is
mediated by virtual excitation of electron-hole pairs due to the HFI between
nuclei and valence electrons
\cite{BloembergenPR1955,ShulmanPR1955,ShulmanPR1957,ShulmanPR1958,SundforsPR1969}%
. When the virtual excitation is caused by the contact HFI, the indirect
coupling has the isotropic exchange form $\hat{H}_{\mathrm{NN}}^{\mathrm{ex}%
}=-B_{n\alpha,m\beta}^{\mathrm{ex}}\hat{\mathbf{I}}_{n\alpha}\cdot
\hat{\mathbf{I}}_{m\beta}$, where $B_{n\alpha,m\beta}^{\mathrm{ex}}$ is
determined by the band structure of the material. When the virtual excitation
of electron-hole pairs involves both the contact and dipolar HFI, the indirect
nuclear spin coupling has the same form as the direct dipolar interaction in Eq.
(\ref{HDD}) except for a multiplicative factor that depends on the
inter-nuclear distance. When the virtual excitation is caused by the dipolar
HFI alone, the indirect coupling is the sum of an isotropic exchange term and
dipole-dipole term. Except for the direct dipolar coupling, experimental
characterization of indirect couplings is very limited.%

\begin{table}[tbp] \centering
\begin{tabular}
[c]{lccc}\hline\hline
& $\ \mathrm{\mu s}^{-1}$ \  & $\ \mathrm{\mu eV}$ \  & \ mK \ \\\hline
Electron Zeeman splitting & $10^{5}$ & $10^{2}$ & $10^{3}$\\
Nuclear Zeeman splitting & $50$ & $0.05$ & $0.5$\\
Hyperfine interaction & $1$ & $10^{-3}$ & $10^{-2}$\\
N-N dipolar interaction & $10^{-4}$ & $10^{-7}$ & $10^{-6}$\\\hline\hline
\end{tabular}
\caption{Characteristic energy scales in an InAs QD with dimensions $35\times35\times6$ nm$^{3}$ under a magnetic field of 1 T \cite{LiuNJP2007}, with $\hbar$=1.}\label{TableII}%
\end{table}%

Due to the vanishing electric dipole moment of the nucleus, the nuclear spin
is not coupled to constant electric fields. However, a nucleus with spin
$I>1/2$ has a finite electric quadrupole moment, so a nuclear spin
$\hat{\mathbf{I}}$ located at $\mathbf{R}$ with quadrupole moment $Q$ is
coupled to the on-site electric field gradient tensor $V_{ij}\equiv
\partial^{2}V(\mathbf{x})/\partial x_{i}\partial x_{j}|_{\mathbf{x}%
=\mathbf{R}}$ through $\hat{H}_{Q}=\sum_{ij=x,y,z}V_{ij}\hat{Q}_{ij}$, where
\cite{AbragamBook1961}%
\[
\hat{Q}_{ij}\equiv\frac{eQ}{6I(2I-1)}\left[  \frac{3}{2}(\hat{I}_{i}\hat
{I}_{j}+\hat{I}_{j}\hat{I}_{i})-\delta_{ij}I(I+1)\right]
\]
is the nuclear spin quadrupole tensor. In the principal axis $OXYZ$ of the
electric field gradient tensor, only diagonal components $V_{XX},V_{YY}%
,V_{ZZ}$ survives. Using the non-axial parameter $\eta\equiv(V_{XX}%
-V_{YY})/V_{XX}$ and Laplace equation $V_{XX}+V_{YY}+V_{ZZ}=0$ allows the
quadrupolar interaction to be simplified to \cite{AbragamBook1961}%
\[
\hat{H}_{Q}=\frac{eQV_{ZZ}}{4I(2I-1)}\left[  3\hat{I}_{Z}^{2}-I(I+1)+\eta
(\hat{I}_{X}^{2}-\hat{I}_{Y}^{2})\right]  .
\]
The quadrupole moments of some relevant isotopes in III-V QDs are listed in
Table I.

In a crystal with cubic symmetry, the electric field gradient tensor obey
$V_{XX}=V_{YY}=V_{ZZ}$, which together with the Laplace equation dictates
vanishing electric field gradient and quadrupolar interaction. Nonzero
quadrupolar interaction could arise from broken cubic symmetry by lattice
distortion due to semiconductor heterostructure, dopants, or defects. The
quadrupolar interactions have important effects on the nuclear spin dynamics
\cite{AbragamBook1961} and hence the auto-correlations of the noises on a
central electron spin coupled to the nuclear spin bath \cite{SinitsynPRL2012}.
Recently Chekhovich \textit{et al. }measured \cite{ChekhovichNatNano2012}
strain-induced quadrupolar interactions in self-assembled QDs and found that
they suppress the nuclear spin flip-flops \cite{ChekhovichNatComm2015}, while
in gate-defined GaAs QDs, the quadrupolar interaction was found to reduce the
electron spin coherence time by causing faster decorrelation of the nuclear
spin noise \cite{BotzemNatComm2016}.

\subsection{Electron spin decoherence in solid-state nano-systems}

The widely studied systems include semiconductor QDs
\cite{KastnerPhysToday1993,LossPRA1998,GuptaPRB1999}, phosphorus and bismuth
donors in silicon \cite{KaneNature1998,TyryshkinJPC2006,GeorgePRL2010}, and
nitrogen-vacancy (NV) centers in diamond
\cite{GruberScience1997,DohertyPhysRep2013}. In these systems, electron or
hole spins act as qubits. At low temperatures, the spin-phonon scattering
processes are largely suppressed
\cite{KhaetskiiPRB2000,KhaetskiiPRB2001,GolovachPRL2004,BulaevPRL2005}, so the
main noise source for electron spin qubits in these systems are the nuclear
spin baths of the host lattice. As a convention, we use $T_{2}^{\ast}$ for the
dephasing time in FID (since FID is usually dominated by inhomogeneous
dephasing), and use $T_{2}$ for the dephasing time under various DD controls,
where inhomogeneous dephasing has been removed.

\subsubsection{Semiconductor quantum dots}

\begin{figure}[ptb]
\includegraphics[width=\columnwidth]{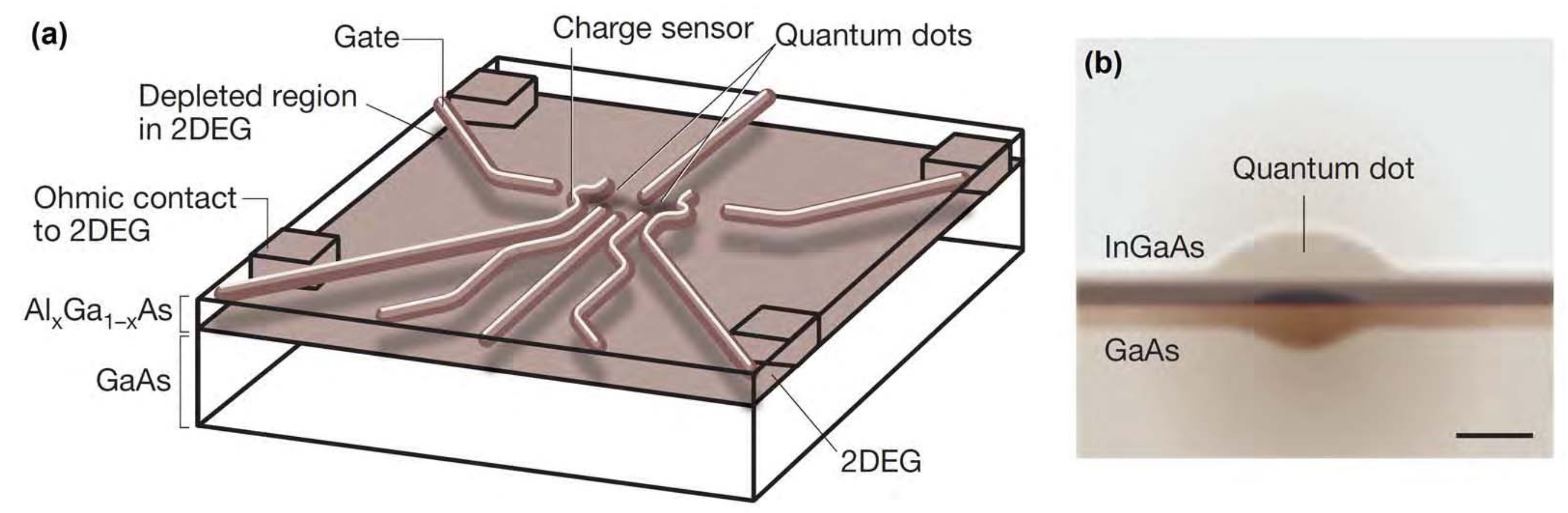}
\caption{(a) A gate-defined
double quantum dots, with 2DEG for two-dimensional electron gas. (b) A
self-assembled quantum dot. Scale bar $\sim$5 nm. Extracted from Fig. 1 of
Ref. \cite{HansonNature2008}.}%
\label{G_QD}%
\end{figure}

Electron spins in QDs are among the earliest candidates for quantum computing
\cite{LossPRA1998,ImamogluPRL1999}. A QD is a semiconductor nanostructure with
size from a few to hundreds of nanometers. The electrons in QDs experience
quantum confinement in all three spatial dimensions, with their energies, wave
functions, and hence spin properties tunable by the QD size and shape
\cite{KouwenhovenRPP2001,ReimannRMP2002,HansonRMP2007}. There are different
ways to fabricate QDs, e.g.,\ gate-defined QDs \cite{KouwenhovenRPP2001}
confine electrons by an electrostatic potential from electric voltages on
lithographically defined metallic gates [Fig. \ref{G_QD}(a)], while
self-assembled QDs \cite{WarburtonNatMater2013} confine electrons with a deep
potential that is created during the random semiconductor growth process [Fig.
\ref{G_QD}(b)]. There are also QDs formed by interface fluctuation in
GaAs/AlGaAs quantum well structures \cite{GammonPRL1991}. The weakly confined
electrons in gate-defined QDs can be controlled electrically at very low
temperatures ($<1$ K), and strongly confined electrons in self-assembled QDs
and interface fluctuation QDs can be controlled optically at a little higher
temperatures ($\sim4$ K).

A critical issue in electron spin qubits in III--V semiconductor QDs is the
inevitable presence of nuclear spins in the semiconductor substrate since all
stable isotopes of the III-V semiconductors have nonzero nuclear spins
\cite{KhaetskiiPRL2002,MerkulovPRB2002}. The thermal noise (see Sec.
\ref{SEC_THERMAL_NOISE} and Sec. \ref{SEC_THERMAL_QUANTUM}) from the nuclear
spin bath leads to rapid inhomogeneous dephasing of the electron spin on a
time scale $T_{2}^{\ast}\sim10$ ns \cite{HansonRMP2007}. When this
inhomogeneous dephasing is removed by Hahn echo, the quantum dynamical noise
from the nuclear spin bath still limits the electron spin dephasing time
$T_{2}$ to a few microseconds \cite{HansonRMP2007}. Fortunately, the nuclear
spin noise has a rather long auto-correlation time $\tau_{c}\sim$ 1 ms ($\sim$
the inverse of nuclear spin interactions, see Table \ref{TableII})
\footnote{Here the nuclear spin noise refers to the nuclear Overhauser field
[i.e., $\mathbf{\hat{h}}\equiv\sum_{n\alpha}{{a_{n\alpha}}{{\hat{\mathbf{I}}%
}_{n\alpha}}}$ in Eq. (\ref{CHFI}) and $\mathbf{\hat{h}}\equiv\sum_{n\alpha
}{{\mathbf{A}_{n\alpha}}\cdot{{\hat{\mathbf{I}}}_{n\alpha}}}$ in Eq.
(\ref{DHFI})]. In a moderate to strong magnetic field, the electron spin
decoherence is usually caused by the fluctuation of the longitudinal component
$\hat{h}_{z}$ along the external magnetic field. The auto-correlation time of
$\hat{h}_{z}$ is determined by the nuclear spin interactions as $\tau_{c}\sim
$1 ms, while that of the transverse components $\hat{h}_{x}$,$\hat{h}_{y}$
could be much shorter, since not only nuclear spin interactions but also the
spread of Larmor precession frequencies of different nuclei contribute to
their decorrelation. Also note that when the nuclear spins are polarized, it
will take a much longer time (from seconds to hours) for the average value of
$\hat{h}_{z}$ to relax to its thermal equilibrium value.}, so it can be
significantly suppressed by various DD sequences, e.g., the multi-pulse CPMG
has extended the $T_{2}$ of a singlet-triplet qubit in gate-defined GaAs
double QDs from $\sim1$ $\mathrm{\mu s}$
\cite{PettaScience2005,GreilichScience2006,KoppensPRL2008} to $\sim$ 1 ms
\cite{BluhmNatPhys2011,MalinowskiArxiv2016}. Recently, silicon-based QDs have
been developed, such as QDs in Si/SiGe heterostructures and gated nanowires
(see \cite{MortonNature2011} for a review). As the only silicon isotope
$^{29}$Si that has non-zero spin is of low natural abundance ($4.7\%$), the
measured electron spin $T_{2}^{\ast}\approx360$ $\mathrm{ns}$ in Si/SiGe
double QDs \cite{MauneNature2012} is longer than in GaAs QDs by more than one
order of magnitude, and further improvements are expected for devices using
isotopically enriched $^{28}$Si.

\subsubsection{Donors in silicon and related materials}

\begin{figure*}[ptb]
\includegraphics[width=1.6\columnwidth]{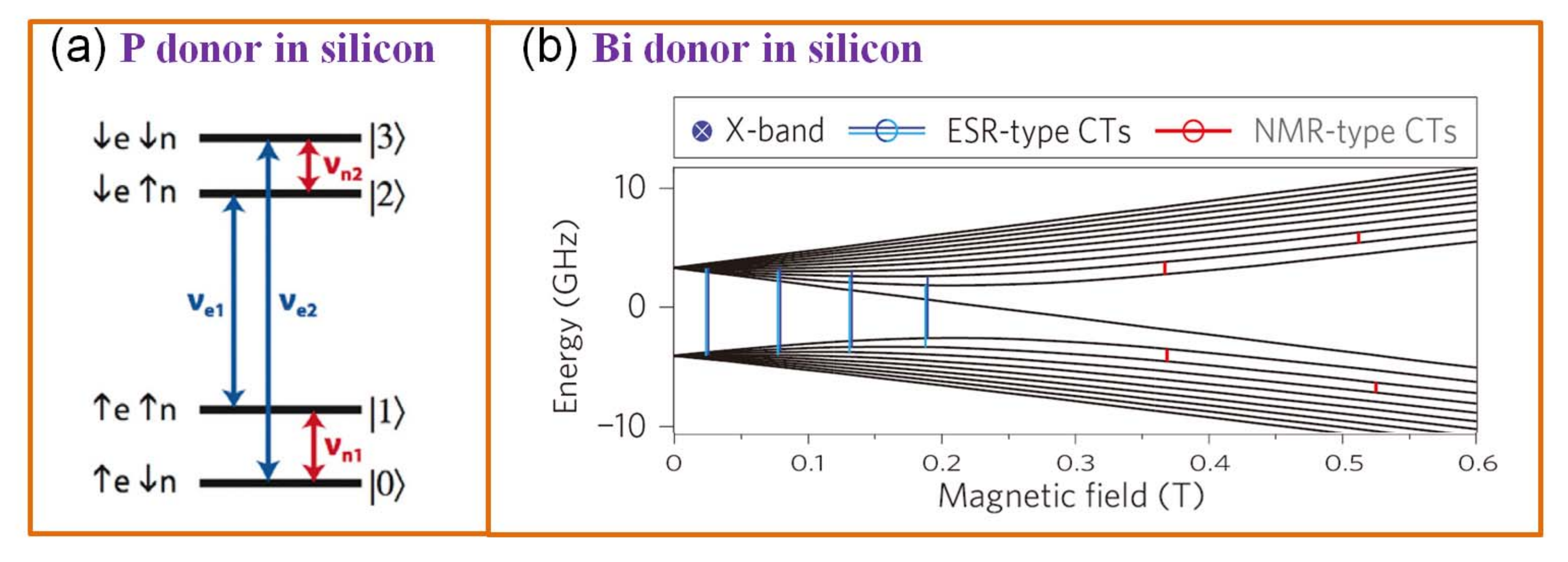}
\centering
\caption{Energy levels of phosphorus and bismuth donors in silicon. (a) P
donor electron spin-1/2 couples to the $^{31}$P nuclear spin-1/2 via a
moderate HFI of strength $A=117.5$ MHz. (b) Bi donor electron spin-1/2 couples
to the $^{209}$Bi nuclear spin-9/2 via strong HFI with a strength$A=1.745$
GHz. Panels (a) and (b)\ are adapted from Refs. \cite{TyryshkinJPC2006} and
\cite{WolfowiczNatNano2013}, respectively. }%
\label{G_DONOR}%
\end{figure*}

As the dominant material in semiconductor industry, silicon provides a
platform to accommodate both quantum and classical information technologies.
Electron and nuclear spins of individual donors in silicon have been proposed
as qubits ever since the early years of solid-state quantum information
\cite{KaneNature1998}. After Kane's infuential proposal, different
architectures have been proposed in which electron spin \cite{VrijenPRA2000}
and orbital \cite{BarrettPRB2003,HollenbergPRB2004}, nuclear spin
\cite{MortonNature2011}, an electron and its donor nuclear spin
\cite{SkinnerPRL2003} are used as qubits.

Silicon has three stable isotopes:\ $^{28}$Si (natural abundance $92.2\%$),
$^{29}$Si (natural abundance $4.7\%$), and $^{30}$Si (natural abundance
$3.1\%$), among which only $^{29}$Si has a nonzero nuclear spin $I=1/2$, in
sharp contrast to III-V group semiconductors where all isotopes have nonzero
nuclear spins. The low concentration of spinful nuclear isotopes and weak
spin--orbit coupling in silicon results in long electron spin coherence times
compared to that of spin qubits in III-V group semiconductor QDs. For high-mobility
two-dimensional electron systems, their $T_{1}$ and $T_{2}$ reach a few
microseconds at low temperature \cite{TyryshkinPRL2005}, limited by phonon
scattering via spin-orbit coupling. When the electron spin is tightly bound to
a donor, the spin-orbit coupling is further suppressed, so at low temperatures
its $T_{1}$ can reach minutes to hours \cite{FeherPR1959}, while its $T_{2}$
is usually limited by $T_{1}$ process at high temperature, or by other donor
electron spins and the sparse $^{29}$Si nuclear spin bath at low temperature
\cite{KlauderPR1962}.

Among all the group-V dopants in silicon, phosphorus donors in natural silicon
($^{\text{nat}}$Si:P) or isotopically purified $^{28}$Si ($^{28}$Si:P) have
been widely studied. Phosphorus has only one stable isotope $^{31}$P with
nuclear spin $I=1/2$ [Fig. \ref{G_DONOR}(a)]. The P donor electron spins was
exhaustively studied almost sixty years ago in the first electron-nuclear
double resonance experiment \cite{FeherPR1959a}. At low donor concentrations,
the electron $T_{1}$ increases dramatically with decreasing temperature,
reaching thousands of seconds at low temperature $\sim1$ K
\cite{FeherPR1959,CastnerPRL1962,TyryshkinPRB2003,MorelloNature2010}, while
the $^{31}$P nuclear spin relaxation time exceeds 10 hours \cite{FeherPR1959a}%
. At low temperature, the extrapolated $T_{2}$ of an isolated $^{28}$P donor
electron spin from spin echo measurements can reach $60$ ms \cite{TyryshkinPRB2003}, comparable with
$T_{1}$ $\sim280$ ms. In $^{\text{nat}}$Si:P system, $T_{2}$ of the P donor
electron spin under spin echo control is limited by the $^{29}$Si nuclear spin
bath to $\sim1$ ms \cite{TyryshkinPRB2003}. Recently unprecedented long
electron spin $T_{2}\approx12$ s \cite{TyryshkinNatMater2012} and nuclear spin
dephasing time up to a few minutes were reported in ultrapure $^{28}$Si
crystals \cite{MortonNature2008,StegerScience2012,MuhonenNatNano2014}. Due to
its exceptional long relaxation and dephasing times, the $^{31}$P nuclear spin
is a good candidate as long-lived quantum memory or combined with the donor
electron spin into a hybrid quantum register [Fig. \ref{G_DONOR}(a)]
\cite{SkinnerPRL2003}.

Recently bismuth donors in silicon (Si:Bi) has attracted much attention as it
has a number of advantages over the P donors in silicon. Bismuth has one
long-lived isotope $^{209}$Bi with nuclear spin $I=9/2$ [Fig. \ref{G_DONOR}%
(b)]. Compared with the Si:P system, the Bi donors in silicon have a much
larger nuclear spin $I=9/2$ and a much stronger on-site HFI $A=1.4754$ GHz
\cite{FeherPR1959a} between the Bi electron spin and the $^{209}$Bi nuclear
spin. The much stronger on-site HFI in Si:Bi strongly mixes the electron and
the nuclear spin even under moderate magnetic field. This enables
electron-nuclear hybrid qubit, where each level consists of nearly equal
superpositions of the electronic and Bi nuclear spin components
\cite{MorleyNatMater2013}. Consequently, the strong magnetic dipolar
interaction between the electron spin and the microwave magnetic field can
induce rapid NMR transitions on the nanosecond time scale, two orders of
magnitude faster than conventional NMR \cite{MorleyNatMater2010,GeorgePRL2010}
and several orders of magnitude faster than the decoherence of the hybrid
qubit $T_{2}\sim0.5$ ms, limited by $^{29}$Si nuclear spins. By tuning the
magnetic field to the \textquotedblleft clock\textquotedblright\ transition
between hybridized levels, whose frequency is insensitive to variations in the
magnetic field to first order, the electron spin coherence time $T_{2}$ of up
to 3 s has been observed \cite{WolfowiczNatNano2013}.

\subsubsection{Nitrogen-vacancy centers~in diamond and related systems}

\begin{figure}[ptb]
\includegraphics[width=\columnwidth]{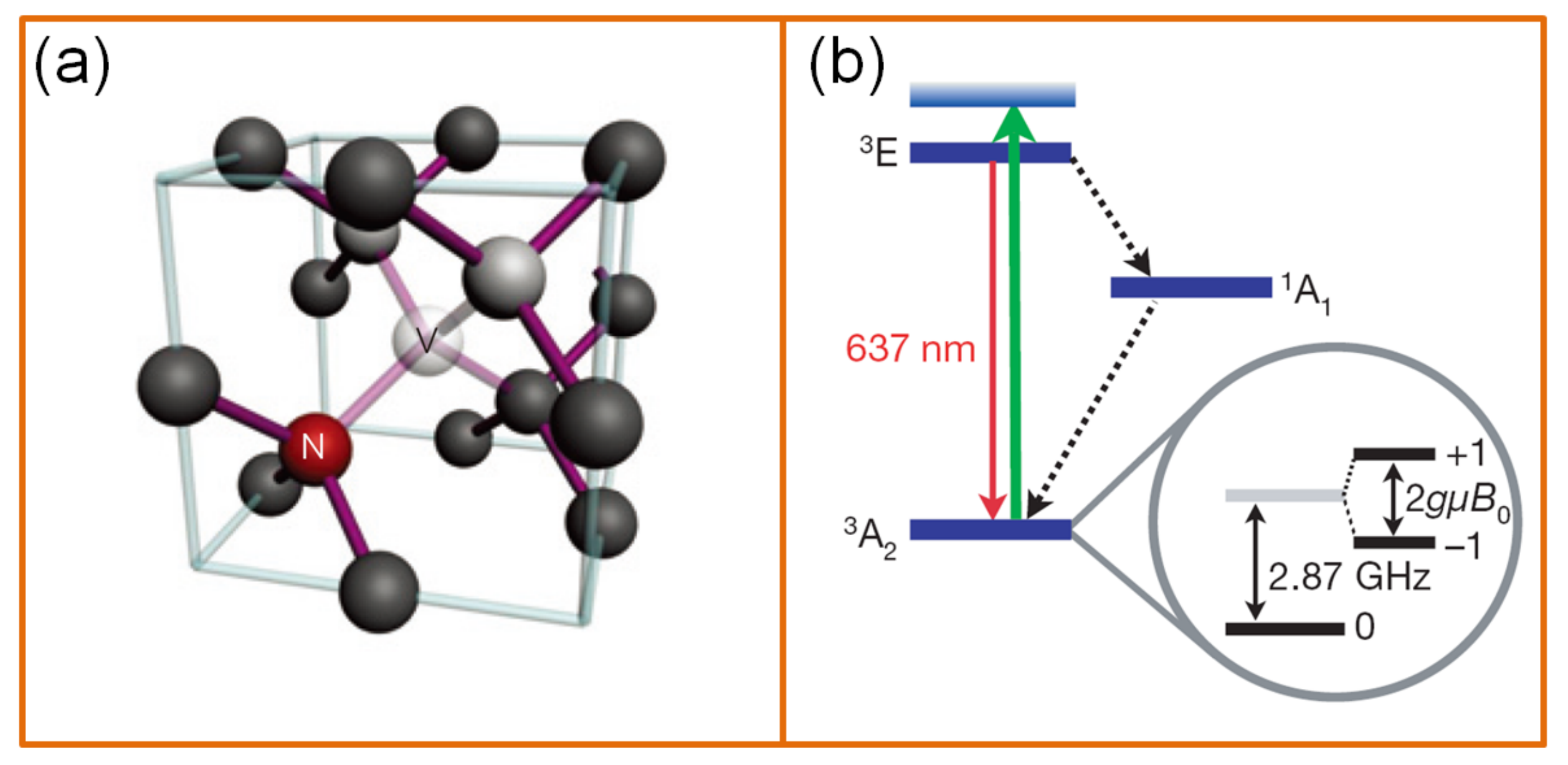}
\caption{(a) Structure of an NV
defect center in the diamond lattice. (b) Energy levels of an NV center in
diamond: green, upward arrow for off-resonant optical transitions, red, downward arrow for
fluorescence, and dashed arrows for non-radiative, spin-flip decay. Panel (a)
is extracted from Ref. \cite{BalasubramanianNatMater2009} and panel (b) is
reproduced from Ref. \cite{MazeNature2008}.}%
\label{G_NV}%
\end{figure}

The negatively charged nitrogen-vacancy (NV) center in diamond consists of a
substitutional nitrogen atom adjacent to a carbon vacancy, which has
C$_{\mathrm{3v}}$ symmetry with the symmetry axis pointing from the nitrogen
to the vacancy (NV axis) [Fig. \ref{G_NV}(a)]. The ground state of the NV
center $^{3}$A$_{2}$ is a spin triplet ($S=1$) with the degenerate $m=\pm1$
doublet states energetically higher than the $m=0$ sublevel by the zero-field
splitting $D_{\mathrm{gs}}=2.87$ GHz [Fig. \ref{G_NV}(b)], where $m$ is the
spin projection along the N-V symmetry axis. Since Gruber \textit{et al.}
observed the magnetic resonance of individual NV centers by optical confocal
microscopy at room temperature \cite{GruberScience1997}, NV centers have been
intensively studied for for quantum information processing
\cite{JelezkoPRL2004a,WrachtrupJPC2006,DuttScience2007,JiangPRL2008} and
quantum sensing
\cite{BalasubramanianNature2008,MazeNature2008,TaylorNatPhys2008,HallPRL2009,HallPNAS2010,RondinRPP2014}%
.

The high Debye temperature of the diamond crystal, the weak spin-orbit
coupling, and low abundance ($\approx1.1\%$) of spinful $^{13}$C isotopes
($I=1/2$) allow very long spin coherence time of the NV center ground state.
The NV electron spin $T_{1}$ can reach a few milliseconds at room temperature
(even as long as minutes at low temperature)
\cite{TakahashiPRL2008,JarmolaPRL2012}. The NV electron spin $T_{2}$ is
usually limited by its coupling to other electron spins and the $^{13}$C
nuclear spins in diamond. In type-Ib diamond samples, the main paramagnetic
centers are nitrogen donors with one unpaired electron spin (the P1 centers).
For typical P1 concentration ($\sim10^{2}$ ppm), these P1 centers limit the NV
electron spin $T_{2}^{\ast}\sim0.1$ $\mathrm{\mu s}$ for FID
\cite{HansonScience2008,LangeScience2010}. In high-purity type-IIa diamond
samples, the NV electron spin $T_{2}^{\ast}$ is limited by HFI with the $^{13}$C
nuclear spin bath to a few microseconds
\cite{KennedyAPL2003,JelezkoPRL2004,JelezkoPRL2004a,HansonPRB2006,ChildressScience2006}%
. When the concentration of the $^{13}$C isotope is reduced by isotropic
purification, the room temperature $T_{2}$ can reach a few milliseconds
\cite{GaebelNature2006,TakahashiPRL2008,BalasubramanianNatMater2009}, limited
by $T_{1}$. Although most of the distant $^{13}$C nuclei weakly coupled to the
NV electron spin serve as a detrimental source of noise that limits the NV
electron spin $T_{2}$, the on-site nitrogen atomic nucleus and a few proximal
$^{13}$C nuclei strongly coupled to the NV center electron spin
\cite{ChildressScience2006} have exceptional long coherence times (exceeding
one second) and can be individually addressed and manipulated through their
HFI with the NV electron spin
\cite{JelezkoPRL2004a,DuttScience2007,NeumannScience2010,PfaffNatPhys2013}.
These nuclear spins serve as a beneficial quantum memory.

The NV center in diamond possesses two distinguishing features compared with
other solid state qubit systems: (i) highly localized electronic states well
isolated from sources of decoherence, leading to millisecond spin coherence
time at room temperature; (ii) a series of optical transitions that allow
high-fidelity optical initialization and readout of the NV electron spin state
under ambient conditions. These exceptional quantum properties have motivated
efforts to search for similar defects in other semiconductors
\cite{WeberPNAS2010}, as they may offer an expanded range of functionality.
First-principle computations and magnetic resonance experiments
\cite{MizuochiPRB2002,SonPRB2003,BaranovJETPL2005,SonPRL2006,WeberPNAS2010,BaranovPRB2011}
suggest several defects in SiC as good candidates, such as Si-C divacancy
\cite{BaranovJETPL2005,KoehlNature2011}, Si and carbon vacancies
\cite{MizuochiPRB2002,OrlinskiPRB2003,BaranovJETP2007,BaranovPRB2011}, and
TV2a center \cite{SonPRB2003}. In particular, the three most common SiC
polytypes (3C-SiC, 4H-SiC, and 6H-SiC) all host optically addressable defect
spin states with long coherence time $\sim$ a few tens of microseconds at room
temperature \cite{FalkNatComm2013}, e.g., $T_{2}^{\ast}\sim1$ $\mathrm{\mu s}$
and $T_{2}\sim$ a few hundred microseconds in 4H-SiC \cite{KoehlNature2011}.
In addition, rare-earth-doped crystals and silicon-vacancy centers in diamond
are receiving increasing interest. A long coherence time $T_{2}=2$ ms close to
the measured $T_{1}=4.5$ ms has been reported for the electron spin of a
single Ce$^{3+}$ ion in yttrium aluminium garnet (YAG) crystal
\cite{SiyushevNatCommun2014}. For electron spins in silicon-vacancy centers in
SiC, $T_{2}^{\ast}>45$ ns and $T_{1}=2.4$ ms have been reported
\cite{PingaultPRL2014,RogersPRL2014}.

\section{Microscopic quantum many-body theories}

\label{SEC_MANYBODY_THEORY}

In previous sections, we have described the central spin decoherence in a
quantum bath using a generic pure dephasing Hamiltonian [Eq. (\ref{HAMIL})].
In this section, we focus on the most relevant issue in quantum
computing:\ the decoherence of a central electron spin in a nanoscale nuclear
spin bath in semiconductor nanostructures, such as quantum dots, donors in
silicon, and diamond NV centers. First we give the microscopic Hamiltonian
relevant for these systems.

\subsection{Microscopic model}

Under moderate to strong external magnetic field (whose axis is defined as the
$z$ axis), the non-secular terms of the HFI between the electron spin and
various intrinsic nuclear spin interactions (as discussed in Sec.
\ref{SEC_NN_INTERACTION}) are suppressed. The total Hamiltonian includes the
electron Zeeman term $\hat{H}_{0}\equiv\omega_{0}\hat{S}_{z}$, the nuclear
Zeeman term (for simplicity, we consider one nuclear spin species with spin
$I$ and gyromagnetic ratio $\gamma_{I}$)
\[
\hat{H}_{Z}\equiv-\gamma_{I}B\sum_{j}\hat{I}_{j}^{z}\equiv\omega_{I}\sum
_{j}\hat{I}_{j}^{z},
\]
the secular part $\hat{S}_{z}\sum_{i}a_{i}\hat{I}_{i}^{z}\equiv\hat{S}_{z}%
\hat{h}_{z}$ of the HFI ($a_{i}$ is the HFI coefficient and $\hat{h}_{z}$ is
widely known as the \textit{nuclear Overhauser field}), the diagonal part
\begin{equation}
\hat{H}_{\mathrm{d}}\equiv\frac{1}{2}\sum_{i\neq j}\lambda_{ij}^{\mathrm{d}%
}\hat{I}_{i}^{z}\hat{I}_{j}^{z} \label{HD}%
\end{equation}
and pair-wise flip-flops part
\begin{equation}
\hat{H}_{\mathrm{ff}}\equiv\sum_{i\neq j}\lambda_{ij}^{\mathrm{ff}}\hat{I}%
_{i}^{+}\hat{I}_{j}^{-} \label{HFF}%
\end{equation}
of intrinsic nuclear spin interactions, and electron-mediated nuclear spin
flip-flop term $2\hat{S}_{z}\tilde{H}_{\mathrm{ff}}$, where
\cite{YaoPRB2006,LiuNJP2007,YaoPRL2007,CywinskiPRB2009}
\begin{equation}
\tilde{H}_{\mathrm{ff}}=\frac{\hat{h}_{x}^{2}+\hat{h}_{y}^{2}}{4\omega_{0}%
}\approx\sum_{i\neq j}\frac{a_{i}a_{j}}{4\omega_{0}}\hat{I}_{i}^{+}\hat{I}%
_{j}^{-}\equiv\sum_{i\neq j}\tilde{\lambda}_{ij}^{\mathrm{ff}}\hat{I}_{i}%
^{+}\hat{I}_{j}^{-}, \label{HFF_TILD}%
\end{equation}
$\hat{h}_{x}$, $\hat{h}_{y}$ are the transverse parts of $\mathbf{\hat{h}%
}\equiv\sum_{i}a_{i}\mathbf{\hat{I}}_{i}$ and in the approximation we have neglected a small correction $\sim
\sum_{i}a_{i}^{2}/\omega_{0}$ to the electron Zeeman splitting. In the
interaction picture with respect to $\hat{H}_{0}+\hat{H}_{Z}$, the total
Hamiltonian assumes the standard pure dephasing form (Eq. (\ref{HAMIL})), where
\cite{YaoPRB2006,LiuNJP2007,YaoPRL2007}
\begin{subequations}
\label{HPM}%
\begin{align}
\hat{H}_{\pm}  &  =\hat{H}_{\mathrm{d}}+\hat{H}_{\mathrm{ff}}\pm\frac{1}%
{2}(2\tilde{H}_{\mathrm{ff}}+\hat{h}_{z})\\
&  =\pm\sum_{j}\frac{a_{j}}{2}\hat{I}_{j}^{z}+\frac{1}{2}\sum_{i\neq j}%
\lambda_{ij}^{\mathrm{d}}\hat{I}_{i}^{z}\hat{I}_{j}^{z}+\sum_{i\neq j}%
(\lambda_{ij}^{\mathrm{ff}}\pm\tilde{\lambda}_{ij}^{\mathrm{ff}})\hat{I}%
_{i}^{+}\hat{I}_{j}^{-}.
\end{align}
This corresponds to $\hat{H}_{B}\equiv\hat{H}_{\mathrm{d}}+\hat{H}%
_{\mathrm{ff}}$ and $\hat{b}\equiv2\tilde{H}_{\mathrm{ff}}+\hat{h}_{z}$. As
listed in Table \ref{TableII}, the HFI $\{a_{i}\}$ are much larger than
various nuclear spin interactions $\lambda_{ij}^{\mathrm{d}},\lambda
_{ij}^{\mathrm{ff}},\tilde{\lambda}_{ij}^{\mathrm{ff}}$. Also note that the
intrinsic nuclear spin interactions $\lambda_{ij}^{\mathrm{d}},\lambda
_{ij}^{\mathrm{ff}}$ are local, i.e., negligible between distant nuclear
spins, while the electron-mediated nuclear spin interactions $\tilde{\lambda
}_{ij}^{\mathrm{ff}}$ are non-local.

\begin{figure}[ptb]
\includegraphics[width=\columnwidth]{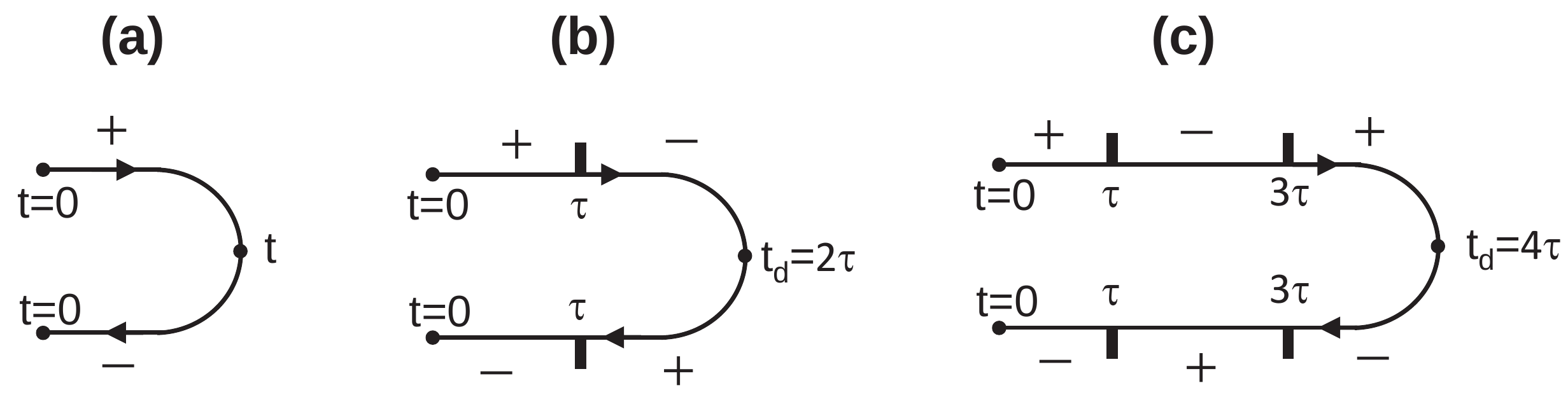}
\caption{Contour
modulation function for different DD sequences: (a) FID, (b) Hahn echo, and
(c) CPMG-2.}%
\label{G_SDD_CONTOUR}%
\end{figure}

The initial state of the nuclear spin bath is the therml state $\hat{\rho}%
_{B}^{\mathrm{eq}}\propto\hat{I}$ [Eq. (\ref{RHOB})], which is maximally mixed
even at very low temperature (e.g., a few Kelvins) due to the small nuclear
Zeeman splitting. Choosing a different initial state of the bath will change the
thermal noise and hence inhomogeneous dephasing, but usually does not
influence the \textquotedblleft true\textquotedblright\ decoherence due to the
quantum noise. For \textquotedblleft true\textquotedblright\ decoherence,
sometimes we may take the initial state of the bath as a pure product state
\end{subequations}
\begin{equation}
|J\rangle=\otimes_{j}|m_{j}\rangle\label{J_PRODUCT}%
\end{equation}
of the Zeeman eigenstate of each bath spin ($\hat{I}_{j}^{z}|m_{j}\rangle
=m_{j}|m_j\rangle$). The central spin decoherence can be written as an integral over the
contour $\mathrm{C}$: $0\rightarrow t_{\mathrm{d}}\rightarrow0$:%
\begin{equation}
L(t_{\mathrm{d}})=\langle\mathcal{T}_{\mathrm{C}}e^{-i\int_{\mathrm{C}}\hat
{H}(z)dz}\rangle, \label{LTD_CONTOUR}%
\end{equation}
where $\langle\cdots\rangle\equiv\langle J|\cdots|J\rangle$ for a pure initial
state or $\langle\cdots\rangle\equiv\operatorname*{Tr}[\hat{\rho}_{B}^{\mathrm{eq}}(\cdots)]$ for a thermal initial state,%
\begin{equation}
\hat{H}(z)=\sum_{j}\omega_{j}(z)\hat{I}_{j}^{z}+\frac{1}{2}\sum_{i\neq
j}\lambda_{ij}^{\mathrm{d}}(z)\hat{I}_{i}^{z}\hat{I}_{j}^{z}+\sum_{i\neq
j}\lambda_{ij}^{\mathrm{ff}}(z)\hat{I}_{i}^{+}\hat{I}_{j}^{-} \label{HZ}%
\end{equation}
is the bath Hamiltonian on the contour, $\omega_{j}(z)\equiv a_{j}s(z)/2$,
$\lambda_{ij}^{\mathrm{ff}}(z)\equiv\lambda_{ij}^{\mathrm{ff}}+s(z)\tilde
{\lambda}_{ij}^{\mathrm{ff}}$, $\lambda_{ij}^{\mathrm{d}}(z)\equiv\lambda
_{ij}^{\mathrm{d}}$, and $s(z)$ is the DD modulation function on the contour
$\mathrm{C}$:\ it start from $+1$ and switch its sign whenever the central
spin is flipped or at $t_{\mathrm{d}}$. FID corresponds to $s(z)\equiv+1$ on
the upper branch and $s(z)\equiv-1$ on the lower branch. Some examples of
$s(z)$ are shown in Fig. \ref{G_SDD_CONTOUR}. For an arbitrary function
$f(z)\equiv f_{+}(t)$ ($z\in$ upper branch) or $f_{-}(t)$ ($z\in$ lower
branch), the contour integral is defined as $\int_{\mathrm{C}}f(z)dz\equiv
\int_{0}^{t_{\mathrm{d}}}f_{+}(t)dt+\int_{t_{\mathrm{d}}}^{0}f_{-}(t)dt$.

\subsection{Noises from spin bath dynamics: general considerations}

The central spin decoherence is the product of inhomogeneous dephasing due to
the thermal noise and \textquotedblleft true\textquotedblright\ decoherence
due to the quantum noise (see Sec. \ref{SEC_QUANTUM_DYN}). The former usually
dominates the FID, but is completely removed by any DD at the echo time, so
only the quantum noise, which is usually independent of the initial state of the bath,
contributes to central spin decoherence under DD.

\subsubsection{Thermal noise}

On the time scale of inhomogeneous dephasing, nuclear spin interactions
$\hat{H}_{\mathrm{d}},$ $\hat{H}_{\mathrm{ff}}$, $\tilde{H}_{\mathrm{ff}}$ can
be neglected and $L(t)\equiv\langle e^{-i\hat{h}_{z}t}\rangle$. For a
sufficiently large number of nuclear spins, $\hat{h}_{z}$ as the sum of many
independent random variables obeys the Gaussian statistics. This gives the
Gaussian inhomogeneous dephasing [cf. Eq. (\ref{L_T2STAR})]
\begin{equation}
L_{\mathrm{inh}}(t)=e^{-(h_{\mathrm{rms}}t)^{2}/2}\equiv e^{-t^{2}%
/(T_{2}^{\ast})^{2}} \label{LINH}%
\end{equation}
on a time scale
\[
T_{2}^{\ast}=\frac{\sqrt{2}}{h_{\mathrm{rms}}}%
\]
with
\[
h_{\mathrm{rms}}^{2}\equiv\langle\hat{h}_{z}^{2}\rangle=\frac{I(I+1)}{3}%
\sum_{i}a_{i}^{2},
\]
which is insensitive to the specific distribution of $\{a_{i}\}$. From Table
I, the typical inhomogeneous dephasing time is estimated as $T_{2}^{\ast}\sim$
a few nanoseconds for a QD containing $N\sim10^{4}$-$10^{6}$ nuclei
\cite{YaoPRB2006,LiuNJP2007,YaoPRL2007}.

\subsubsection{Quantum noises from nuclear spin clusters}

Quantum noises are determined by the quantum fluctuations of the baths.
According to Eq. (\ref{HPM}), the elementary excitations of the bath are
flip-flops of bath spin pairs. On a short time scale, the flip-flops of
different pairs are nearly independent. On a longer time scale, the successive
flip-flop of different pairs involving a common bath spin generates
\textit{correlated} fluctuation of larger and larger clusters. When central
spin decoherence time is relatively long (e.g., for a small $^{13}$C nuclear
spin bath in diamond NV centers), or when the correlated fluctuation of small
clusters are reduced by DD control, the correlated fluctuations of larger
clusters become important.

In recent years, microscopic quantum many-body theories have been developed to
quantitatively describe the correlated fluctuations of nuclear spin clusters
and the induced electron spin decoherence in nanoscale nuclear spin baths. The
pair-correlation approximation \cite{YaoPRB2006,LiuNJP2007,YaoPRL2007} and
density matrix cluster expansion \cite{WitzelPRB2005,WitzelPRB2006} are the
two first quantum many-body theories, which have been independently developed
and are equivalent in the leading order. The former treated the flip-flop of
different pairs as independent and provides a transparent physical picture for
central spin decoherence, but neglects the correlated fluctuation of larger
clusters. The latter provides a convenient way to include the leading-order
effect of correlated fluctuation of larger clusters, but may not converge to
the exact results for relatively small baths. The subsequent theory, the
linked-cluster expansion (LCE) \cite{SaikinPRB2007}, accurately accounts for
the fluctuations due to successively higher-order interactions among the
nuclear spins through Feynman diagrams of successively higher order, but
becomes increasingly inefficient at higher orders and may not converge for a
relatively small spin bath. When each nuclear
spin is coupled to all the other nuclear spins [see Eq. (\ref{HFF_TILD}) for
an example], a large-$N$ expansion ($N$ is the number of nuclear spins) of LCE
is possible (so-called ring diagram approximation
\cite{CywinskiPRB2009,CywinskiPRL2009,CywinskiPRB2010}), which turns out to be
equivalent to the semi-classical noise model
\cite{BluhmNatPhys2011,NederPRB2011}. For a simple and accurate account for the
correlated fluctuation of large clusters, the cluster-correlation expansion
(CCE) has been developed \cite{YangPRB2008a,YangPRB2009a}, which covers the
validity ranges of previous theories, produces the exact results even for
relatively small baths, and has successfully predicted and explained a series
of experimental results for various solid state systems.

In the following subsections, we will review these many-body theories. First
we introduce the LCE, which accounts for various fluctuation processes through
Feynman diagrams. Then we introduce the ring diagram approximation as a
partial summation of an infinite number of certain Feynman diagrams. Next CCE
is introduced as an expansion method corresponding to an infinite summation of
all the Feynman diagrams. Finally, we will give a conceptual understanding of
CCE as a systematic method to treat the correlated fluctuation of larger spin
clusters in a canonical quantum spin system (while the cluster expansion and
disjoint cluster approximation \cite{MazePRB2008,HallPRB2014} can be regarded as certain
approximations to the CCE), thus it can be used to calculate not only central
spin decoherence but also other quantities such as the quantum noise spectrum
of the spin bath.

Before discussing the different many-body theories, we emphasize that in cluster expansion and CCE the term
\textquotedblleft cluster" refers to a group of physical bath spins, e.g., a
three-spin cluster contains three different bath spins. The
\textquotedblleft linked-cluster expansion" (LCE) is a diagrammatic expansion
with respect to the number of interaction lines in a Feynman diagram, e.g., a
third-order Feynman diagram contains three interactions lines, but not
necessarily contain three different bath spins. By contrast, the CCE theory
and the cluster expansion theory are expansions with respect to the number of
bath spins, e.g., a three-spin cluster contains three different bath spins.
Nevertheless, there is a close connection between the number of bath spins
contained in a cluster and the order of Feynman diagrams. This allows us to
establish a connection between the CCE and the LCE (to be discussed shortly).

\subsection{Linked-cluster expansion}

\label{SEC_LCE_RDT}

Linked-cluster expansion (LCE) is a standard many-body technique to evaluate
the average of a general time-ordered exponential, as defined by its Taylor
expansion:
\begin{align}
&  \langle\mathcal{T}_{\mathrm{C}}e^{-i\int_{\mathrm{C}}\hat{O}(z)dz}%
\rangle\nonumber\\
&  \equiv\sum_{n=0}^{\infty}\frac{(-i)^{n}}{n!}\int_{\mathrm{C}}dz_{1}%
\cdots\int_{\mathrm{C}}dz_{n}\langle\mathcal{T}_{\mathrm{C}}[\hat{O }%
(z_{1})\cdots\hat{O}(z_{n})]\rangle, \label{TC_EXP}%
\end{align}
where the average $\langle\cdots\rangle$ is carried out over a
\textit{non-interacting} ensemble of bosons, fermions, or spin systems
\cite{SaikinPRB2007}, and $\hat{O}(z)$ consists of bosonic (or fermionic)
field operators or spin operators in the \textit{interaction picture}. An
example is the contour Hamiltonian in Eq. (\ref{HZ}), where the spin operators
in the interaction picture are $\hat{I}_{i}^{z}(z)\equiv\hat{I}_{i}^{z}$ and
$\hat{I}_{i}^{\pm}(z)\equiv\hat{I}_{i}^{\pm}$.

LCE dictates that the expansion in Eq. (\ref{TC_EXP}) can be reduced to an
exponential function of linked diagrams - hence the name of LCE. When $\hat
{O}(z)=\tilde{\varphi}$ is a classical Gaussian random variable or when
$\hat{O}(z)=\sum_{m}(\alpha_{m}\hat{c}_{m}+\beta_{m}\hat{c}_{m}^{\dagger})$ is
a bosonic field operator, there is only one linked diagram corresponding to
$(-1/2)\langle\tilde{\varphi}^{2}\rangle$ or $(-1/2)\langle\hat{O}^{2}\rangle
$, so LCE reduces to Eq. (\ref{EXP_GAU}) or Eq. (\ref{LCE1}). Here we
introduce the LCE for spin baths relevant for central spin decoherence in
nuclear spin baths, thus the average $\langle\cdots\rangle\equiv
\operatorname*{Tr}[\hat{\rho}_{\mathrm{NI}}(\cdots)]$ refers to a
non-interacting spin bath state%
\begin{align*}
\hat{\rho}_{\mathrm{NI}}  &  =\frac{e^{-\beta\hat{H}_{\mathrm{NI}}}%
}{\operatorname*{Tr}e^{-\beta\hat{H}_{\mathrm{NI}}}},\\
\hat{H}_{\mathrm{NI}}  &  =\sum_{i}\omega_{i}\hat{I}_{i}^{z}.
\end{align*}
The spin operators in the interaction picture are taken to be $\hat{I}_{i}^{\pm}(z)=e^{\pm i\varepsilon_{i}z}\hat{I}_{i}^{\pm}$ and $\hat{I}_{i}^{z}(z)=\hat{I}_{i}^{z}$, where in general $\varepsilon_{i}$ could
be different from $\omega_{i}$.

\subsubsection{LCE for spin baths}

\begin{figure}[ptb]
\includegraphics[width=\columnwidth,clip]{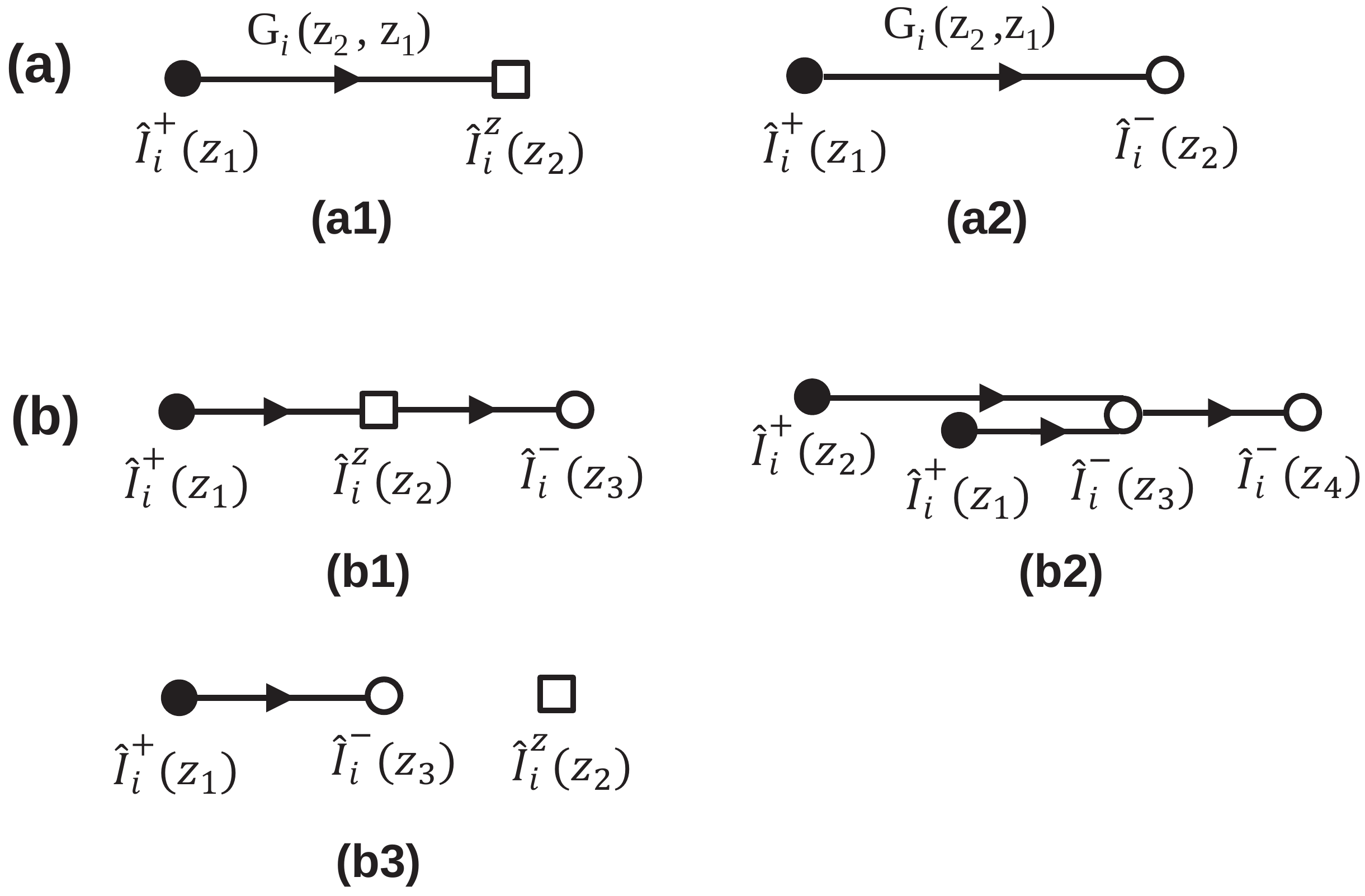}
\caption{Diagrammatic
representation of the contraction of (a) two and (b) three spin operators. The
spin operators $\hat{I}_{+},$ $\hat{I}_{-}$, and $\hat{I}_{z}$ correspond to a
filled circle, an empty circle, and an empty square. }%
\label{G_LCE_WICKTHEOREM}%
\end{figure}

The first key ingredient of LCE for a spin bath is the concept of contraction
\cite{VaksJETP1968}, defined between a spin raising operator $\hat{I}_{i}%
^{+}(z_{1})$ and an arbitrary spin operator $\hat{I}_{j}^{\alpha}(z_{2})$ in
the interaction picture:%
\begin{align}
\lbrack\hat{I}_{j}^{\alpha}(z_{2})]^{\bullet}[\hat{I}_{i}^{+}(z_{1}%
)]^{\bullet}  &  \equiv\lbrack\hat{I}_{i}^{+}(z_{1})]^{\bullet}[\hat{I}%
_{j}^{\alpha}(z_{2})]^{\bullet}\nonumber\\
&  \equiv\delta_{i,j}G_{i}(z_{2},z_{1})e^{i\varepsilon_{i}(z_{1}-z_{2})}[\hat
{I}_{j}^{\alpha},\hat{I}_{i}^{+}](z_{2}).
\end{align}
where $z$ is the time on the contour $\mathrm{C}$, $G_{i}(z_{2},z_{1}%
)=\theta(z_{2}-z_{1})[1+\bar{n}(\omega_{i})]+\theta(z_{1}-z_{2})\bar{n}(\omega_{i})$ is
the contour Green's function, $\theta(z)$ is the Heaviside step function on the contour, and
$\bar{n}(\omega)\equiv1/(e^{\beta\omega}-1)$ is the Bose-Einstein distribution
function. The
contraction can be visualized by Feynman diagrams as sketched in Fig.
\ref{G_LCE_WICKTHEOREM}(a). The contraction of $\hat{I}_{i}^{+}(z_{1})$ and
$\hat{I}_{i}^{\alpha}(z_{2})$ is represented by an arrow going from $\hat
{I}_{i}^{+}(z_{1})$ to $\hat{I}_{i}^{\alpha}(z_{2})$:\ the arrow itself
represents $e^{i\varepsilon_{i}(z_{1}-z_{2})}G_{i}(z_{2},z_{1})$, while the
commutator $[\hat{I}_{i}^{+},\hat{I}_{i}^{\alpha}](z_{\alpha})$ is to be taken
at the end of the arrow. Since $[\hat{I}_{z},\hat{I}_{+}]=\hat{I}_{+}$ and
$[\hat{I}_{-},\hat{I}_{+}]=-2\hat{I}_{z}$, the contraction of $\hat{I}_{i}%
^{+}(z_{1})$ and $\hat{I}_{i}^{z}(z_{2})$ [or $\hat{I}_{i}^{-}(z_{2})$]
eliminates $\hat{I}_{i}^{+}(z_{1})$ and converts $\hat{I}_{i}^{z}(z_{2})$ [or
$\hat{I}_{i}^{-}(z_{2})$] to $\hat{I}_{i}^{+}(z_{2})$ [or $-2\hat{I}_{i}%
^{z}(z_{2})$], reducing the number of spin operators by one. Note that the
contraction of spin operators is quite different from the contraction of
bosonic or fermionic field operators: the latter is just a c-number, while the
former is still a spin operator that should be used in subsequent
contractions. For example, as shown in Fig. \ref{G_LCE_WICKTHEOREM}(b1), the
contraction of $\hat{I}_{i}^{+}(z_{1})$ and $\hat{I}_{i}^{z}(z_{2})$ produces
$\hat{I}_{i}^{+}(z_{2})$, which in turn contracts with $\hat{I}_{i}^{-}%
(z_{3})$ and produces $(-2)\hat{I}_{i}^{z}(z_{3})$. Another example is shown
in Fig. \ref{G_LCE_WICKTHEOREM}(b2): the contraction of $\hat{I}_{i}^{+}%
(z_{1})$ [or $\hat{I}_{i}^{+}(z_{2})$] and $\hat{I}_{i}^{-}(z_{3})$ produces
$(-2)\hat{I}_{i}^{z}(z_{3})$, then $\hat{I}_{i}^{z}(z_{3})$ contracts with
$\hat{I}_{i}^{+}(z_{2})$ [or $\hat{I}_{i}^{+}(z_{1})$] to produce $\hat{I}%
_{i}^{+}(z_{3})$, which in turn contracts with $\hat{I}_{i}^{-}(z_{4})$ to
produce $(-2)\hat{I}_{i}^{z}(z_{4})$. Here the order of the contraction does
not change the result.

The second key ingredient is the Wick's theorem for spin operators
\cite{VaksJETP1968,YangPRB1974,SaikinPRB2007,YangPRB2008a,MaNatCommun2014}.
Lets consider an arbitrary contour time-ordered product of spin operators in
the interaction picture (spin operators commute inside the $\mathcal{T}%
_{\mathrm{C}}$ product)
\begin{equation}
\langle\mathcal{T}_{\mathrm{C}}[\hat{I}_{i}^{\alpha}(z_{1})\cdots\hat{I}%
_{k}^{\gamma}(z_{n})]\rangle=\operatorname*{Tr}\{\hat{\rho}_{\mathrm{NI}%
}\mathcal{T}_{\mathrm{C}}[\hat{I}_{i}^{\alpha}(z_{1})\cdots\hat{I}_{k}%
^{\gamma}(z_{n})]\}. \label{TC_PRODUCT}%
\end{equation}
Wick's theorem states that the contour time-ordered product of spin operators
$\mathcal{T}_{\mathrm{C}}[\cdots]$ in Eq. (\ref{TC_PRODUCT}) can be replaced
by the sum of all possible \textit{fully contracted} products containing only
$\hat{I}_{z}$ operators. If $\mathcal{T}_{\mathrm{C}}[\cdots]$ contains
different numbers of $\hat{I}_{+}$ and $\hat{I}_{-}$ operators, then Eq.
(\ref{TC_PRODUCT}) vanishes. For example, the diagram in Fig.
\ref{G_LCE_WICKTHEOREM}(a1) vanishes since it only contains one $\hat{I}_{+}$
operator but no $\hat{I}_{-}$ operator, while all the other diagrams in Fig.
\ref{G_LCE_WICKTHEOREM} containing equal numbers of $\hat{I}_{+}$ and $\hat
{I}_{-}$ operators are fully contracted products. As another example,
$\mathcal{T}_{\mathrm{C}}[\hat{I}_{i}^{+}(z_{1})\hat{I}_{i}^{z}(z_{2})\hat
{I}_{i}^{-}(z_{3})]$ has two possible fully contracted products: a connected
diagram $([\hat{I}_{i}^{+}(z_{1})]^{\bullet}[\hat{I}_{i} ^{z}(z_{2})]^{\bullet})^{\circ}(\hat{I}_{i}^{-}(z_{3}))^{\circ}$[Fig. \ref{G_LCE_WICKTHEOREM}(b1)] and a disconnected diagram $[\hat{I}_{i}^{+}(z_{1})]^{\bullet}[\hat{I}_{i}^{-}(z_{3})]^{\bullet
}\times\hat{I}_{i}^{z}(z)$  [Fig.
\ref{G_LCE_WICKTHEOREM}(b3)].

\begin{figure}[ptb]
\includegraphics[width=\columnwidth]{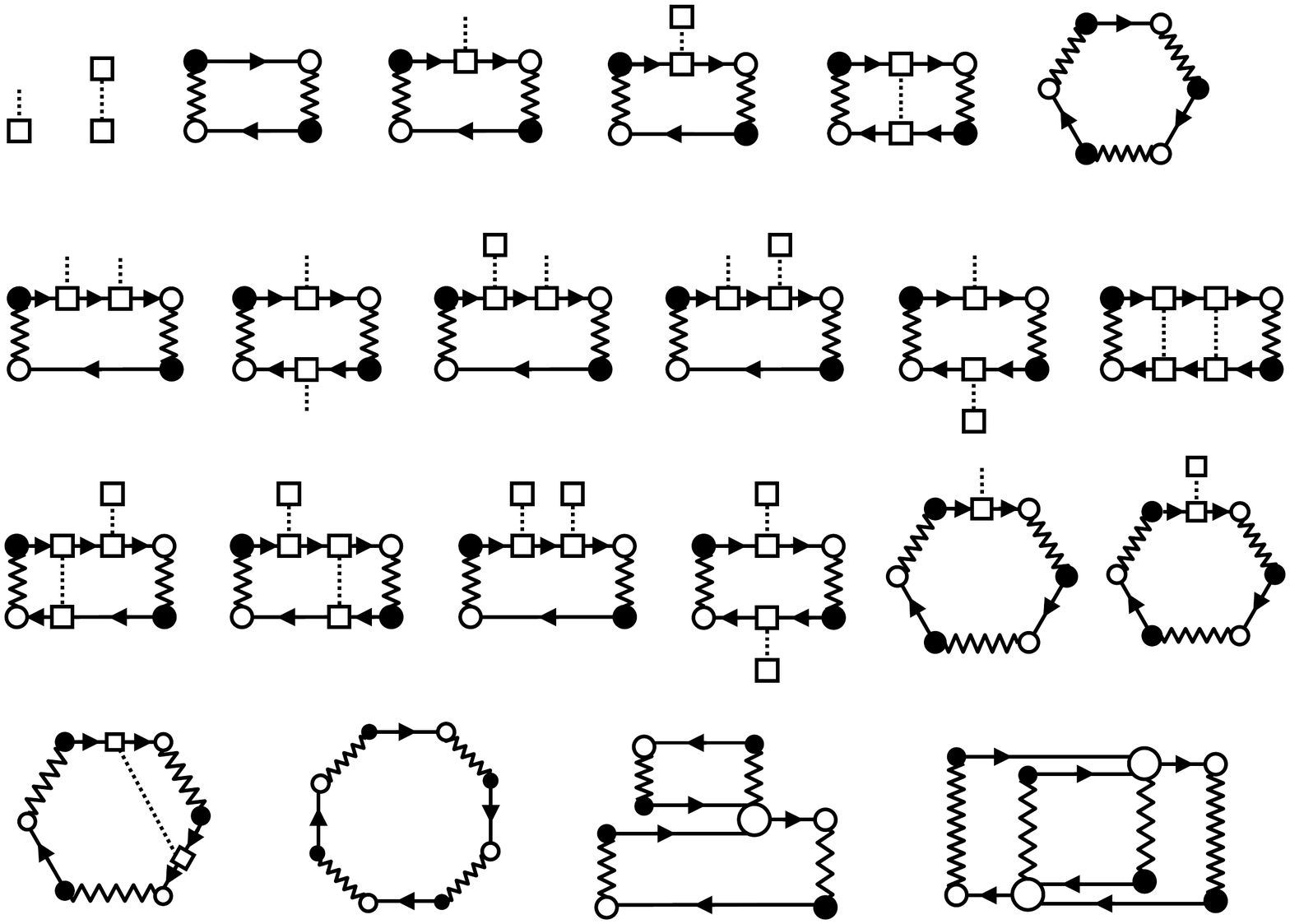}
\caption{Topologically
inequivalent connected diagrams up to the 4th order for the Hamiltonian in
Eq.~(\ref{HZ}). Here dotted lines connected to a single empty square denote
$\omega_{j}(z)$, dotted lines connected to two empty squares denotes
$\lambda_{ij}^{\mathrm{d}}(z)$, and wavy lines denote $\lambda_{ij}%
^{\mathrm{ff}}(z)$, e.g., the first (second) diagram represents the first
(second) term of $\hat{H}(z)$. Reproduced from Ref. \cite{YangPRB2008a}.}%
\label{G_LCE_DIAGRAM}%
\end{figure}

By appling Wick theorem to each $\mathcal{T}_{\mathrm{C}}$ product, Eq.
(\ref{TC_EXP}) can be decomposed as the sum of fully contracted products or
equivalently diagrams, including connected ones and disconnected ones. The LCE
theorem states that all these diagrams can be resummed into an exponential
form \cite{AbrikosovBook1963}:%
\begin{equation}
\langle\mathcal{T}_{\mathrm{C}}e^{-i\int_{\mathrm{C}}\hat{O}(z)dz}%
\rangle=\langle e^{\hat{\pi}}\rangle, \label{LCE_ENSEMBLE}%
\end{equation}
where $\hat{\pi}$ represents the sum of all the \textit{connected} diagrams
contained in Eq. (\ref{TC_EXP}). Taking the contour bath Hamiltonian in Eq.
(\ref{HZ}) as an example, all the topologically inequivalent connected
diagrams up to the 4th order of the bath interactions are shown in
Fig.~\ref{G_LCE_DIAGRAM}.

For a spin-1/2 bath, the average $\left\langle J\left\vert \cdots\right\vert
J\right\rangle $ over a pure product state $|J\rangle$ [Eq. (\ref{J_PRODUCT})]
(an eigenstates of $\hat{H}_{\mathrm{NI}}$) can be taken as the average over
an zero-temperature ensemble $\operatorname*{Tr}[\hat{\rho}_{\mathrm{NI}%
}\cdots]$, with $\omega_{i}<0$ (or $>0$) for $\left\vert m_{i}\right\rangle
=\left\vert \uparrow\right\rangle $ (or $\left\vert \downarrow\right\rangle
$). Therefore the \textquotedblleft true\textquotedblright\ decoherence caused
by a bath in the pure state $|J\rangle$ can be written as
\cite{SaikinPRB2007,YangPRB2008a,MaNatCommun2014}:%
\begin{equation}
L(t_{\mathrm{d}})=\langle J|\mathcal{T}_{\mathrm{C}}e^{-i\int_{\mathrm{C}}%
\hat{H}(z)dz}|J\rangle=e^{\pi}, \label{LCE_PURE}%
\end{equation}
where
\begin{equation}
\pi\equiv\langle J|\hat{\pi}|J\rangle=\langle J|\mathcal{T}_{\mathrm{C}%
}e^{-i\int_{\mathrm{C}}\hat{H}(z)dz}|J\rangle_{\mathrm{connected}}
\label{PI_DEF}%
\end{equation}
is the sum of all connected Feynman diagrams contained in $\langle
J|\mathcal{T}_{\mathrm{C}}e^{-i\int_{\mathrm{C}}\hat{H}(z)dz}|J\rangle$. Here
each constituent diagram of $\pi$ is a c-number, obtained from the constituent
diagram of $\hat{\pi}$ by replacing $\hat{I}_{i}^{z}$ with $\langle J|\hat
{I}_{i}^{z}|J\rangle=m_{i}$. When bath spins are higher than 1/2, the average
$\left\langle J\left\vert \cdots\right\vert J\right\rangle $ over a pure
product state $|J\rangle$ [Eq. (\ref{J_PRODUCT})] can still be taken as an
zero-temperature ensemble $\operatorname*{Tr}[\hat{\rho}_{\mathrm{NI}}\cdots]$
as long as each bath spin is mapped to a composite of pseudo-spin-1/2's
\cite{YangPRB2008a}.

The LCE has been applied to the phosphorus donor electron spin in a $^{29}$Si
nuclear spin bath \cite{SaikinPRB2007,MaNatCommun2014}. The connected Feynman
diagrams have been evaluated up to the fourth order of the bath interactions.
The results agree reasonably with the experimental data
\cite{TyryshkinJPC2006}. The FID is dominated by the leading-order flip-flop
process of nuclear spin pairs (the third diagram in Fig. \ref{G_LCE_DIAGRAM}).
Under higher-order DD, it is necessary to take higher-order diagrams into
account, but the tedious diagram counting and evaluation makes it difficult to
go to very high orders.

\subsubsection{Ring diagram approximation}

\begin{figure}[ptb]
\includegraphics[width=\columnwidth,clip]{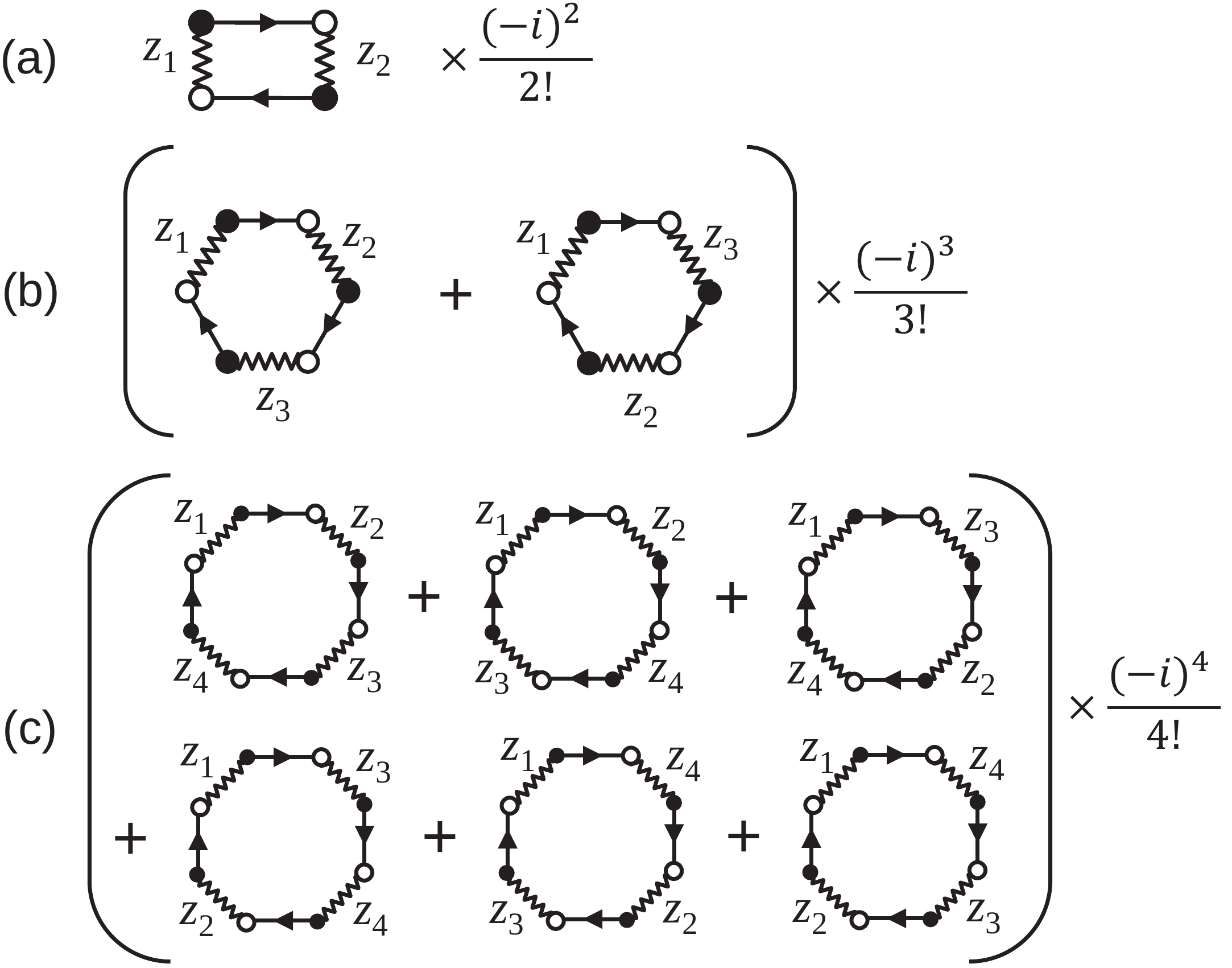}
\caption{Ring diagrams
containing up to (a) two, (b) three, and (c) four nucelar spins.}%
\label{G_RDT}%
\end{figure}

The difficulty in counting and evaluating higher-order Feynman diagrams in LCE
could be greatly simplified under a relatively weak magnetic field
\cite{CywinskiPRB2009,CywinskiPRL2009,CywinskiPRB2010}, where the
electron-mediated nuclear spin interactions $\tilde{\lambda}_{ij}%
^{\mathrm{ff}}$ dominates over the intrinsic interactions $\lambda
_{ij}^{\mathrm{ff}}$ and $\lambda_{ij}^{\mathrm{d}}$. In this case
$\lambda_{ij}^{\mathrm{ff}}$ and $\lambda_{ij}^{\mathrm{d}}$ in Eq. (\ref{HZ})
can be dropped, the total Hamiltonian becomes $\hat{H}=\hat{S}_{z}\hat{b}$
with $\hat{b}\equiv\hat{h}_{z}+2\tilde{H}_{\mathrm{ff}}$ [Eq. (\ref{HFF_TILD}%
)], and the decoherence $L(t_{\mathrm{d}})=\langle e^{-i\hat{b}\int_{0}^{t_{\mathrm{d}}}s(t)dt}\rangle$ [cf. Eq.
(\ref{LTD_CONTOUR})] is completely removed by any DD at the echo time (in this
case decoherence comes from the flip-flop between nuclei of different species
\cite{CywinskiPRB2009,CywinskiPRL2009,CywinskiPRB2010}). The \textquotedblleft
true\textquotedblright\ decoherence due to quantum noises in FID
\cite{CywinskiPRB2009,CywinskiPRL2009,CywinskiPRB2010},
\[
L_{\mathrm{dyn}}(t)=\langle J|e^{-i\hat{b}t}|J\rangle=\langle J|\mathcal{T}%
e^{-2i\int_{0}^{t}\tilde{H}_{\mathrm{ff}}(t^{\prime})dt^{\prime}}%
|J\rangle\approx\langle\mathcal{T}e^{-2i\int_{0}^{t}\tilde{H}_{\mathrm{ff}}%
(t^{\prime})dt^{\prime}}\rangle
\]
is described by the simplified \textquotedblleft contour\textquotedblright%
\ Hamiltonian$\allowbreak\allowbreak\allowbreak$
\[
\tilde{H}_{\mathrm{ff}}(t)=\sum_{i\neq j}\tilde{\lambda}_{ij}^{\mathrm{ff}%
}\hat{I}_{i}^{+}\hat{I}_{j}^{-}e^{i(a_{i}-a_{j})t/2},
\]
thus all the Feynman diagrams (Fig. \ref{G_LCE_DIAGRAM}) involving
$\lambda_{ij}^{\mathrm{d}}(z)$ and $\omega_{j}(z)$ vanish. Since each nuclear
spin couples to all the other nuclear spins with comparable strength, the
value of a Feynman diagram involving $m$ different nuclear spins is of the
order $O(N^{m}$) with $N$ being the number of nuclear spins. This enables a
leading-order expansion with respect to $1/N$
\cite{SaikinPRB2007,CywinskiPRB2009,CywinskiPRL2009,CywinskiPRB2010}:\ among
all the Feynman diagrams containing the same number of interaction lines, it
suffices to keep only those diagrams involving the maximal number of nuclear
spins, i.e. all the ring diagrams contained in $\langle\mathcal{T}%
e^{-2i\int_{0}^{t}\tilde{H}_{\mathrm{ff}}(t^{\prime})dt^{\prime}}\rangle$ (Fig.
\ref{G_RDT}). This is the ring diagram approximation
\cite{CywinskiPRB2009,CywinskiPRL2009,CywinskiPRB2010}. For $t\ll$ inverse
HFI, it gives a power-law decay \footnote{Here we have removed a phase factor
$e^{it/T_{\mathrm{dyn}}}$, which is an artifact of using $2\tilde
{H}_{\mathrm{ff}}=2\sum_{i\neq j}\tilde{\lambda}_{ij}^{\mathrm{ff}}\hat{I}%
_{i}^{+}\hat{I}_{j}^{-}$ instead of the more accurate expression $2\tilde
{H}_{\mathrm{ff}}=(\hat{h}_{x}^{2}+\hat{h}_{y}^{2})/(2\omega_{0})$ [see Eq.
(\ref{HFF_TILD})]:\ the latter contains a small correction $(1/2\omega
_{0})\sum_{i}a_{i}^{2}[I(I+1)-(\hat{I}_{i}^{z})^{2}]\approx1/T_{\mathrm{dyn}}$
to the electron Zeeman splitting.}
\begin{equation}
L_{\mathrm{dyn}}(t)\approx\frac{1}{1+it/T_{\mathrm{dyn}}}, \label{LDYN_RDT}%
\end{equation}
which is insensitive to the specific distribution of the HFI coefficients $\{a_{i}\}$, on a time-scale%
\begin{equation}
T_{\mathrm{dyn}}\equiv\frac{\omega_{0}}{h_{\mathrm{rms}}^{2}}. \label{TDYN_RDT}%
\end{equation}
For $t\gg T_{\mathrm{dyn}}$, the denominator $1+it/T_{\mathrm{dyn}}\approx
it/T_{\mathrm{dyn}}$ gives rise to a $\pi/2$ phase shift of the electron
Larmor precession. Power-law behavior and a long-time $\pi/4$ phase shift has
also been observed in the single-spin Rabi oscillation decay
\cite{KoppensPRL2007,HansonScience2008}. For long times $t\gg$ inverse HFI, it
gives an exponential decay \cite{LiuNJP2007,CoishPRB2008} on a time scale that
depends sensitively on the distribution of $\{a_{i}\}$. The relevance of the
power-law decay and the exponential decay depends on the magnetic field. For
weak fields such that $T_{\mathrm{dyn}}\ll$ inverse HFI, most of the coherence
decay follows the short-time power-law behavior in Eq. (\ref{LDYN_RDT}). By
contrast, in the opposite limit $T_{\mathrm{dyn}}\gg$ inverse HFI, most of the
coherence decays exponentially. Under spin echo, the calculated decoherence
under weak magnetic fields agrees with the experiment in gated GaAs QDs
\cite{KoppensPRL2008}. At slightly stronger magnetic fields, characteristic
oscillations with frequencies equal to the differences of the nuclear Zeeman
frequencies of different nuclear species is predicted \cite{CywinskiPRL2009,CywinskiPRB2009} and subsequently
observed experimentally \cite{BluhmNatPhys2011}.

Essentially, the ring diagram approximation assumes that all the spin
operators in $\tilde{H}_{\mathrm{ff}}(z)$ commute with each other
\cite{CywinskiPRB2009,CywinskiPRL2009}. This involves an $O(1/N)$ error
\cite{BarnesPRB2011} and is equivalent to a semi-classical treatment of the
quantum noise. A detailed discussion can be found in Ref. \cite{NederPRB2011}
and \cite{CywinskiAPPA2011}. Taking the \textquotedblleft
true\textquotedblright\ decoherence in the FID as an example, on a time scale
$t\ll$ inverse HFI,
\[
L_{\mathrm{dyn}}(t)=\langle e^{-2i{\tilde{H}}_{\mathrm{ff}}t}\rangle=\langle
e^{-i(\hat{h}_{x}^{2}+\hat{h}_{y}^{2})t/(2\omega_{0})}\rangle.
\]
Regarding ${\hat{h}}_{x}$ and ${\hat{h}}_{y}$ as independent, classical
quasi-static Gaussian noise obeying the distribution $P(h)=e^{-h^{2}%
/(2h_{\mathrm{rms}}^{2})}/(\sqrt{2\pi}h_{\mathrm{rms}})$ gives
\begin{equation}
L_{\mathrm{dyn}}(t)\approx\int P(h_{x})dh_{x}\int P(h_{y})dh_{y}%
\ e^{-i({h}_{x}^{2}+{h}_{y}^{2})t/(2\omega_{0})},\nonumber
\end{equation}
which reproduces Eqs. (\ref{LDYN_RDT}) and (\ref{TDYN_RDT}).

\subsection{Cluster-correlation expansion}

The key idea of CCE is to factorize Eq. (\ref{LTD_CONTOUR}) into the product
of cluster-correlation terms, each of which accounting for the irreducible,
correlated fluctuations in a given bath spin cluster. For a finite-time
evolution as in the central spin decoherence problem, a convergent result is
obtained by truncating the expansion up to a certain cluster size. The
two-spin cluster truncation of the CCE corresponds to the pair-correlation
approximation \cite{YaoPRB2006,LiuNJP2007,YaoPRL2007}. When the central spin
decoherence comes from the contribution of a large number of cluster
correlation terms and the contribution from each individual term is small, as
the usual case for relatively large baths, CCE coincides with the cluster
expansion \cite{WitzelPRB2005,WitzelPRB2006}. For small baths, however,
central spin decoherence may be dominated by the coherent dynamics of a few
cluster correlation terms. In this case, CCE converges to the exact results
while the cluster expansion does not.

CCE has been applied to electron spin decoherence for phosphorus donor in
silicon (Si:P) \cite{WitzelPRL2010}, bismuth donors in silicon (Si:Bi)
\cite{GeorgePRL2010,BalianPRB2012,MorleyNatMater2013}, radical spins in
malonic acid crystals~\cite{DuNature2009}, and diamond NV
center~\cite{ZhaoPRB2012}. The calculated results agree well with the
experimental data, including the decoherence time scale, the temporal profile,
and its dependence on the magnetic field. The CCE also provides convincing
theoretical demonstrations of atomic-scale sensing of distant nuclear spin
clusters~\cite{ZhaoNatNano2011} and anomalous decoherence
effects~\cite{ZhaoPRL2011}. Both effects have been observed subsequently
\cite{HuangNatCommun2011,ShiNatPhys2014} and these experiments are well
explained by CCE calculations.

CCE can be understood in two different viewpoints. First, CCE is a re-grouping
and infinite summation of all the LCE diagrams \cite{YangPRB2008a}. Second,
CCE is a systematic method to treat the irreducible, correlated fluctuation of
successively larger spin clusters in a canonical ensemble, so CCE can also be
used to calculate other quantities such as the quantum noise spectrum. The
first understanding applies to a pure product state $|J\rangle$ of the bath
[Eq. (\ref{J_PRODUCT})]. The second understanding leads to two formulations:
CCE for a general non-interacting bath state has a simpler form
\cite{YangPRB2009a}, but CCE for a pure product state of a general spin bath
has better convergence \cite{YangPRB2008a}.

\subsubsection{CCE as infinite summation of LCE diagrams}

\label{SEC_CCE_LCE}

\begin{figure}[ptb]
\includegraphics[width=\columnwidth]{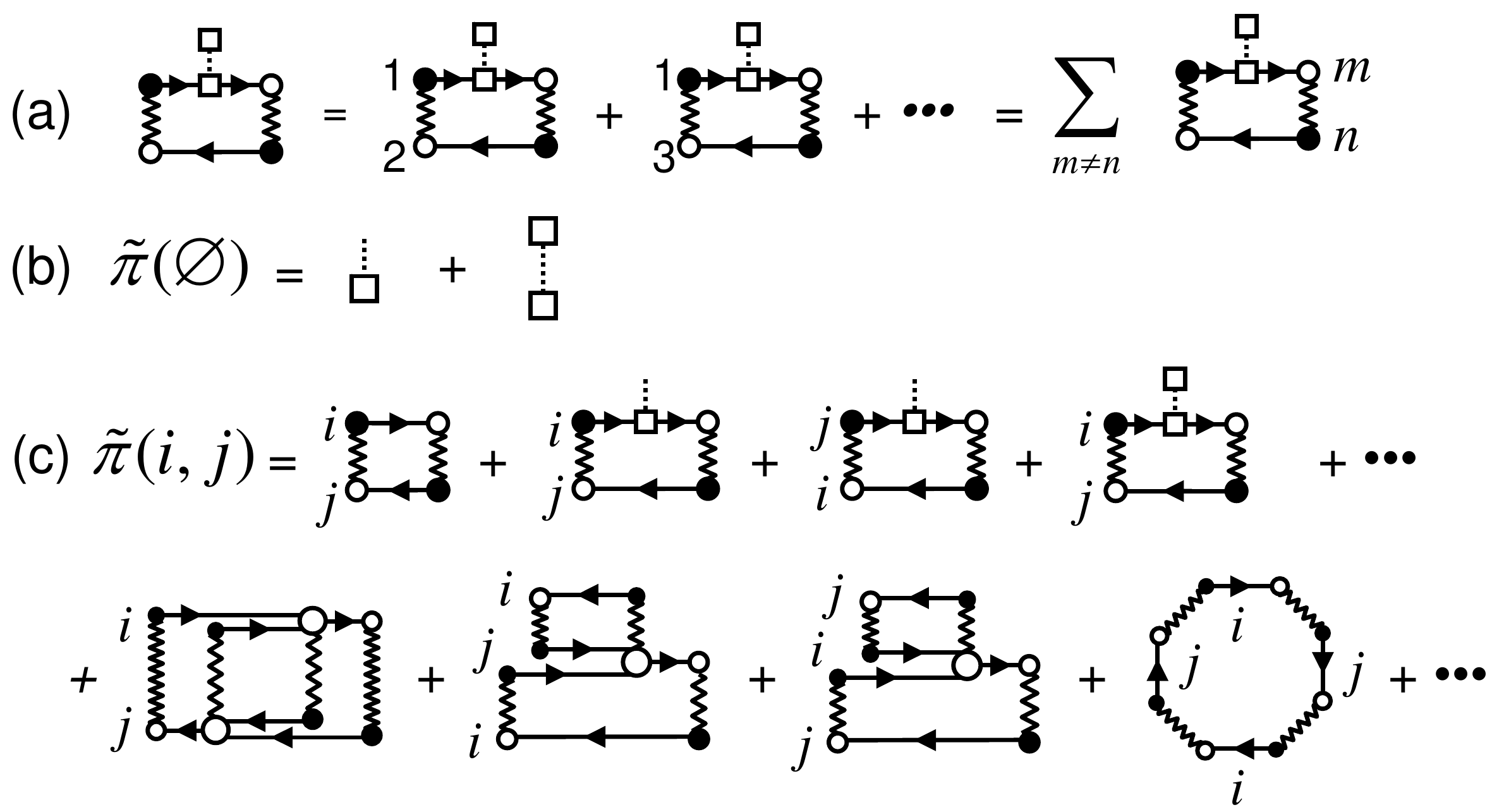}
\caption{(a) Expansion of a
third-order connected diagram into diagrams involving the flip-flops of
different clusters of spins. (b) and (c) show the diagrams contained in
$\tilde{\pi}(\varnothing)$ and $\tilde{\pi}(i,j)$, respectively. Reproduced
from Ref. \cite{YangPRB2008a}.}%
\label{G_CCE_LCE}%
\end{figure}

For a pure product state $|J\rangle$ [Eq. (\ref{J_PRODUCT})] of the spin bath,
the connection between LCE [Eqs. (\ref{LCE_PURE}) and (\ref{PI_DEF})] to CCE
is based on the observation that each connected LCE diagram can be expanded as
the sum of diagrams involving the flip-flops of different clusters of spins.
As an example shown in Fig.~\ref{G_CCE_LCE}(a) for a spin-1/2 bath, a
third-order LCE diagram involving the flip-flop of a spin pair contains
diagrams for spin clusters $(1,2)$, $(1,3)$, $\cdots$, where the numbers stand
for the indices of the spins that have been flipped. Thus all the connected
diagrams can be classified according to the spin clusters instead of the
interaction orders. For an arbitrary (empty or non-empty) cluster
$\mathcal{C}$, we define the \textit{cluster-correlation term }$\tilde{\pi
}({\mathcal{C}})$ as the sum of all connected diagrams in which all and only
the spins in cluster $\mathcal{C}$ have been flipped. For instance, some of
the diagrams constituting $\tilde{\pi}(\varnothing)$ and $\tilde{\pi}(i,j)$
for a spin-1/2 Hamiltonian in Eq.~(\ref{HZ}) are shown in
Figs.~\ref{G_CCE_LCE}(b) and \ref{G_CCE_LCE}(c), respectively.

With these $\{\tilde{\pi}\}$ functions, the exact LCE is expressed as
\begin{equation}
\pi=\sum\limits_{\mathcal{C}\subseteq\{1,2,\cdots,N\}}\tilde{\pi}\left(
{\mathcal{C}}\right)  , \label{LCE_EXPANSION}%
\end{equation}
i.e., the sum of cluster-correlation terms for all spin clusters (including
the empty cluster $\varnothing$) in the $N$-spin bath. In particular, the
infinite summation of all the connected diagrams for a certain cluster
${\mathcal{C}}$ and all its subsets
\begin{equation}
\pi\left(  {\mathcal{C}}\right)  \equiv\sum_{{\mathcal{C}}^{\prime}%
\subseteq{\mathcal{C}}}\tilde{\pi}\left(  {\mathcal{C}}^{\prime}\right)  ,
\label{PI_PITILDE}%
\end{equation}
is just equal to $\pi$ [Eq.~(\ref{PI_DEF})] with all the terms involving the
flip-flop of spins outside the cluster ${\mathcal{C}}$ dropped, or,
equivalently, with the bath Hamiltonian $\hat{H}(z)=\hat{H}(\hat{\mathbf{I}%
}_{1},\cdots,\hat{\mathbf{I}}_{N})$ replaced with $H(\{\hat{\mathbf{I}}%
_{j\in\mathcal{C}}\},\{\langle J|\hat{\mathbf{I}}_{j\notin\mathcal{C}%
}|J\rangle\})$ in which the spins outside the cluster are mean-field averaged.
Thus we have
\begin{equation}
e^{\pi\left(  {\mathcal{C}}\right)  }=\langle J|\mathcal{T}_{\mathrm{C}%
}e^{-i\int_{\mathrm{C}}\hat{H}(\{\hat{\mathbf{I}}_{j\in\mathcal{C}%
}\},\{\langle J|\hat{\mathbf{I}}_{j\notin\mathcal{C}}|J\rangle\})dz}|J\rangle.
\label{LC_DEF}%
\end{equation}
For small clusters $\mathcal{C}$, the Hamiltonian $\hat{H}(\{\hat{\mathbf{I}%
}_{j\in\mathcal{C}}\},\{\langle J|\hat{\mathbf{I}}_{j\notin\mathcal{C}%
}|J\rangle\})$ only contains spin operators inside the cluster $\mathcal{C}$,
thus $\pi(\mathcal{C})$ can be calculated from Eq. (\ref{LC_DEF}) by direct
diagonalization. This, in turn, allows $\{\tilde{\pi}(\mathcal{C})\}$ to be
extracted recursively from Eq. (\ref{PI_PITILDE}):
\begin{equation}
\tilde{\pi}(\mathcal{C})=\pi\left(  {\mathcal{C}}\right)  -\sum_{{\mathcal{C}%
}^{\prime}\subset{\mathcal{C}}}\tilde{\pi}\left(  {\mathcal{C}}^{\prime
}\right)  , \label{PITILDE_C}%
\end{equation}
e.g., $\tilde{\pi}(\varnothing)=\pi(\varnothing)$ and $\tilde{\pi}%
(i)=\pi\left(  i\right)  -\tilde{\pi}(\varnothing)$.

For a cluster $\mathcal{C}$ containing $|\mathcal{C}|$ bath spins, each
diagram in $\tilde{\pi}(\mathcal{C})$ consists of at least $|\mathcal{C}|$
off-diagonal interaction lines that connect all the spins in cluster
$\mathcal{C}$ into a linked cluster (see Fig.~\ref{G_CCE_LCE} for examples),
thus $\tilde{\pi}(\mathcal{C})\sim(\lambda_{\mathrm{ff}}t_{\mathrm{d}%
})^{|\mathcal{C}|}$, where $\lambda_{\mathrm{ff}}$ is the typical value of the
off-diagonal interactions $\lambda_{ij}^{\mathrm{ff}}(z)$. On a time scale
$t_{\mathrm{d}}\ll$ $1/\lambda_{\mathrm{ff}}$, cluster-correlation terms of
large clusters are small, so Eq. (\ref{LCE_EXPANSION}) can be truncated, e.g.,
keeping cluster-correlation terms containing up to $M$ bath spins gives the
$M$th-order truncated CCE (CCE-$M$ for short):%
\[
\pi^{(M)}=\sum\limits_{|\mathcal{C}|\leq M}\tilde{\pi}\left(  {\mathcal{C}%
}\right)  =\tilde{\pi}(\varnothing)+\tilde{\pi}^{(1)}+\cdots+\tilde{\pi}%
^{(M)},
\]
where $\tilde{\pi}^{(M)}\equiv%
{\textstyle\sum_{|\mathcal{C}|=M}}
\tilde{\pi}\left(  {\mathcal{C}}\right)  $ is the total contribution from all
$M$-spin cluster-correlation terms. If each bath spin interacts, on average,
with $q$ spins, then the number of linked clusters containing $m$ bath spins
is $\sim Nq^{m-1}$ and $\tilde{\pi}^{(m)}\sim(N/q)(q\lambda_{\mathrm{ff}%
}t_{\mathrm{d}})^{m}$. Therefore, a sufficient condition for convergence of
CCE is
\begin{equation}
t_{\mathrm{d}}<\frac{1}{q\lambda_{\mathrm{ff}}}. \label{CCE_SINGLE_CONVERGE}%
\end{equation}
which is usually much longer than the electron spin decoherence time.

\subsubsection{CCE without LCE: general non-interacting bath state}

\label{SEC_CCE_ENSEMBLE}

The CCE formalism described below can be directly used to calculate the
average of a general time-ordered exponential over an arbitrary
non-interacting ensemble $\hat{\rho}_{\mathrm{NI}}=\otimes_{i=1}^{N}\hat{\rho
}_{i}$ ($\hat{\rho}_{i}$ for the $i$th bath spin), i.e., Eq.
(\ref{LTD_CONTOUR}) with $\hat{H}(z)$ being a general spin bath Hamiltonian
[not necessarily Eq. (\ref{HZ})] and $\langle\mathcal{\cdots}\rangle
\equiv\operatorname*{Tr}[\hat{\rho}_{\mathrm{NI}}(\cdots)]$. However, for
clarity we consider Eq. (\ref{LTD_CONTOUR}) with $\hat{H}(z)$ given by Eq.
(\ref{HZ}), but the initial state of the bath is a general non-interacting state
$\hat{\rho}_{\mathrm{NI}}$.

The first step is to define the \textit{cluster term}
\begin{equation}
L(\mathcal{C})\equiv\langle\mathcal{T}_{\mathrm{C}}e^{-i\int_{\mathrm{C}}%
\hat{H}_{\mathcal{C}}(z)dz}\rangle, \label{LC_CCE_ENSEMBLE}%
\end{equation}
where $\hat{H}_{\mathcal{C}}(z)$ is obtained from $\hat{H}(z)$ by dropping all
the bath spins outside cluster $\mathcal{C}$, e.g., $\hat{H}_{\{i\}}%
(z)=\omega_{i}(z)\hat{I}_{i}^{z}$ and $\hat{H}_{\{i,j\}}(z)=\omega_{i}%
(z)\hat{I}_{i}^{z}+\omega_{j}(z)\hat{I}_{j}^{z}+\lambda_{ij}^{\mathrm{d}%
}(z)\hat{I}_{i}^{z}\hat{I}_{j}^{z}+\lambda_{ij}^{\mathrm{ff}}(z)(\hat{I}%
_{i}^{+}\hat{I}_{j}^{-}+h.c.)$. The key observation is that when a cluster
$\mathcal{C}$ can be divided into two subsets $\mathcal{C}_{1}$ and
$\mathcal{C}_{2}$ such that $\hat{H}_{\mathcal{C}}(z)$ does not contain any
interaction between the two subsets, $L(\mathcal{C})$ is
factorizable:$\ L(\mathcal{C})=L(\mathcal{C}_{1})L(\mathcal{C}_{2})$. This
allows singling out the irreducible, correlated fluctuations of different
clusters by defining a hierarchy of \textit{cluster-correlation terms}
$\{\tilde{L}(\mathcal{C})\}$:
\begin{subequations}
\label{CCE_ENSEMBLE_DEF}%
\begin{align}
\tilde{L}(i)  &  \equiv L(i),\\
\tilde{L}(i,j)  &  \equiv\frac{L(i,j)}{\tilde{L}(i)\tilde{L}(j)},\\
&  \cdots\nonumber\\
\tilde{L}(\mathcal{C})  &  \equiv\frac{L(\mathcal{C})}{\prod_{\mathcal{C}%
^{\prime}\subset C}\tilde{L}(\mathcal{C}^{\prime})}.
\end{align}
\ The central spin coherence is expressed \textit{exactly} as the product of
all possible cluster-correlation terms:%
\end{subequations}
\begin{equation}
L=\prod_{\mathcal{C}\subseteq\{1,2,\cdots,N\}}\tilde{L}(\mathcal{C})=\left(
\prod_{i}L(i)\right)  \left(  \prod_{i\neq j}\tilde{L}(i,j)\right)  \cdots.
\label{CCE_EXACT}%
\end{equation}

Since $\hat{H}(z)$ contains pair-wise interactioins $\lambda_{ij}$ (including
$\lambda_{ij}^{\mathrm{d}}$ and $\lambda_{ij}^{\mathrm{ff}}$), the
pair-correlation term $\ln\tilde{L}(i,j)$ is \textit{at least} first order in
$\lambda_{ij}t_{\mathrm{d}}$ since $L(i,j)=L(i)L(j)$ and hence $\ln\tilde
{L}(i,j)=0$ when $\lambda_{ij}=0$. By mathematical induction, it can be proved
that the cluster-correlation term $\ln\tilde{L}(\mathcal{C})$ vanishes if the
interactions contained in $\hat{H}_{\mathcal{C}}(z)$ cannot connect all the
spins in group $\mathcal{C}$ into a linked cluster. Consequently, in the
Taylor expansion of $\ln\tilde{L}(\mathcal{C})$ with respect to the bath
interaction times the evolution time $t_{\mathrm{d}}$, the bath interaction
coefficients contained in every term must: (i) connect all the spins in group
$\mathcal{C}$ into a linked cluster, (ii) ensure that each spin inside cluster
$\mathcal{C}$ is flipped an even number ($0$, $2$, $4$, $\cdots$) of times,
and, if the initial state of the bath $\hat{\rho}_{\mathrm{NI}}\propto\hat{I}$ is
maximally mixed, (iii) ensure that every bath spin inside cluster
$\mathcal{C}$ appear an even number ($2,4,6,\cdots$) of times. Condition (i)
alone ensures that $\ln\tilde{L}(\mathcal{C})$ is at least $(|\mathcal{C}%
|-1)$th order, while conditions (ii) and (iii) usually make the order of
$\ln\tilde{L}(\mathcal{C})$ even higher.

\begin{figure}[ptb]
\includegraphics[width=\columnwidth,clip]{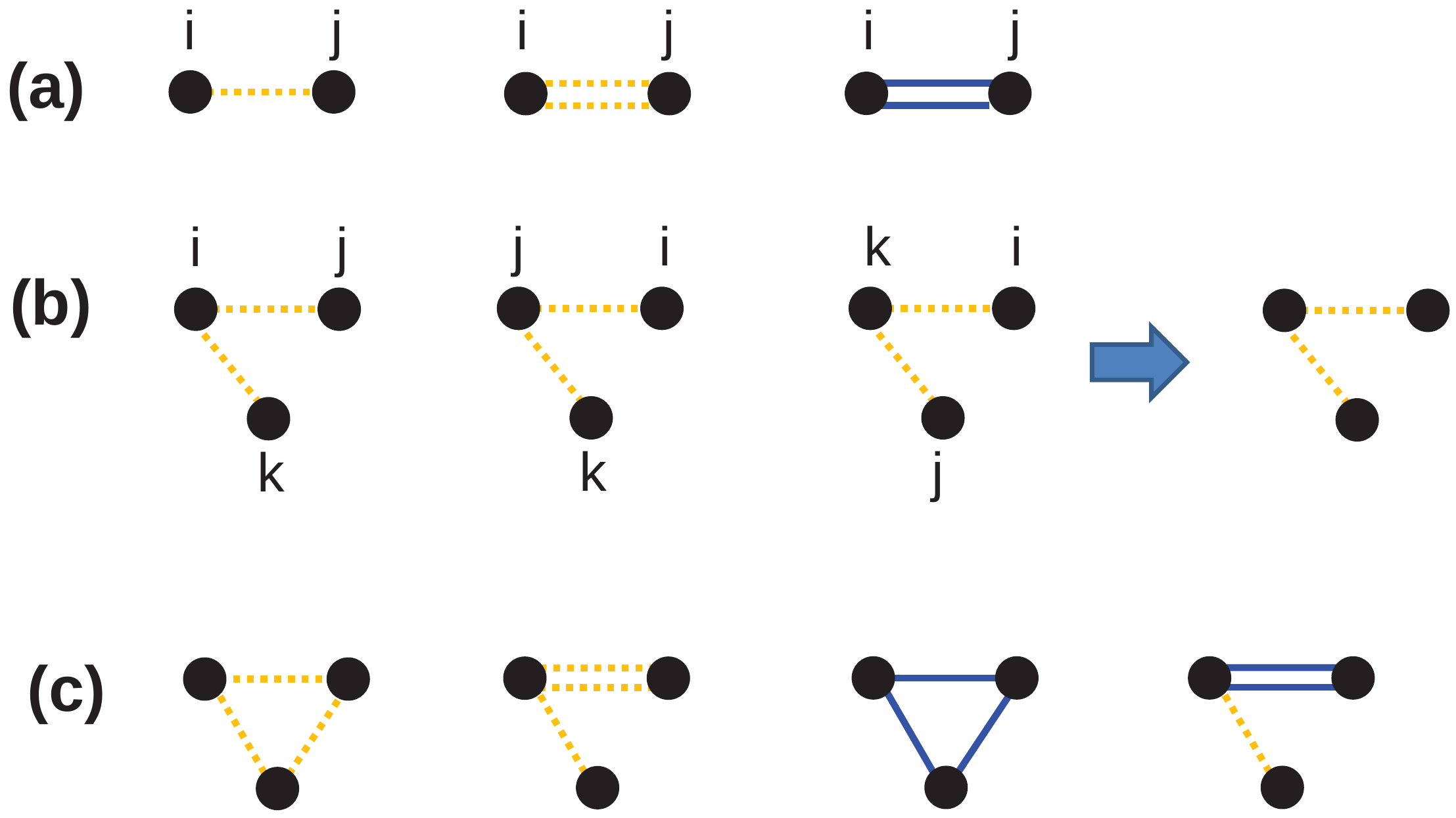}
\caption{Diagrammatic representation of the lowest-order processes
contributing to ensemble cluster-correlation terms for (a) $\tilde{L}(i,j)$
and (b),(c) $\tilde{L}(i,j,k)$. Solid (Dashed) line for off-diagonal
(diagonal) spin-spin interaction. }%
\label{G_CCE_ENSEMBLE_CLUSTER}%
\end{figure}

For example, the Taylor expansion of $\ln\tilde{L}(i,j)$ with respect to
$\lambda_{ij}^{\mathrm{d}}t_{\mathrm{d}}$ and $\lambda_{ij}^{\mathrm{ff}%
}t_{\mathrm{d}}$ satisfying conditions (i) and (ii) reads:
\[
\ln\tilde{L}(i,j)=c_{1}\lambda_{ij}^{\mathrm{d}}t_{\mathrm{d}}+c_{2}%
(\lambda_{ij}^{\mathrm{d}}t_{\mathrm{d}})^{2}+c_{3}(\lambda_{ij}^{\mathrm{ff}%
}t_{\mathrm{d}})^{2}+O(t_{\mathrm{d}}^{3}),
\]
as shown diagrammatically in Fig. \ref{G_CCE_ENSEMBLE_CLUSTER}(a). In the
first and second terms/diagrams, spin $i$ and spin $j$ are not flipped. In the
third term/diagram, each spin is flipped twice. If the initial state of the bath is
maximally mixed, then condition (iii) dictates that the first term/diagram
vanishes since in this term/diagram each spin only appears once. In this case,
the leading Taylor expansion of $\ln\tilde{L}(i,j)$ is second order, including
two terms:\ $(\lambda_{ij}^{\mathrm{d}}t_{\mathrm{d}})^{2}$ and $(\lambda
_{ij}^{\mathrm{ff}}t_{\mathrm{d}})^{2}$. Similarly, according to conditions
(i) and (ii), the Taylor expansion of $\ln\tilde{L}(i,j,k)$ is at least second
order. The lowest, second order expansion includes three terms $(\lambda
_{ij}^{\mathrm{d}}t_{\mathrm{d}})(\lambda_{ik}^{\mathrm{d}}t_{\mathrm{d}})$,
$(\lambda_{ij}^{\mathrm{d}}t_{\mathrm{d}})(\lambda_{jk}^{\mathrm{d}%
}t_{\mathrm{d}})$, $(\lambda_{ki}^{\mathrm{d}}t_{\mathrm{d}})(\lambda
_{kj}^{\mathrm{d}}t_{\mathrm{d}})$, as shown diagrammatically in Fig.
\ref{G_CCE_ENSEMBLE_CLUSTER}(b). As these terms/diagrams are obtained from the
first one by interchanging $i,j,k$, they can be represented by a single
diagram [the last diagram in Fig. \ref{G_CCE_ENSEMBLE_CLUSTER}(b)]. The third
order expansion are shown diagrammatically in Fig.
\ref{G_CCE_ENSEMBLE_CLUSTER}(c): the first diagram denotes $(\lambda
_{ij}^{\mathrm{d}}t_{\mathrm{d}})(\lambda_{jk}^{\mathrm{d}}t_{\mathrm{d}%
})(\lambda_{ki}^{\mathrm{d}}t_{\mathrm{d}})$, the second diagram denotes
$(\lambda_{ij}^{\mathrm{d}}t_{\mathrm{d}})^{2}(\lambda_{ik}^{\mathrm{d}%
}t_{\mathrm{d}})$ and other terms obtained by interchanging $i,j,k$, the third
diagram denotes $(\lambda_{ij}^{\mathrm{ff}}t_{\mathrm{d}})(\lambda
_{jk}^{\mathrm{ff}}t_{\mathrm{d}})(\lambda_{ki}^{\mathrm{ff}}t_{\mathrm{d}})$,
and the fourth diagram denotes $(\lambda_{ij}^{\mathrm{ff}}t_{\mathrm{d}}%
)^{2}(\lambda_{ik}^{\mathrm{d}}t_{\mathrm{d}})$ and other terms obtained by
interchanging $i,j,k$. If the initial state of the bath is maximally mixed, then
condition (iii) further dictates that all the second order diagrams in Fig.
\ref{G_CCE_ENSEMBLE_CLUSTER}(b) and the second and fourth diagrams in Fig.
\ref{G_CCE_ENSEMBLE_CLUSTER}(c) should vanish. In this case, $\ln\tilde
{L}(i,j,k)$ is third order, with the leading order term being $(\lambda
_{ij}^{\mathrm{d}}t_{\mathrm{d}})(\lambda_{jk}^{\mathrm{d}}t_{\mathrm{d}%
})(\lambda_{ki}^{\mathrm{d}}t_{\mathrm{d}})$ and $(\lambda_{ij}^{\mathrm{ff}%
}t_{\mathrm{d}})(\lambda_{jk}^{\mathrm{ff}}t_{\mathrm{d}})(\lambda
_{ki}^{\mathrm{ff}}t_{\mathrm{d}})$. For an arbitrary cluster $\mathcal{C}$,
$\ln\tilde{L}(\mathcal{C})$ is at least $(|\mathcal{C}|-1)$th order, or at
least $|\mathcal{C}|$th order if the initial state of the bath is maximally mixed.

On a time scale $t_{\mathrm{d}}\ll$ $1/\lambda$ ($\lambda$ is the typical
value of bath spin interactions), cluster-correlation terms for large clusters
are small, so the exact CCE [Eq. (\ref{CCE_EXACT})] can be truncated, e.g.,
the $M$th-order truncated CCE (CCE-$M$ for short):%
\begin{equation}
L^{(M)}=\prod_{|\mathcal{C}|\leq M}\tilde{L}(\mathcal{C})=\tilde{L}%
^{(1)}\tilde{L}^{(2)}\cdots\tilde{L}^{(M)}, \label{CCE_M_ENSEMBLE}%
\end{equation}
where $\tilde{L}^{(m)}\equiv\prod_{|\mathcal{C}|=m}\tilde{L}(\mathcal{C})$ is
the contribution of $m$-spin cluster-correlation terms. For example, CCE-1,
$L^{(1)}=\tilde{L}^{(1)}=L(1)L(2)\cdots L(N)$, provides a good description for
the FID, which is dominated by inhomogeneous dephasing. CCE-2 gives
$L^{(2)}(1,2,\cdots,N)=\tilde{L}^{(1)}\tilde{L}^{(2)}$, which is just the
pair-correlation approximation \cite{YaoPRB2006,LiuNJP2007,YaoPRL2007}. By
going to higher-order truncations, the contribution from cluster-correlation
terms of larger clusters can be included accurately and systematically. For
small truncation size $M,$ the cluster-correlation terms can be easily
calculated by exact numerical diagonalization. If each bath spin interacts, on
average, with $q$ spins, then the number of connected size-$m$ clusters is
$\sim Nq^{m-1}$. Since $\ln\tilde{L}(\mathcal{C})\sim(\lambda t_{\mathrm{d}%
})^{|\mathcal{C}|-1}$ ($\lambda$ is the typical bath interaction), the
contribution from $m$-spin correlation is $\ln\tilde{L}^{(m)}\sim N(q\lambda
t_{\mathrm{d}})^{m-1}$. Therefore, a sufficient condition for the convergence
of CCE is
\begin{equation}
t_{\mathrm{d}}<\frac{1}{q\lambda}. \label{CCE_ENSEMBLE_CONVERGE}%
\end{equation}

When the spin bath can be divided into many non-overlapping clusters
$\mathcal{C}_{1},\mathcal{C}_{2},\cdots$ such that the interactions between
different clusters are small, we can regard each cluster as an effective bath
spin and apply the CCE formalism to these effective spins. Namely,
\begin{subequations}
\label{CCE_SUBSETS_DEF}%
\begin{align}
\tilde{L}(\mathcal{C}_{i})  &  \equiv L(\mathcal{C}_{i}),\\
\tilde{L}(\mathcal{C}_{i},\mathcal{C}_{j})  &  \equiv\frac{L(\mathcal{C}%
_{i},\mathcal{C}_{j})}{\tilde{L}(\mathcal{C}_{i})\tilde{L}(\mathcal{C}_{j}%
)},\\
&  \cdots,\nonumber
\end{align}
and the central spin coherence is expanded exactly as%
\end{subequations}
\begin{equation}
L=\left(  \prod_{i}L(\mathcal{C}_{i})\right)  \left(  \prod_{i\neq j}\tilde
{L}(\mathcal{C}_{i},\mathcal{C}_{j})\right)  \cdots, \label{CCE_SUBSETS}%
\end{equation}
similar to Eqs. (\ref{CCE_ENSEMBLE_DEF})\ and (\ref{CCE_EXACT}). When the
correlation between different subsets are neglected, Eq. (\ref{CCE_SUBSETS})
reduces to the disjoint cluster approximation \cite{MazePRB2008}: $L=\prod
_{i}L(\mathcal{C}_{i})$.

\subsubsection{CCE without LCE: pure bath state}

\label{SEC_CCE_PURE}

The CCE formalism for general non-interacting bath states treats both the
off-diagonal interactions $\lambda_{ij}^{\mathrm{ff}}$ and diagonal
interactions $\lambda_{ij}^{\mathrm{d}}$ as perturbations. However, for a pure
product bath state $|J\rangle$ [Eq. (\ref{J_PRODUCT})], the diagonal
interactions alone do not cause non-trivial evolution of the bath. In other
words, the essential spin bath dynamics starting from a product state
$|J\rangle$ is the spin flip-flop by the off-diagonal interactions: the
diagonal interactions can be exactly taken into account and only the
off-diagonal interactions need to be treated as perturbations. This goal has
been achieved by the CCE formalism in Sec. \ref{SEC_CCE_LCE} with the
assistance of diagrammatic LCE \cite{SaikinPRB2007}. Here we present an
alternative formulation of this approach without relying on the LCE.

We are interested in the \textquotedblleft true\textquotedblright%
\ decoherence
\[
L_{J}\equiv\langle J|\mathcal{T}_{\mathrm{C}}e^{-i\int_{\mathrm{C}}\hat
{H}(z)dz}|J\rangle,
\]
where contour Hamiltonian $\hat{H}(z)=H(\hat{\mathbf{I}}_{1},\cdots
,\hat{\mathbf{I}}_{N})$ is written as an explicit function of all the bath
spin operators. The CCE formalism described below applies to a general spin
bath Hamiltonian, but we consider $\hat{H}(z)$ given in Eq. (\ref{HZ}) for the
sake of clarity. The idea of CCE is to single out the contributions from
irreducible, correlated fluctuations from successively larger clusters. First,
replacing all bath spin operators in $H(\hat{\mathbf{I}}_{1},\cdots
,\hat{\mathbf{I}}_{N})$ with their mean-field averages $\hat{\mathbf{I}}%
_{j}\rightarrow\langle J|\hat{\mathbf{I}}_{j}|J\rangle$ gives the c-number
mean-field Hamiltonian $H(\langle J|\hat{\mathbf{I}}_{1}|J\rangle
,\cdots,\langle J|\hat{\mathbf{I}}_{N}|J\rangle)$ and hence the mean-field
contribution without involving the flip of any bath spins
\begin{equation}
L_{J}(\varnothing)\equiv e^{-i\int_{\mathrm{C}}H(\langle J|\hat{\mathbf{I}%
}_{1}|J\rangle,\cdots,\langle J|\hat{\mathbf{I}}_{N}|J\rangle)dz}.
\label{LJPHI}%
\end{equation}
Since $L_{J}(\varnothing)$ is trivially evaluated, hereafter focus is put on
the decoherence caused by the dynamic fluctuation of bath spins:%
\[
\delta L_{J}\equiv\frac{L_{J}}{L_{J}(\varnothing)}=\langle J|\mathcal{T}%
_{\mathrm{C}}e^{-i\int_{\mathrm{C}}\delta\hat{H}dz}|J\rangle,
\]
where
\[
\delta\hat{H}\equiv H(\hat{\mathbf{I}}_{1},\cdots,\hat{\mathbf{I}}%
_{N})-H(\langle J|\hat{\mathbf{I}}_{1}|J\rangle,\cdots,\langle J|\hat
{\mathbf{I}}_{N}|J\rangle)
\]
is the bath Hamiltonian with the mean-field part removed. To proceed, the
decoherence due to the dynamical fluctuation of a non-empty cluster
$\mathcal{C}$ is defined as%
\begin{equation}
\delta L_{J}(\mathcal{C})\equiv\frac{L_{J}(\mathcal{C})}{L_{J}(\varnothing
)}\equiv\langle J|\mathcal{T}_{\mathrm{C}}e^{-i\int_{\mathrm{C}}\delta\hat
{H}_{\mathcal{C}}dz}|J\rangle, \label{DLC}%
\end{equation}
where
\begin{equation}
\delta\hat{H}_{\mathcal{C}}\equiv H(\{\hat{\mathbf{I}}_{j\in\mathcal{C}%
}\},\{\langle J|\hat{\mathbf{I}}_{j\notin\mathcal{C}}|J\rangle\})-H(\langle
J|\hat{\mathbf{I}}_{1}|J\rangle,\cdots,\langle J|\hat{\mathbf{I}}_{N}%
|J\rangle) \label{DHC}%
\end{equation}
is the fluctuation part of the Hamiltonian of cluster $\mathcal{C}$, obtained
from $\delta\hat{H}$ by replacing bath spin operators outside cluster
$\mathcal{C}$ with their mean-field averages $\hat{\mathbf{I}}_{j\notin
\mathcal{C}}\rightarrow\langle J|\hat{\mathbf{I}}_{j\notin\mathcal{C}%
}|J\rangle$. By definition, $\delta\hat{H}_{\mathcal{C}}$ does not contain any
bath spin operators outside cluster $\mathcal{C}$.

The key observation is that if cluster $\mathcal{C}$ can be divided into two
subsets $\mathcal{C}_{1}$ and $\mathcal{C}_{2}$ such that (A) $\hat{H}(z)$
does not contain any interaction between $\mathcal{C}_{1}$ and $\mathcal{C}%
_{2}$, or (B) $\hat{H}(z)$ contains no spin-flip terms for the spins of one
subset (say $\mathcal{C}_{1}$), then $\delta L_{J}(\mathcal{C})$ can be
factorized as $\delta L_{J}(\mathcal{C})=\delta L_{J}(\mathcal{C}_{1})\delta
L_{J}(\mathcal{C}_{2})$. This is because condition (A) leads to $\delta\hat
{H}_{\mathcal{C}}=\delta\hat{H}_{\mathcal{C}_{1}}+\delta\hat{H}_{\mathcal{C}%
_{2}}$, i.e., the dynamical fluctuation of $\mathcal{C}_{1}$ and
$\mathcal{C}_{2}$ are independent, while condition (B) allows all operators
inside $\mathcal{C}_{1}$ to be replaced with their mean-field averages (i.e.,
spins in $\mathcal{C}_{1}$ has no dynamical fluctuation), so that $\delta
L_{J}(\mathcal{C}_{1})=1$ and $\delta L_{J}(\mathcal{C})=\delta L_{J}%
(\mathcal{C}_{2})$. This motivates the following definition of a hierarchy of
\textit{cluster-correlation terms}, in a way similar to the previous
subsection:
\begin{subequations}
\label{CCE_SINGLE_DEF}%
\begin{align}
\delta\tilde{L}_{J}(i)  &  \equiv\delta L_{J}(i),\\
\delta\tilde{L}_{J}(i,j)  &  \equiv\frac{\delta L_{J}(i,j)}{\delta\tilde
{L}_{J}(i)\delta\tilde{L}_{J}(j)},\\
&  \cdots\nonumber\\
\delta\tilde{L}_{J}(\mathcal{C})  &  \equiv\frac{\delta L_{J}(\mathcal{C}%
)}{\prod_{\mathcal{C}^{\prime}\subset C}\delta\tilde{L}_{J}(\mathcal{C}%
^{\prime})}.
\end{align}
The decoherence is expressed \textit{exactly} as the product of all possible
cluster-correlation terms:%
\end{subequations}
\begin{equation}
\delta L_{J}=\prod_{\mathcal{C}\subseteq\{1,2,\cdots,N\}}\delta\tilde{L}%
_{J}(\mathcal{C}). \label{CCE_SINGLE_EXACT}%
\end{equation}

So-defined cluster-correlation $\delta\tilde{L}_{J}(\mathcal{C})$ vanishes if
the interactions contained in $\delta\hat{H}_{\mathcal{C}}$ cannot\textit{
}connect all the spins in group $\mathcal{C}$ into a linked cluster (i.e.,
cluster $\mathcal{C}$ consists of two subsets with independent dynamical
fluctuation), or if the off-diagonal interaction terms contained in
$\delta\hat{H}_{\mathcal{C}}$ do not flip certain spins inside cluster
$\mathcal{C}$ (i.e., these spins have no dynamical fluctuation). Therefore,
the cluster-correlation term $\delta\tilde{L}_{J}(\mathcal{C})$ accounts for
the \textit{irreducible}, \textit{fully correlated} \textit{dynamical}
fluctuation of all spins in cluster $\mathcal{C}$. Consequently, in the Taylor
expansion of $\ln\delta\tilde{L}_{J}(\mathcal{C})$ with respect to
$\lambda_{ij}^{\mathrm{ff}}t_{\mathrm{d}}$ and $\lambda_{ij}^{\mathrm{d}%
}t_{\mathrm{d}}$, the interaction coefficients contained in every term must
(i) connect all the spins in group $\mathcal{C}$ into a linked cluster, and
(ii) ensure that every spin in cluster $\mathcal{C}$ is flipped an even
($2,4,6,\cdots$) number of times.

\begin{figure}[ptb]
\includegraphics[width=\columnwidth,clip]{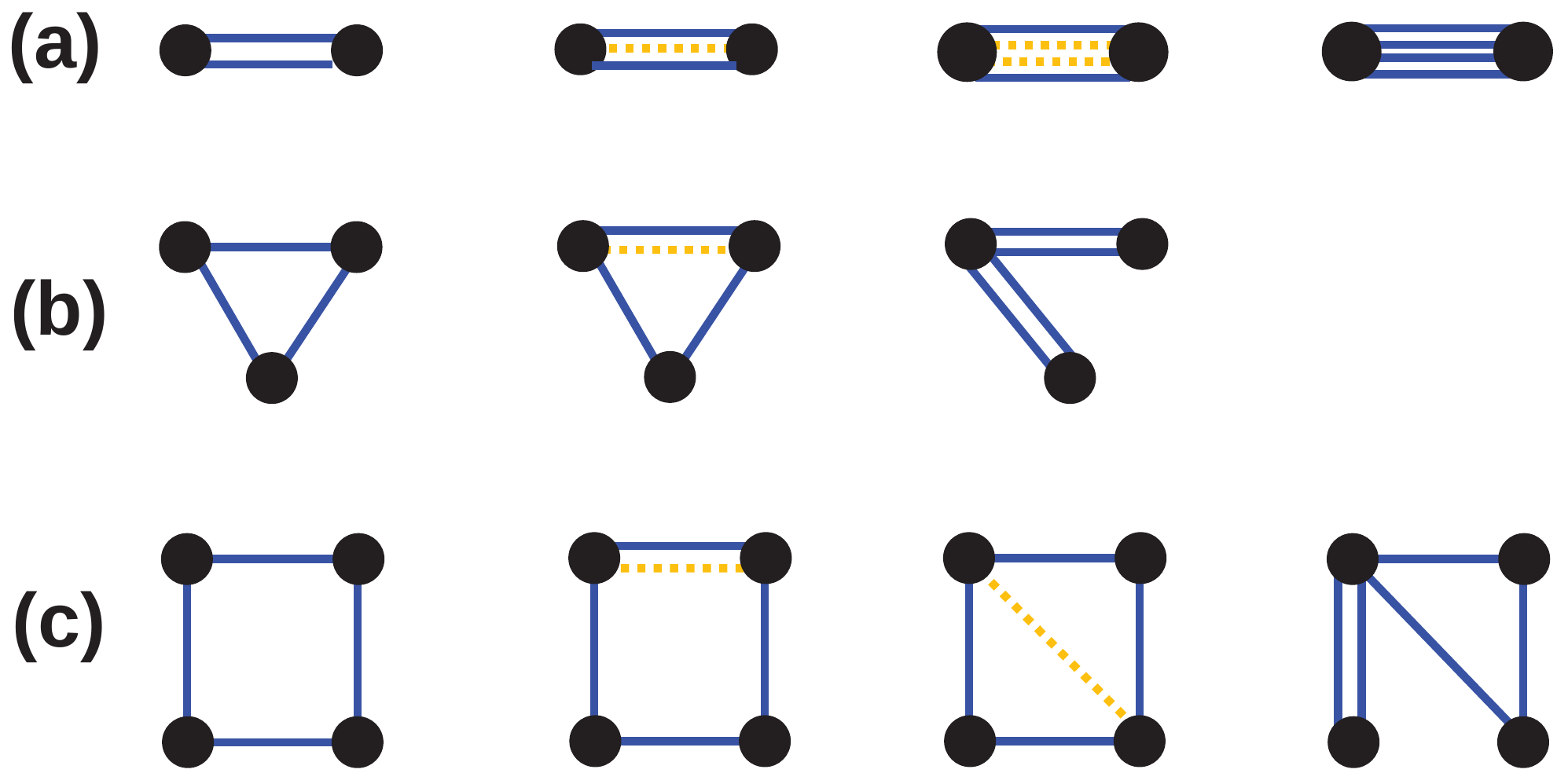}
\caption{Diagrammatic representation of the lowest order processes
contributing to cluster-correlation terms (for a pure initial state of the bath) of
and (a) $\delta\tilde{L}_{J}(i,j)$, (b) $\delta\tilde{L}_{J}(i,j,k)$, and (c)
$\delta\tilde{L}_{J}(i,j,k,l)$. Solid (Dashed) line for off-diagonal
(diagonal) spin-spin interaction. }%
\label{G_CCE_SINGLE_CLUSTER}%
\end{figure}

Conditions (i) and (ii)\ ensure that $\ln\delta\tilde{L}_{J}(\mathcal{C})$ is
\textit{at least} $|\mathcal{C}|$th order in $(\lambda_{\mathrm{ff}%
}t_{\mathrm{d}})$, where $\lambda_{\mathrm{ff}}$ is the typical off-diagonal
bath interactions. Taking the Taylor expansion of $\ln\delta\tilde{L}%
_{J}(i,j)$ as an example, the first few terms in the expansion are shown
diagrammatically in Fig. \ref{G_CCE_SINGLE_CLUSTER}(a), including the lowest,
second-order term\ $(\lambda_{ij}^{\mathrm{ff}}t_{\mathrm{d}})^{2}$ (first
diagram), the third-order term $(\lambda_{ij}^{\mathrm{d}}t_{\mathrm{d}%
})(\lambda_{ij}^{\mathrm{ff}}t_{\mathrm{d}})^{2}$ (second diagram), and the
fourth-order terms $(\lambda_{ij}^{\mathrm{d}}t_{\mathrm{d}})^{2}(\lambda
_{ij}^{\mathrm{ff}}t_{\mathrm{d}})^{2}$ (third diagram) and $(\lambda
_{ij}^{\mathrm{ff}}t_{\mathrm{d}})^{4}$ (fourth diagram). Similarly, the
Taylor expansion of $\ln\delta\tilde{L}_{J}(i,j,k)$ is shown diagrammatically
in Fig. \ref{G_CCE_SINGLE_CLUSTER}(b), including the lowest, third-order term
$(\lambda_{ij}^{\mathrm{ff}}t_{\mathrm{d}})(\lambda_{jk}^{\mathrm{ff}%
}t_{\mathrm{d}})(\lambda_{ki}^{\mathrm{ff}}t_{\mathrm{d}})$ (first diagram),
the fourth-order term $(\lambda_{ij}^{\mathrm{d}}t_{\mathrm{d}})(\lambda
_{ij}^{\mathrm{ff}}t_{\mathrm{d}})(\lambda_{jk}^{\mathrm{ff}}t_{\mathrm{d}%
})(\lambda_{ki}^{\mathrm{ff}}t_{\mathrm{d}})$ and other terms obtained by
interchanging$\ i,j,k$ (second diagram), the fourth-order term $(\lambda
_{ij}^{\mathrm{ff}}t_{\mathrm{d}})^{2}(\lambda_{jk}^{\mathrm{ff}}%
t_{\mathrm{d}})^{2}$ and other terms obtained by interchanging$\ i,j,k$ (third
diagram). The first few terms in the Taylor expansion of $\ln\delta\tilde
{L}_{J}(i,j,k,l)$ are shown in Fig. \ref{G_CCE_SINGLE_CLUSTER}(c), including
the lowest, fourth-order term (first diagram) and a few fifth-order terms
(other diagrams). For the Taylor expansion of $\ln\delta\tilde{L}%
_{J}(\mathcal{C})$ for a general cluster $\mathcal{C}$, the lowest-order term
is the ring diagram formed by $|\mathcal{C}|$ off-diagonal interaction lines,
such as the first diagram in Figs. \ref{G_CCE_SINGLE_CLUSTER}%
(a)-\ref{G_CCE_SINGLE_CLUSTER}(c), thus $\ln\delta\tilde{L}_{J}(\mathcal{C})$
is $|\mathcal{C}|$th order in $\lambda_{\mathrm{ff}}t_{\mathrm{d}}$.

\begin{figure}[ptb]
\includegraphics[width=0.8\columnwidth]{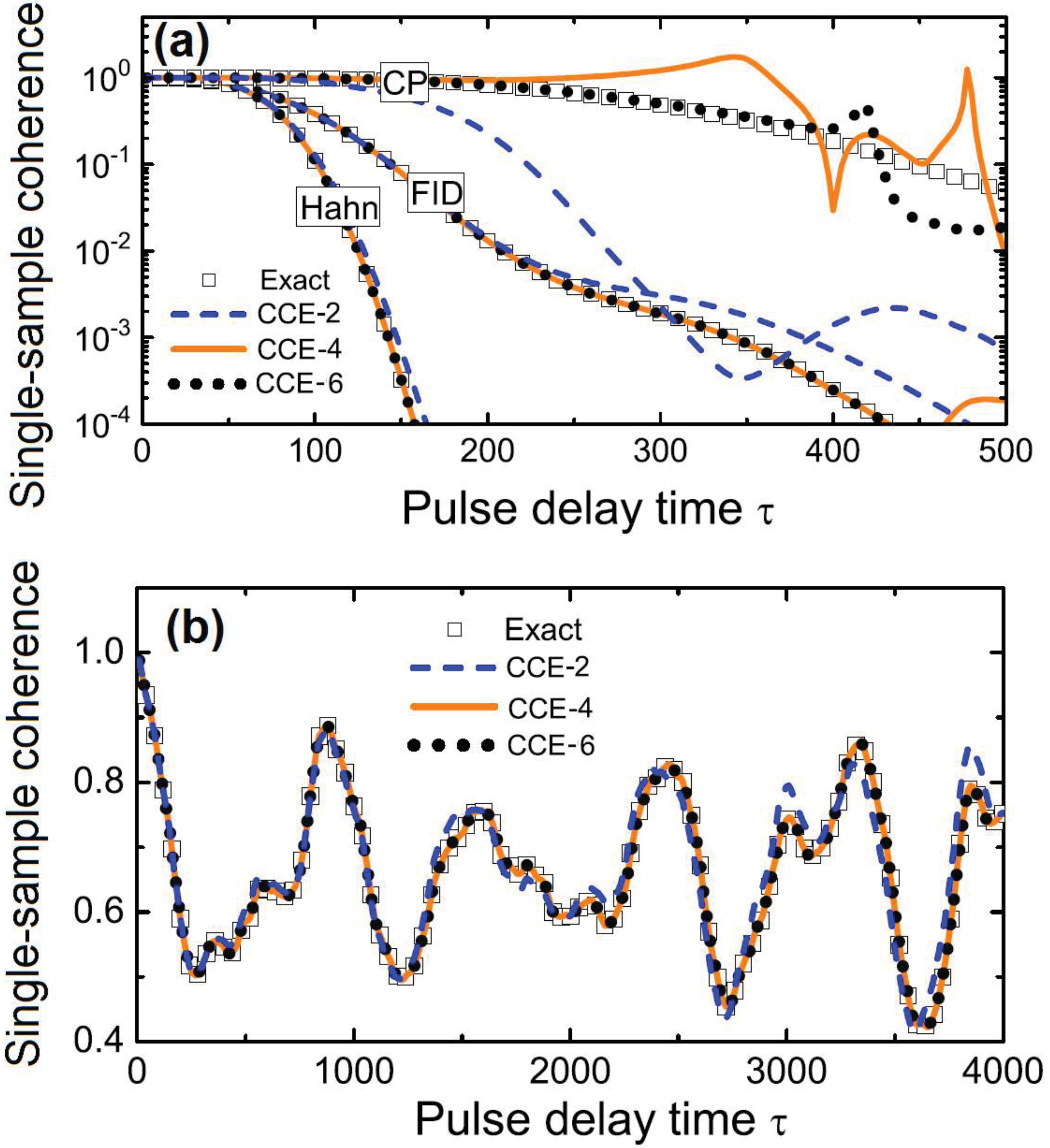}
\caption{\textquotedblleft True\textquotedblright\ decoherence from CCE
compared with the exact solutions for one-dimensional spin-1/2 XY model
consisting of $N=100$ bath spins. (a)\ FID, Hahn echo, and under CPMG-2
control. The level splitting of bath spins vary smoothly with location. (b)
FID. The level splitting of bath spins varies randomly with location.
Reproduced from Ref. \cite{YangPRB2008a}.}%
\label{G_CCE_EXACT}%
\end{figure}

On a time scale $t_{\mathrm{d}}\ll$ $1/\lambda_{\mathrm{ff}}$, the exact CCE
[Eq. (\ref{CCE_SINGLE_EXACT})] can be truncated, e.g., the $M$th-order
truncated CCE (CCE-$M$ for short) is:%
\begin{equation}
\delta L_{J}^{(M)}=\prod_{|\mathcal{C}|\leq M}\delta\tilde{L}_{J}%
(\mathcal{C}). \label{CCE_PURE_M}%
\end{equation}
For a relatively small truncation size $M,$ the cluster-correlation terms can
be calculated by exact numerical diagonalization. CCE-1 gives $\delta
L_{J}^{(1)}=\delta\tilde{L}_{J}(1)\cdots\delta\tilde{L}_{J}(N)$, corresponding
to the independent precession of individual bath spin in the mean-field
produced by other bath spins. CCE-2 accounts for the irreducible bath spin
correlations up to pairs and is equivalent to the pair-correlation
approximation \cite{YaoPRB2006,LiuNJP2007,YaoPRL2007}. Going to successively
higher order truncations allows systematic inclusion of successively
higher-order irreducible correlations in the spin bath evolution and accurate
description of the decoherence under various DD controls [Fig.
\ref{G_CCE_EXACT}(a)].

In terms of $\tilde{\pi}(\mathcal{C})$ in Sec. \ref{SEC_CCE_LCE}, we have
$e^{\tilde{\pi}(\varnothing)}=L_{J}(\varnothing)$ and $e^{\tilde{\pi
}(\mathcal{C})}=\delta\tilde{L}_{J}(\mathcal{C})$ for non-empty $\mathcal{C}$,
thus a sufficient condition for convergence is still Eq.
(\ref{CCE_SINGLE_CONVERGE}). However, in some cases, the convergence could
even go well beyond. One such scenario is a disordered spin bath with highly
non-uniform or even random spin splitting energies for different spins. In
this case the disorder induced localization effect would bound the size of
irreducible fully correlated clusters up to a critical size $M_{0}$, such that
CCE-$M_{0}$ would converge to the exact results on any time scales [see Fig.
\ref{G_CCE_EXACT}(b) for an example].

The random spin splitting of bath spins inside a cluster $\mathcal{C}$ may
come from their random couplings to the central spin or to the mean-field
averages of other bath spins outside cluster $\mathcal{C}$, e.g., in
$\delta\hat{H}_{\mathcal{C}}$, the spin splitting of the $j$th spin inside
$\mathcal{C}$ consists of two non-uniform parts: $s(z)a_{j}/2$
due to HFI and $\sum_{k\notin\mathcal{C}}\lambda_{jk}^{\mathrm{d}}\langle
J|\hat{I}_{k}^{z}|J\rangle$ due to coupling to external bath spins. This
observation makes it possible to modify the CCE to improve its convergence
\cite{WitzelPRB2012}. The hybrid CCE \cite{WitzelPRB2012} is the same as the
CCE in Sec. \ref{SEC_CCE_ENSEMBLE} except for a different definitions of
$\hat{H}_{\mathcal{C}}(z)$ and hence $L(\mathcal{C})$ [Eq.
(\ref{LC_CCE_ENSEMBLE})]: instead of dropping all the bath spins outside
cluster $\mathcal{C}$ from the bath Hamiltonian $\hat{H}(z)$, now $\hat
{H}_{\mathcal{C}}(z)$ is obtained from $\hat{H}(z)$ by dropping the terms that
flip the bath spins outside cluster $\mathcal{C}$. Thus $L(\mathcal{C})$ in
hybrid CCE is connected to the cluster-correlation terms $\delta
L_{J}(\mathcal{C})$ in Sec. \ref{SEC_CCE_PURE} via
\begin{equation}
L(\mathcal{C})=\sum_{J}P_{J}L_{J}(\varnothing)\delta L_{J}(\mathcal{C}%
)=\sum_{J}P_{J}L_{J}(\mathcal{C}). \label{LC_MODIFIED}%
\end{equation}
The mean fields from external bath spins randomizes the splitting of bath
spins inside $\mathcal{C}$ and improves the convergence, although it is not
obvious whether or not the hybrid CCE converges to the exact results: here
$L(\mathcal{C})$ is no longer factorizable even if $\mathcal{C}$ can be
divided into two subsets with no inter-subset interactions, so $\ln\tilde
{L}(\mathcal{C})$ [as defined in Eq. (\ref{CCE_ENSEMBLE_DEF})] does not vanish
even if the interactions contained in $\hat{H}_{\mathcal{C}}(z)$ cannot
connect all the spins in group $\mathcal{C}$ into a linked cluster. In
practice, the ensemble average in Eq. (\ref{LC_MODIFIED}) has exponential
complexity. The algorithm to deal with this issue and detailed discussion of
its applications to Si:P, silicon QDs,\ and NV centers can be found in Ref.
\cite{WitzelPRB2012}.

Finally, when the spin bath consists of many non-overlapping subsets
$\mathcal{C}_{1},\mathcal{C}_{2},\cdots$ with weak inter-subset interactions,
we can regard each subset as an effective bath spin and apply the CCE
formalism [Eqs. (\ref{CCE_SINGLE_DEF}) and (\ref{CCE_SINGLE_EXACT})] to these
effective spins, in the same way as Eqs. (\ref{CCE_SUBSETS_DEF}) and
(\ref{CCE_SUBSETS}).

\subsubsection{CCE for quantum noise auto-correlation function}

\label{SEC_CCE_CORRELATION}

Recently, the idea of CCE has been adapted \cite{WitzelPRB2014,MaPRB2015} to
calculate the auto-correlation $\langle\hat{b}(t)\hat{b}\rangle$ of the
quantum noise $\hat{b}(t)\equiv e^{i\hat{H}_{B}t}\hat{b}e^{-i\hat{H}_{B}t}$
driven by an interacting bath Hamiltonian $\hat{H}_{B}$, where the noise
operator $\hat{b}=\sum_{j}\hat{b}_{j}$ is the sum of operators of individual
spins (e.g., $\hat{b}_{j}=\mathbf{a}_{j}\cdot\hat{\mathbf{I}}_{j}$) and
$\langle\cdots\rangle\equiv\operatorname*{Tr}[\hat{\rho}_{B}(\cdots)]$ is the
ensemble average in a product bath state $\hat{\rho}_{B}=\otimes_{j}\hat{\rho
}_{j}$. The first step is to define the quantum noise from a spin cluster,
\[
\hat{b}_{\mathcal{C}}(t)\equiv e^{i\hat{H}_{\mathcal{C}}t}\hat{b}%
_{\mathcal{C}}e^{-i\hat{H}_{\mathcal{C}}t}%
\]
where $\hat{b}_{\mathcal{C}}=\sum_{j\in\mathcal{C}}\hat{b}_{j}$ and $\hat
{H}_{\mathcal{C}}$ is the Hamiltonian of a cluster $\mathcal{C}$, obtained
from the bath Hamiltonian by dropping all bath spins except for those in
cluster $\mathcal{C}$. The second step is to define the noise auto-correlation%
\begin{equation}
C_{\{1,2,\cdots,N\}}(t)\equiv\langle\hat{b}(t)\hat{b}\rangle-\langle\hat
{b}(t)\rangle\langle\hat{b}\rangle\label{CCE_CORR1}%
\end{equation}
and the contribution from a cluster $\mathcal{C}$:
\begin{equation}
C_{\mathcal{C}}(t)\equiv\langle\hat{b}_{\mathcal{C}}(t)\hat{b}_{\mathcal{C}%
}\rangle-\langle\hat{b}_{\mathcal{C}}(t)\rangle\langle\hat{b}_{\mathcal{C}%
}\rangle, \label{CCE_CORR2}%
\end{equation}
where $N$ is the number of bath spins. If a cluster $\mathcal{C}$ consists of
two subsets $\mathcal{C}_{1}$ and $\mathcal{C}_{2}$ and $\hat{H}_{\mathcal{C}%
}$ does not contain any interaction between these two subsets, then the noise
auto-correlation from this cluster is additive: $C_{\mathcal{C}}%
(t)=C_{\mathcal{C}_{1}}(t)+C_{\mathcal{C}_{2}}(t)$. This motivates the
definition of a hierarchy of cluster-correlation terms:%
\begin{equation}
\tilde{C}_{\mathcal{C}}(t)\equiv C_{\mathcal{C}}(t)-\sum_{\mathcal{C}^{\prime
}\subset\mathcal{C}}\tilde{C}_{\mathcal{C}^{\prime}}(t), \label{CCE_CORR_EXP}%
\end{equation}
e.g., $\tilde{C}_{i}(t)\equiv C_{i}(t)$, $\tilde{C}_{\{i,j\}}(t)\equiv
C_{\{i,j\}}(t)-\tilde{C}_{i}(t)-\tilde{C}_{j}(t)$, etc. By definition, the
pair-correlation $\tilde{C}_{\{i,j\}}(t)$ vanishes when there is no
interaction between spin $i$ and spin $j$, thus $\tilde{C}_{\{i,j\}}(t)$ is at
least first order in the bath spin interactions. Similarly, $\tilde
{C}_{\mathcal{C}}(t)$ vanishes when the interactions contained in $\hat
{H}_{\mathcal{C}}$ cannot connect the spins in group $\mathcal{C}$ into a
linked cluster, thus $\tilde{C}_{\mathcal{C}}(t)$ is at least $(|\mathcal{C}%
|-1)$th order in the bath spin interactions. Finally, the noise
auto-correlation is approximated by truncating the expansion, e.g., keeping
cluster-correlation terms containing up to $M$ spins gives (CCE-$M$ for
short):%
\begin{equation}
C_{\{1,2,\cdots,N\}}^{(M)}(t)=\sum_{\mathcal{C},|\mathcal{C}|\leq M}\tilde
{C}_{\mathcal{C}}(t). \label{CCE-M_COR}%
\end{equation}
Since $C_{\mathcal{C}}(t)$ and hence $\tilde{C}_{\mathcal{C}}(t)$ for small
$|\mathcal{C}|$ can be easily calculated by exact numerical diagonalization,
Eq. (\ref{CCE-M_COR}) provides a systematic approach to calculate the
auto-correlation up to successively higher orders of inter-spin correlation,
e.g., for a non-interacting spin bath, the cluster contributions $\tilde
{C}_{\mathcal{C}}(t)=0$ for $|\mathcal{C}|\geq2$, so CCE-1 gives the exact
result: $C(1,2,\cdots,N)=\sum_{i}\tilde{C}(i)=\sum_{i}C(i)$, even in the
presence of rapid single-spin dynamics such as that induced by the anisotropic
HFI (see Sec. \ref{SEC_ANIHFI}).

We notice that the original formulation \cite{WitzelPRB2014,MaPRB2015} of the
CCE of noise auto-correlation uses a slightly different definition:
\begin{align}
C_{\{1,2,\cdots,N\}}^{\prime}(t)  &  \equiv\langle\hat{b}(t)\hat{b}%
\rangle-\langle\hat{b}^{2}\rangle,\\
C_{\mathcal{C}}^{\prime}(t)  &  \equiv\langle\hat{b}_{\mathcal{C}}(t)\hat
{b}_{\mathcal{C}}\rangle-\langle\hat{b}_{\mathcal{C}}^{2}\rangle
\end{align}
instead of Eqs. (\ref{CCE_CORR1}) and (\ref{CCE_CORR2}). In this case, even
when a cluster $\mathcal{C}$ consists of two subsets $\mathcal{C}_{1}$ and
$\mathcal{C}_{2}$ and $\hat{H}_{\mathcal{C}}$ does not contain any interaction
between these two subsets, the cluster term $C_{\mathcal{C}}^{\prime}(t)$ does
not reduce to $C_{\mathcal{C}_{1}}^{\prime}(t)+C_{\mathcal{C}_{2}}^{\prime
}(t)$ due to the existence of cross-correlation terms $(\langle\hat
{b}_{\mathcal{C}_{1}}(t)\rangle-\langle\hat{b}_{\mathcal{C}_{1}}%
\rangle)\langle\hat{b}_{\mathcal{C}_{2}}\rangle$ and $(\langle\hat
{b}_{\mathcal{C}_{2}}(t)\rangle-\langle\hat{b}_{\mathcal{C}_{2}}%
\rangle)\langle\hat{b}_{\mathcal{C}_{1}}\rangle$. Thus this formulation is
essentially a short-time perturbative expansion of $W(t)$ around $t=0$
\cite{WitzelPRB2014}. For a maximally mixed bath state $\hat{\rho}_{B}%
\propto\hat{I}$, the cross-correlation terms vanish, thus $C_{\mathcal{C}%
}^{\prime}(t)=C_{\mathcal{C}}(t)-C_{\mathcal{C}}(0)$, i.e., the original
formulation coincides with Eqs. (\ref{CCE_CORR1})-(\ref{CCE_CORR_EXP}).

\subsubsection{Numerical techniques}

\begin{figure}[ptb]
\includegraphics[width=\columnwidth]{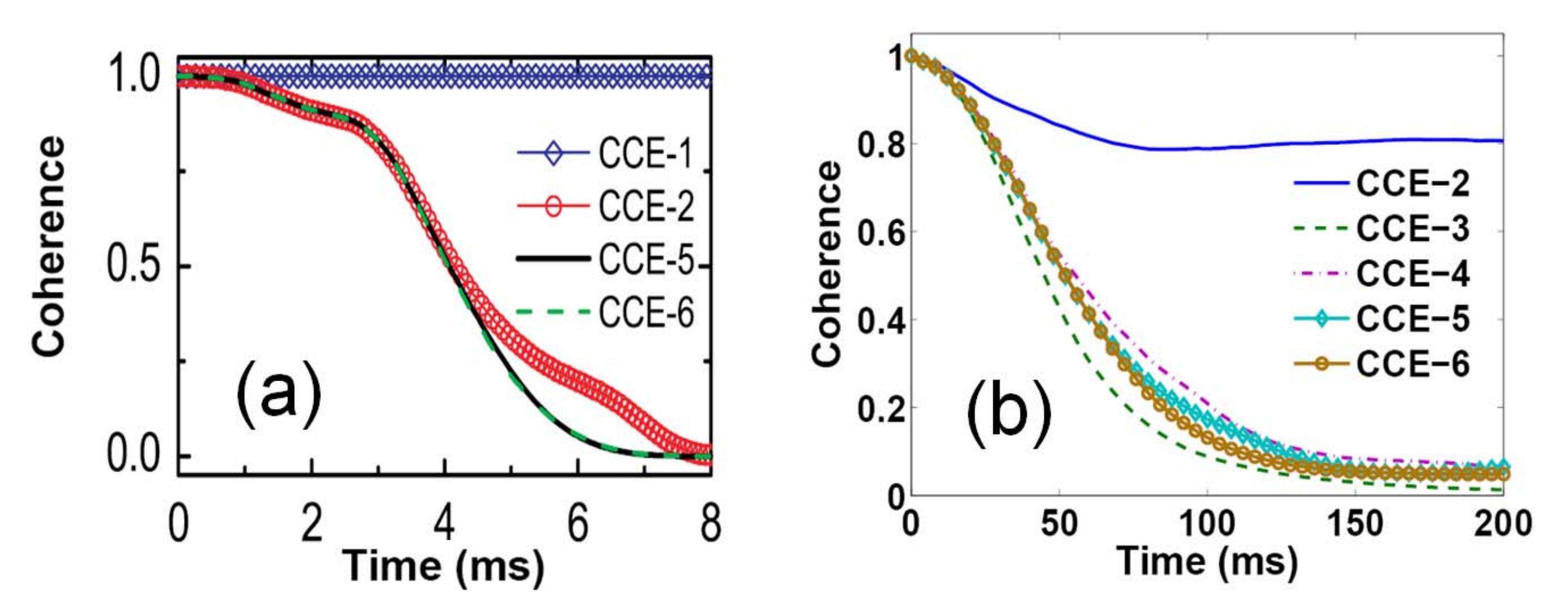}
 \caption{Convergence of
CCE in realistic nuclear spin baths. (a) Decoherence of the NV electron spin
transition $|+1\rangle\leftrightarrow|-1\rangle$ under CPMG-5 control and
$B=0.3$ T along the NV axis, caused by the $^{13}$C nuclear spins with natural
abundance 1.1\%. (b) Decay of Hahn echo of the Bi donor electron spin near the
``clock'' transition ($B=B_{\mathrm{CT}}+0.3$ mT with $B_{\mathrm{CT}}=79.9$
mT), caused by the $^{29}$Si nuclear spins with natural abundance 4.7\%. Panal
(a) is extracted from Ref. \cite{ZhaoPRL2011} and panal (b) is extracted from
the Supplementary Information of Ref. \cite{MaPRB2015}.}%
\label{CCE NV Bi}%
\end{figure}

The CCE calculations converge rapidly for the electron spin decoherence in
nuclear spin baths with short-ranged dipolar interactions, such as the
decoherence of the electron spin of an NV center caused by the $^{13}$C
nuclear spins with natural abundance 1.1\% in diamond
\cite{ZhaoPRL2011,ZhaoNatNano2011,ZhaoPRB2012} and the decoherence of the
donor (phosphorus or bismuth) electron spin caused by the $^{29}$Si nuclear
spins with natural abundance 4.7\% in silicon
\cite{WitzelPRL2007,GeorgePRL2010,MaNatCommun2014,BalianPRB2014,MaPRB2015,BalianPRB2015}%
, as shown in Fig. \ref{CCE NV Bi}. Here we summarize the main steps of
performing the CCE calculations in realistic systems and provide the numerical
tricks in each step \cite{WitzelPRB2012,ZhaoPRB2012,MaNatCommun2014,MaPRB2015}.

Step 1:\ choosing the initial state of the bath. Since the maximally mixed thermal
bath state and a \textit{typical }pure product bath state sampled from the
thermal ensemble give similar \textquotedblleft true\textquotedblright%
\ decoherence (see Sec. \ref{SEC_QUANTUM_DYN}), especially for a relatively
large bath (such as the $^{29}$Si nuclear spin bath in silicon), it is
preferrable to calculate the \textquotedblleft true\textquotedblright%
\ decoherence with the CCE method for a pure bath state (Sec.
\ref{SEC_CCE_PURE}), which has faster convergence
\cite{WitzelPRB2012,MaPRB2015} than that for a general noninteracting bath
state (\ref{SEC_CCE_ENSEMBLE}). The thermal noise is relevant only for the FID
and amounts to a multiplicative factor $L_{\mathrm{inh}}(t)$ [Eq. (\ref{LINH})].

Step 2: determining the size of the nuclear spin bath. The bath spins can be
chosen as those within a certain threshold distance $R_{\mathrm{c}}$ away from
the central spin (or threshold HFI strength), which is gradually increased
till convergence. Typically $R_{\mathrm{c}}\sim4$ nm (including $N\sim500$
bath spins) for the natural $^{13}$C nuclear spin bath in diamond
\cite{ZhaoPRL2011,ZhaoNatNano2011} and $R_{\mathrm{c}}\sim8$ nm (including
$N\sim5000$ bath spins) for the natural $^{29}$Si nuclear spin bath in silicon
\cite{MaNatCommun2014,MaPRB2015}.

Step 3: defining effective bath spins. A cluster of bath spins that are fully
linked via very strong interactions is identified as a large effective spin,
so the bath is divided into many \textit{non-overlapping}, strongly linked
clusters $\mathcal{C}_{1},\mathcal{C}_{2},\cdots$, or equivalently many
effective spins. Then subsequent steps and the CCE all apply to these
effective spins [cf. Eqs. (\ref{CCE_SUBSETS_DEF}) and (\ref{CCE_SUBSETS})].

Step 4: selecting contributing clusters. For a given truncation size $M$, it
is not necessary to keep all the clusters containing $M$ effective spins,
since only clusters with fully correlated fluctuations contribute to central
spin decoherence. For nuclear spins coupled through short-range dipolar
interactions and within the central spin decoherence time, significant
inter-spin correlation develops only among a few nearest neighbours,
especially among those that are fully linked through sufficiently strong
interactions. Thus it suffices to keep clusters whose diameter is smaller than
a certain upper cutoff $d_{\mathrm{c}}$, which is gradually increased till
convergence. Typically $d_{\mathrm{c}}\sim$ 1 nm for both the natural $^{13}$C
nuclear spin bath in diamond and the natural $^{29}$Si nuclear spin bath in
silicon \cite{ZhaoNatNano2011,MaNatCommun2014,MaPRB2015}. A lower cutoff
$\lambda_{\min}$ in the cluster connectivity strength $\lambda_{\mathcal{C}}$
(defined as the smallest interaction necessary to complete the full
connectivity of the cluster $\mathcal{C}$) is also preferred
\cite{WitzelPRB2012}.

Finally, Monte Carlo sampling technique can be used when there are too many
contributing clusters.

\subsection{Real-space cluster expansion}

As one of the first quantum many-body theories for central spin decoherence,
the density matrix cluster expansion \cite{WitzelPRB2005,WitzelPRB2006}
provides a convenient method to include multi-spin correlations, in the spirit
of the virial expansion for interacting gases in grand canonical ensembles. In
terms of $L(\mathcal{C})$ defined in Eq. (\ref{LC_CCE_ENSEMBLE}) of Sec.
\ref{SEC_CCE_ENSEMBLE}, cluster expansion defines the (irreducible)
cluster-correlation terms $\{W(\mathcal{C})\}$ by \textit{substracting} all
reducible parts:
\begin{subequations}
\label{CE_DEF}%
\begin{align}
W(i)  &  \equiv L(i),\\
W(i,j)  &  \equiv L(i,j)-W(i)W(j),\\
W(i,j,k)  &  \equiv L(i,j,k)-W(i)W(j)W(k)-W(i)W(j,k)\\
&  \quad-W(j)W(i,k)-W(k)W(i,j),\\
&  \cdots\nonumber\\
W(\mathcal{C})  &  \equiv L(\mathcal{C})-\sum_{\{\mathcal{C}_{\alpha}\}}%
\prod_{\mathcal{C}_{\alpha}}W(\mathcal{C}_{\alpha}),
\end{align}
where in the last line the sum runs over all possible partitions of the
cluster $\mathcal{C}$ into non-overlapping and non-empty subsets
$\mathcal{C}_{1},\mathcal{C}_{2},\cdots$. The central spin coherence can be
expressed \textit{exactly} in terms of these cluster-correlation terms:
\end{subequations}
\begin{equation}
L=W(1,2,\cdots,N)+\sum_{\{\mathcal{C}_{\alpha}\}}\prod_{\mathcal{C}_{\alpha}%
}W(\mathcal{C}_{\alpha}), \label{CE_EXACT}%
\end{equation}
where the sum in the last line runs over all possible partitions of all bath
spins $\{1,2,\cdots,N\}$ into non-overlapping and non-empty subsets
$\mathcal{C}_{1},\mathcal{C}_{2},\cdots$.\footnote{For translationally
invariant spin baths such that $W(i)=W_{1},$ $W(i,j)=W_{2},\cdots$, Eq.
(\ref{CE_EXACT}) simplifies to%
\[
L_{N}=N!\sum_{m_{1},m_{2},\cdots}\frac{(W_{1}/1!)^{m_{1}}}{m_{1}!}\frac
{(W_{2}/2!)^{m_{2}}}{m_{2}!}\cdots
\]
subjected to the constraint $\sum_{l}lm_{l}=N$. The quantity $\sum_{N}\xi
^{N}L_{N}/N!=\exp(\sum_{l}\xi^{l}W_{l}/l!)$ corresponds to the virial
expansion of interacting identical gases in grand canonical ensembles.}

The cluster-correlation terms $\{W(\mathcal{C})\}$ in cluster expansion has
very similar properties as the cluster-correlation terms $\tilde
{L}(\mathcal{C})$ in CCE (see Sec. \ref{SEC_CCE_ENSEMBLE}), e.g.,
$W(\mathcal{C})$ vanishes if the interactions contained in $\hat
{H}_{\mathcal{C}}(z)$ cannot connect all the spins in group $\mathcal{C}$ into
a linked cluster, so $W(\mathcal{C})$ is at least $(|\mathcal{C}|-1)$th order
in $(\lambda t_{\mathrm{d}})$, where $\lambda$ is the typical interaction
strength in the bath. Keeping cluster-correlation terms containing up to $M$
spins gives the $M$th-order truncated cluster expansion (CE-$M$ for short):%
\begin{equation}
L^{(M)}=\sum_{\{\mathcal{C}_{\alpha}\},|\mathcal{C}_{\alpha}|\leq M}%
\prod_{\mathcal{C}_{\alpha}}W(\mathcal{C}_{\alpha}), \label{CE-M}%
\end{equation}
where the sum runs over all possible partitions of the bath into
non-overlapping non-empty clusters $\mathcal{C}_{1},\mathcal{C}_{2},\cdots$ of
size up to $M$. In the cluster expansion for interacting gases in grand
canonical ensembles with translational symmetry, the evaluation of a truncated
cluster expansion reduces to the calculation of a finite number of cluster
terms $\{W(\mathcal{C}_{\alpha})\}$ with $|\mathcal{C}_{\alpha}|\leq M$, which
can be easily done by exact numerical diagonalization. For a finite-size spin
bath or for a bath without translational symmetry, however, it is very
difficult to calculate the sum in Eq. (\ref{CE-M}) even for a small $M$.

When all the cluster terms $\{W(\mathcal{C})\}$ are individually small, Eq.
(\ref{CE-M}) can be approximated by a factorized form by adding some
overlapping terms that are higher-order small quantities. For example, the contour
Hamiltonian in Eq. (\ref{HZ}) gives $W(i)=L(i)=1$ at the echo time of DD
control, so CE-$M$ can be approximated by%
\begin{equation}
\bar{L}^{(M)}=\prod_{1<|\mathcal{C}_{\alpha}|\leq M}[1+W(\mathcal{C}_{\alpha
})]\approx\prod_{1<|\mathcal{C}_{\alpha}|\leq M}e^{W(\mathcal{C}_{\alpha})}.
\label{CE-M1}%
\end{equation}
Comparing the factorized form in Eq. (\ref{CE-M1}) to the exact CE-$M$ in Eq.
(\ref{CE-M}), the error $L_{\mathrm{err}}^{(M)}\equiv\bar{L}^{(M)}-L^{(M)}$
\begin{align}
L_{\mathrm{err}}^{(M)}  &  \equiv\bar{L}^{(M)}-L^{(M)}\nonumber\\
&  =\sum_{i<j<k}W(i,j)W(j,k)+\sum_{i<j<k<l}W(i,j,k)W(k,l)+\cdots,
\end{align}
contains the products of all possible cluster terms sharing at least one spin.
Such overlapping terms are higher-order small quantities and hence Eq. (\ref{CE-M1}) is
justified when each individual cluster term for $|\mathcal{C}_{\alpha}|>1$ is
small, e.g., for large spin baths, where the number of contributing clusters
is large and hence the contribution from each individual cluster remains small
within the time scale of decoherence.

The error $L_{\mathrm{err}}^{(M)}$ from the overlapping terms becomes relevant
for small spin baths, where the coherent dynamics of a small number of
multi-spin clusters dominating the decoherence may persist well beyond the
bath spin flip-flop time, such that the small-term condition is no longer
satisfied. In this case the cluster expansion may not converge to the exact results.

\begin{figure}[ptb]
\includegraphics[width=\columnwidth]{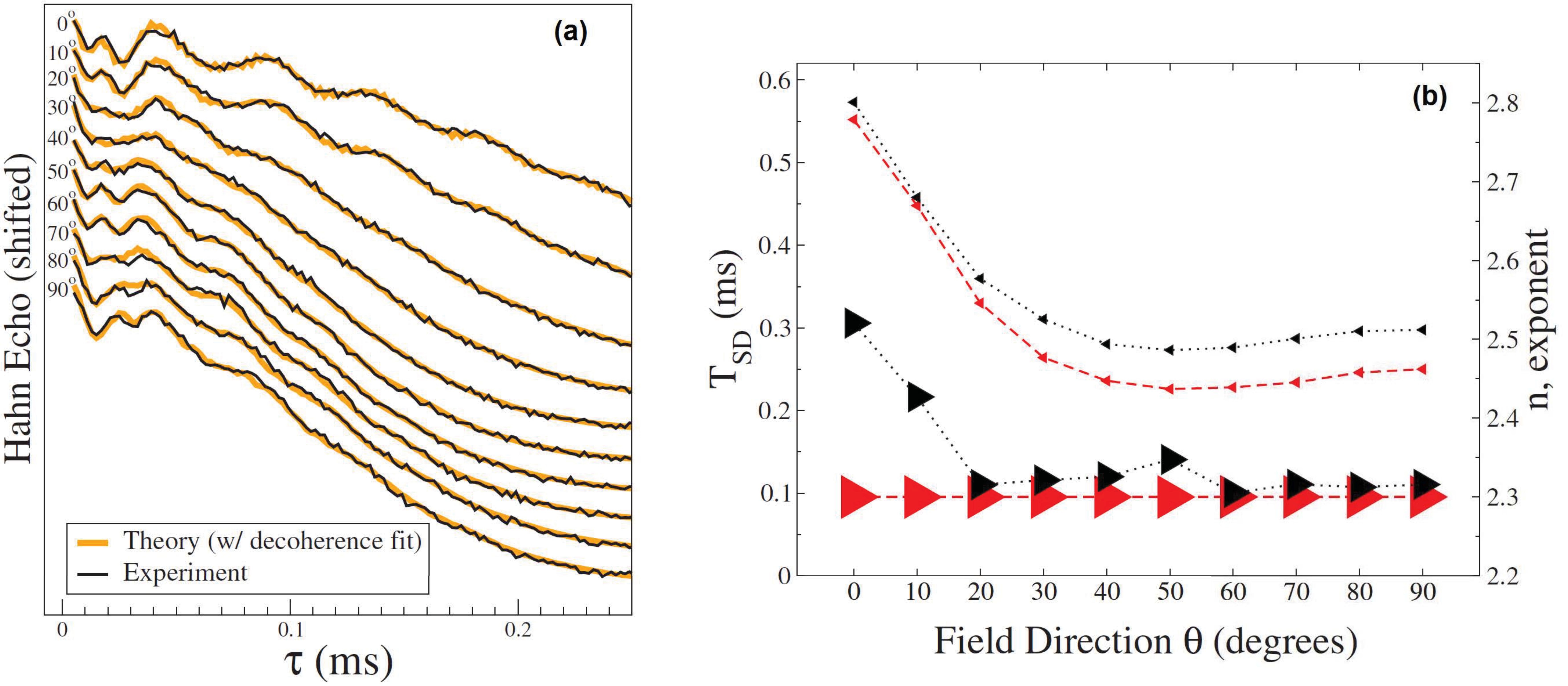}
\caption{Spin echo decay in
Si:P for ten different magnetic field orientations ranging from [001] to
[110]. Apart from a few common fitting parameters for all curves, each curve
are fitted with two parameters: the decoherence time $T_{\mathrm{SD}}$ due to
$^{29}$Si nuclei and its exponent $n$. These fitting parameters (black
triangles) are compared with the cluster expansion calculations (red
triangles) in (b). Right (Left) triangles correspond to $n$ ($T_{\mathrm{SD}}%
$). The fitted $n$ deviates from theory only at small magnetic field angles,
where the nearest-neighbor dipolar flip-flop interactions approach zero.
Reproduced from Ref. \cite{WitzelPRB2007b}.}%
\label{G_CE_SIP}%
\end{figure}

The factorized CE-$M$ [Eq. (\ref{CE-M1})] has been applied to electron and/or
nuclear spin decoherence in Si:P\ (caused by $^{29}$Si nuclei with natural
abundance $4.7\%$)
\cite{WitzelPRB2005,WitzelPRB2006,WitzelPRB2007a,WitzelPRB2007b}, Si:Bi
\cite{GeorgePRL2010}, GaAs QDs (caused by $^{69}$Ga, $^{71}$Ga, and $^{75}$As
nuclei), and Si:SiGe QDs (caused by $^{73}$Ge and $^{29}$Si nuclei)
\cite{WitzelPRB2006,WitzelPRB2012a}. For electron spin echo in Si:P, cluster
expansion provides a complete understanding of the experimentally measured
decay profile
\cite{TyryshkinPRB2003,AbePRB2004,FerrettiPRB2005,TyryshkinJPC2006,AbePRB2010}
(see Fig. \ref{G_CE_SIP} for an example), including the envelope modulation by
strong anisotropic HFI with a few proximal $^{29}$Si nuclei (as discussed in
Sec. \ref{SEC_ANIHFI}), the dependence on the magnetic field orientations and
$^{29}$Si abundance, and the transition of the electron spin resonance lineshape from Gaussian
(for $^{29}$Si abundance $f\geq f_{0}$) to Lorentzian (for $f\leq1.2\%$),
which arises from ensemble averaging of the inhomogeneous dephasing
$e^{-(t/T_{2}^{\ast})^{2}}$ of each individual donors over the distribution of
$T_{2}^{\ast}$ \cite{AbragamBook1961,DobrovitskiPRB2008}. Cluster expansion
also shows that many-pulse CPMG could prolong the electron spin coherence time
in Si:P and GaAs QDs by factors of 4--10 \cite{WitzelPRL2007} and that in
Si:Bi system, the hybridization of the Bi donor electron spin with $^{209}$Bi
nuclear spin could significantly changes the decay of the electron spin Hahn
echo caused by the $^{29}$Si nuclei \cite{GeorgePRL2010}. For $^{31}$P donor
nuclear spins in GaAs and Si~\cite{WitzelPRB2007a}, cluster expansion gives
negligible decoherence on the time scale of 100 $\mathrm{\mu s}$ in GaAs:P and
1-2 ms in Si:P under CPMG sequences with 2-4 $\pi$-pulses, indicating the
promising role of $^{31}$P nuclear spin as a long-lived quantum memory.

\subsection{Limitations of the many-body theories and possible extension}

Despite the unprecedented understanding of the central spin decoherence under
many experimental conditions, the available many-body theories are still
subjected to several limitations. First, the theories in Sec.
\ref{SEC_MANYBODY_THEORY} are restricted to the pure dephasing model [Eq.
(\ref{HAMIL1}) or (\ref{HAMIL})], which is justified when the central spin
splitting $\gg$ bath spin splitting. A possible extension is to generalize the
idea of CCE to spin relaxation, e.g., the evolution of $\langle\hat{S}%
_{z}(t)\rangle$ can be calculated by applying the CCE formalism to
$L(t)\equiv\langle\hat{S}_{z}(t)\rangle/\langle\hat{S}_{z}(0)\rangle$. Second,
for fast convergence of these theories, the size of the contributing bath spin
clusters (i.e., those with appreciable correlated fluctuations) should be
relatively small within the central spin decoherence time, so that their
contributions can be obtained by exact diagonalization or other methods.
Therefore, these theories also require short-range interactions between bath
spins [i.e., small $q$ in Eq. (\ref{CCE_ENSEMBLE_CONVERGE}) or
Eq.\ (\ref{CCE_SINGLE_CONVERGE})], which in turn necessitates a large central
spin splitting. Otherwise (e.g., in weak magnetic fields
\cite{CoishPRB2004,CoishPRB2008,CywinskiPRB2009,CoishPRB2010} or near the
optimal work points \cite{MaPRB2015,BalianPRB2015}) the successive flip-flops
of the central spin with different bath spins may rapidly induce long-range
correlations in the bath, beyond the description of existing theories.

For very small central spin spliting, the intrinsic bath spin interactions can
be neglected, so the coupled system is described by the central spin model%
\begin{equation}
\hat{H}=\omega_{0}\hat{S}_{z}+\omega_{I}\sum_{i}\hat{I}_{i}^{z}+\hat
{\mathbf{S}}\cdot\hat{\mathbf{h}}, \label{CSM}%
\end{equation}
where $\hat{\mathbf{h}}\equiv\sum_{i}a_{i}\hat{\mathbf{I}}_{i}$. This model
allows the central spin and the bath spins to exchange spin angular momentum,
a feature that is absent from pure dephasing models. The central spin
evolution due to Eq. (\ref{CSM}) has been studied by a great diversity of
approaches, including semi-classical models that treat the bath spins as
classical stochastic variables
\cite{MerkulovPRB2002,SemenovPRB2003,ErlingssonPRB2004,AlHassaniehPRL2006},
exact analytical solutions for uniform HFI
\cite{MelikidzePRB2004,KozlovJETP2007,BortzPRB2007} or fully polarized spin
baths \cite{KhaetskiiPRL2002,KhaetskiiPRB2003}, and direct numerical modelling
\cite{SchliemannJPC2003,CywinskiPRB2010,DobrovitskiPRE2003,DobrovitskiPRL2003,ZhangPRB2006}
and Bethe ansatz solutions \cite{FaribaultPRL2013,FaribaultPRB2013} for small
baths containing a few tens of spins. For large baths, non-Markovian master
equations
\cite{CoishPRB2004,CoishPRB2008,CoishPRB2010,BreuerPRB2004,FischerPRA2007,FerraroPRB2008}
and equation of motion approaches \cite{DengPRB2006,DengPRB2008} have been
used, but they require strong magnetic fields under which central spin
relaxation is suppressed while pure dephasing is usually dominated by the
intrinsic nuclear spin interactions.

Recently, central spin decoherence on a time scale $\ll$ inverse of the HFI
has been treated by the time-dependent density matrix renormalization group
\cite{StanekPRB2013,StanekPRB2014} and resumming the time-convolutionless
master equation \cite{BarnesPRL2012}. The former shows that for large baths,
the central spin dynamics is well described by the semi-classical model (Eq.
(\ref{HAMIL_SEMI})) with a classical Gaussian noise $\mathbf{\tilde{b}}(t)$, or
by treating both the central spin and the bath spins as classical vectors
subjected random initial orientations. The latter shows that the central spin
dynamics depend on the HFI coefficients $\{a_{i}\}$ only through $\sum
_{i}a_{i}^{2}$ and hence can be approximated by a exactly solvable model with
uniform HFI, consistent with the energy-time uncertainty relations
\cite{BarnesPRB2011}. The central spin dynamics on longer time scales, which
depend sensitively on the specific distribution of $\{a_{i}\}$, remains an
open issue.

\section{Quantum decoherence effects}

\label{SEC_EFFECTS}

According to the idea of CCE, the contribution of bath dynamics to the central
spin decoherence is the product of \textit{irreducible, correlated
fluctuations} from bath spin clusters of different sizes [see Eqs.
(\ref{CCE_M_ENSEMBLE}) and (\ref{CCE_PURE_M})]. In this section we discuss
some quantum decoherence effects caused by these fluctuations. In a relatively
weak magnetic field or for the FID, the decoherence is dominated by the
fluctuation of single-spin clusters, thus CCE-1 gives a good approximation
\cite{ZhaoPRB2012,LiuSciRep2012}. In a strong magnetic field or under Hahn
echo control, the fluctuation of single-spin clusters is frozen or its effect
is suppressed by DD, and the correlated fluctuation of nuclear spin pairs
dominates, thus CCE-2 is usually sufficient
\cite{ZhaoPRB2012,ZhaoPRL2011,BalianPRB2014}. Under high-order DD
\cite{ZhaoPRB2012,MaNatCommun2014} or near the optimal working points (e.g. for the electronic-nuclear hybrid spin qubit in Si:Bi system) \cite{MaPRB2015,BalianPRB2015}, multi-spin correlation becomes
pronounced, so higher-order truncation of CCE are needed to get convergent results.

\subsection{Single spin fluctuation}

In a relatively weak magnetic field, each individual nuclear spin has a large
quantum fluctuation since the Zeeman energy and the HFI are comparable and do
not commute with each other, while the pairwise nuclear flip-flop processes
have much weaker effect as the dipolar interaction between nuclear spins is
usually much weaker than the energy cost of the pairwise flip-flop due to HFI
gradient. In this case, we can assume the central spin $\hat{\mathbf{S}}$
($S=1/2$) is coupled to a bath of non-interacting nuclear spins $\{\hat
{\mathbf{I}}_{j}\}$ ($I$=1/2 for simplicity), described by the pure dephasing
Hamiltonian,
\begin{equation}
\hat{H}=\hat{S}_{z}\sum_{j}\mathbf{h}_{j}^{b}\cdot\hat{\mathbf{I}}_{j}%
+\sum_{j}\mathbf{h}_{j}^{B}\cdot\hat{\mathbf{I}}_{j}, \label{H_SINGLE}%
\end{equation}
where $\hat{b}\equiv\sum_{j}\mathbf{h}_{j}^{b}\cdot\hat{\mathbf{I}}_{j}$ with
$\mathbf{h}_{j}^{b}$ being the HFI coupling, and the intrinsic bath
Hamiltonian $\hat{H}_{B}=\sum_{j}\mathbf{h}_{j}^{B}\cdot\hat{\mathbf{I}}_{j}$
with $\mathbf{h}_{j}^{B}$ being the external magnetic field. The bath
Hamiltonian conditioned on the central spin state is $\hat{H}_{\pm}=\sum
_{j}\mathbf{h}_{j}^{(\pm)}\cdot\hat{\mathbf{I}}$ describing the bath spin
precession around the fields $\mathbf{h}_{j}^{(\pm)}=\mathbf{h}_{j}^{B}%
\pm\mathbf{h}_{j}^{b}/2$.

The single-spin fluctuation causes two possible effects. For isotropic HFI, we
have $[\hat{b},\hat{H}_{B}]=0$, so the thermal noise from bath spins leads to
inhomogeneous dephasing of the central spin in FID. For anisotropic HFI [Eq.
(\ref{DHFI})], we have $[\hat{b},\hat{H}_{B}]\neq0$, so the quantum noise from
bath spins gives rise to modulation effects in central spin decoherence (see
Appendix \ref{APP Bloch} for the Bloch vector representation of single nuclear
spin dynamics).

\subsubsection{Isotropic HFI: inhomogeneous dephasing}

For isotropic HFI (e.g., for conduction electron confined in a III-V
semiconductor QD) \cite{YaoPRB2006,LiuNJP2007}, $\mathbf{h}_{j}^{b}$ and
$\mathbf{h}_{j}^{B}$ are both along the $z$ axis and Eq. (\ref{H_SINGLE})
reduces to $\hat{H}=\hat{S}_{z}\sum_{j}{h}_{j}^{b}\hat{{I}}_{j}^{z}+\sum
_{j}{h}_{j}^{B}\hat{{I}}_{j}^{z}$. In this case, $\hat{b}$ commutes with the
bath Hamiltonian $\hat{H}_{B}$, so the noise is static and leads to
Gaussian inhomogeneous dephasing for the FID (c.f. Eq. (\ref{LINH})),%
\[
L_{\mathrm{FID}}(t)=\prod_{j}\cos(h_{j}^{b}t/2)\approx e^{-(t/T_{2}^{\ast
})^{2}},
\]
where the inhomogeneous dephasing time $T_{2}^{\ast}=\sqrt{2}/h_{\mathrm{rms}%
}$ with $h_{\mathrm{rms}}=(1/2)\sqrt{\sum_{j}{(h_{j}^{b})}^{2}}$ being the
root-mean-square fluctuation of the noise field \cite{LiuSciRep2012}. DD can
largely remove the effect of single-spin clusters. So one would have to go to
higher order correlations of the nuclear spins to correctly describe the
``true" decoherence.

\subsubsection{Anisotropic HFI: decoherence envelope modulation}

\label{SEC_ANIHFI}

For anisotropic HFI [Eq. (\ref{DHFI})], the noise field $\mathbf{h}_{j}^{b}$
deviates from the direction of $\mathbf{h}_{j}^{B}$ (pointed along $z$ axis),
then even non-interacting bath spins could cause nontrivial electron spin
decoherence. The anisotropic HFI can exist for donors (or QDs) in silicon and
NV centers in diamond (see Sec. \ref{SEC_HFI}). The FID is entirely determined
by the magnitudes $h_{j}^{(\pm)}$ of the fields $\mathbf{h}_{j}^{(\pm)}$ and
their relative angle $\Theta_{j}$:
\begin{align}
&  L_{\mathrm{FID}}(t)=\prod_{j}L_{j}^{\mathrm{FID}}%
(t)\nonumber\label{FID_ENSEMBLE}\\
=  &  \prod_{j}\left[  \cos^{2}\frac{\Theta_{j}}{2}\cos\frac{(h_{j}%
^{(+)}-h_{j}^{(-)})t}{2}+\sin^{2}\frac{\Theta_{j}}{2}\cos\frac{(h_{j}%
^{(+)}+h_{j}^{(-)})t}{2}\right]  .
\end{align}
We decompose the anisotropic HFI as $\mathbf{h}_{j}^{b}=h_{j}^{b,z}%
\mathbf{e}_{z}+\mathbf{h}_{j}^{b,\perp}$. In the short-time limit,
$L_{\mathrm{FID}}(t)$ shows Gaussian decay: $L_{\mathrm{FID}}(t)\approx
e^{-\sum_{j}(h_{j}^{b,z})^{2}t^{2}/8}$ for $h_{j}^{b}\ll h_{j}^{B}$ and
$L_{\mathrm{FID}}(t)\approx e^{-\sum_{j}(h_{j}^{b})^{2}t^{2}/8}$ for
$h_{j}^{b}\gg h_{j}^{B}$ \cite{ZhaoPRB2012}. In a longer time-scale, since
$\sin\Theta_{j}=(h_{j}^{b,\perp}h_{j}^{B})/(h_{j}^{(+)}h_{j}^{(-)})$, we have
$|\sin\Theta|\ll1$ for $h_{j}^{b}\gg h_{j}^{B}$ or $h_{j}^{b}\ll h_{j}^{B}$,
thus in a strong external magnetic field ($h_j^b\ll h_j^B$), the $j$th nuclear spin contributes a dominant slow oscillation $\sim
\cos(h_{j}^{b,z}t/2)$ modulated by a small-amplitude, fast oscillation
$\sim\cos(h_{j}^{B}t)$ to the central spin decoherence \cite{SaikinPRB2003}.
In QDs or shallow donors with a large nuclear spin bath, the rapid
inhomogeneous dephasing usually makes this effect invisible. In diamond NV
centers with a rather small nuclear spin bath, however, the electron spin
decoherence is usually dominated by a few strongly coupled bath spins. In this
case, the modulation effects is manifested as the deviation of the decoherence
away from Gaussian profile, which has been observed experimentally
\cite{LiuSciRep2012}.

\begin{figure}[ptb]
\includegraphics[width=\columnwidth]{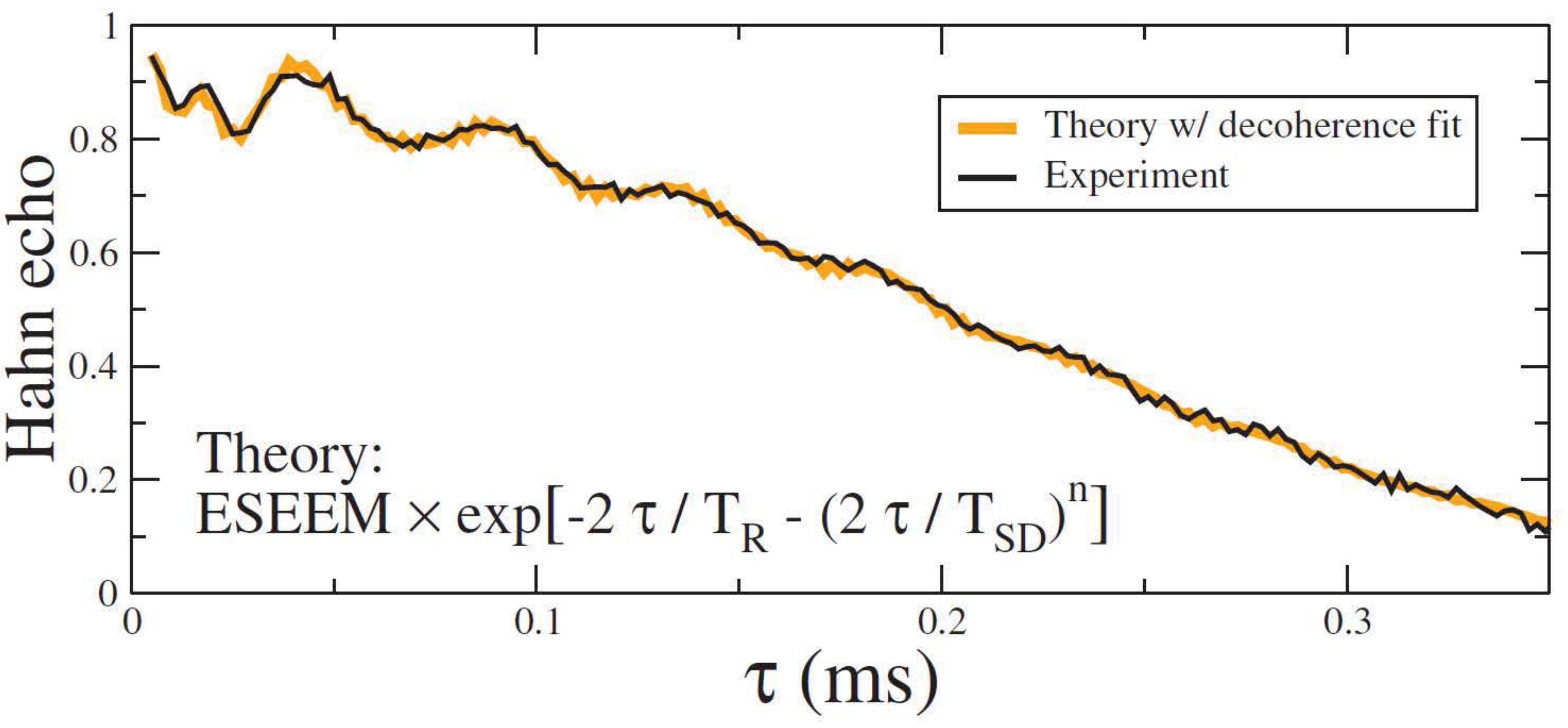}
\caption{Theoretical and
experimental results for electron spin echo decay. The theoretical curve is
the product of spin echo envelope modulation due to anisotropic HFI induced
bifurcated evolution of individual nuclear spins, the non-Markovian decay
$e^{-(2\tau/T_{\mathrm{SD}})^{n}}$ due to nuclear spin flip-flop dynamics, and
Markovian decay $e^{-2\tau/T_{2}}$. Reproduced from Ref. \cite{WitzelPRB2007b}%
.}%
\label{G_ESEEM_SI}%
\end{figure}

The Hahn echo
\begin{equation}
L_{\mathrm{H}}(2\tau)=\prod_{j}\left[  1-2\sin^{2}\Theta_{j}\sin^{2}%
\frac{h_{j}^{(+)}\tau}{2}\sin^{2}\frac{h_{j}^{(-)}\tau}{2}\right]
\label{LH_SINGLESPIN}%
\end{equation}
shows non-Gaussian decay in the short time limit: $L_{\mathrm{H}}%
(2\tau)\approx e^{-|\mathbf{h}_{j}^{B}\times\mathbf{h}_{j}^{b}|^{2}\tau^{4}%
/8}$. On a longer time-scale, the second term in Eq. (\ref{LH_SINGLESPIN})
gives rise to modulations with amplitude $\sim\sin^{2}\Theta_{j}$ on the
electron spin echo decay (called electron spin echo envelope modulation, see
Fig. \ref{G_ESEEM_SI} for an example), where $\sin\Theta_{j}$ is called the
modulation depth parameter \cite{WitzelPRB2007b}. The modulation depth is
appreciable for those nuclei with the HFI and Zeeman energy comparable, i.e.,
$h_{j}^{b}\sim h_{j}^{B}$. When the magnetic field orientation (defined as the
$z$ axis) is chosen such that $\mathbf{h}_{j}^{b}$ is perpendicular to
$\mathbf{h}_{j}^{B}$ and hence $h_{j}^{(+)}=h_{j}^{(-)}=h_{j}$, periodic
restoration of spin coherence can be achieved at $\sin(h_{j}\tau/2)=0$
\cite{WitzelPRB2007b}.

\subsection{Pair-correlation effect}

\label{SEC_PCA}

In the strong magnetic field regime, the individual nuclear spin fluctuation
are suppressed (apart from a trivial inhomogeneous dephasing for the FID), so
central spin decoherence is caused by the correlated fluctuation of larger
nuclear spin clusters. On a short time-scale compared with the inverse nuclear
spin interactions, the correlated fluctuation is mainly from the nuclear spin
pair dynamics, so it can be described by CCE-2 or equivalently the
pair-correlation approximation \cite{YaoPRB2006,LiuNJP2007,YaoPRL2007}.

Here we focus on \textit{\textquotedblleft true\textquotedblright} decoherence
and take the initial state of the nuclear spin bath as $|J\rangle$ [Eq.
(\ref{J_PRODUCT})]. For the Hamiltonian in Eq. (\ref{HPM}), we have
$\delta\tilde{L}_{J}(i)=1$ and hence
\[
L_{J}=\prod_{\{i,j\}}L_{J}(i,j)
\]
up to a trivial phase factor, where%
\[
L_{J}(i,j)\equiv\langle J|\mathcal{T}_{\mathrm{C}}e^{-i\int_{\mathrm{C}}%
\hat{H}(\hat{\mathbf{I}}_{i},\hat{\mathbf{I}}_{j},\{\langle J|\hat{\mathbf{I}%
}_{m\notin\{i,j\}}|J\rangle\})dz}|J\rangle
\]
is the decoherence due to the nuclear spin pair $\{i,j\}$, whose effective
Hamiltonian $\hat{H}(\hat{\mathbf{I}}_{i},\hat{\mathbf{I}}_{j},\{\langle
J|\hat{\mathbf{I}}_{k\notin\{i,j\}}|J\rangle\})$ is obtained from the total
Hamiltonian by replacing all spin operators outside cluster $\mathcal{C}$ by
their mean-field averages.

There are $N(N-1)$ pairs in the bath, as labelled by $k\equiv(i,j)$. The
initial state of the $k$th pair is mapped to the spin-down state of a spin-1/2
pseudospin $\hat{\boldsymbol{\sigma}}_{k}$: $|\Downarrow\rangle_{k}%
\equiv|m\rangle_{i}|n\rangle_{j}$, while the flip-flopped state is mapped to
the spin-up state of this pseudo-spin:\ $|\Uparrow\rangle_{k}\equiv
|m+1\rangle_{i}|n-1\rangle_{j}$. Therefore, the flip-flop dynamics of each
nuclear spin pair are mapped to the flip dynamics of the pseudo-spins starting
from the initial state $|J\rangle=\otimes_{k}|\Downarrow\rangle_{k}$. Here the
flip of the $k$th pseudo-spin gives a state $|J,k\rangle$ that is
energetically higher than $|J\rangle$ by an amount $\langle J,k|\hat{H}_{\pm
}|J,k\rangle-\langle J|\hat{H}_{\pm}|J\rangle=D_{k}\pm Z_{k}$, while the
transition amplitude from $|J\rangle$ to $|J,k\rangle$ is\ $\langle
J,k|\hat{H}_{\pm}|J\rangle=B_{k}\pm A_{k}$, with $D_{k}$ from the diagonal
nuclear spin interaction $\hat{H}_{\mathrm{d}}$, $B_{k}\propto\lambda
_{ij}^{\mathrm{ff}}$ from the nuclear spin flip-flop interaction $\hat
{H}_{\mathrm{ff}}$, $A_{k}\propto\tilde{\lambda}_{ij}^{\mathrm{ff}}$ from the
electron spin mediated nuclear spin interaction $\tilde{H}_{\mathrm{ff}}$, and
$Z_{k}=(a_{i}-a_{j})/2$ the energy cost of a pair flip due to the diagonal HFI
$\hat{S}_{z}\hat{h}_{z}$. Thus $B_{k}$ and $D_{k}$ are nonzero only for
neighboring nuclear spins (i.e., local pairs), while $A_{k}$ remains nonzero
even for non-local pairs, but is suppressed under a strong magnetic field. The
$k$th pseudospin is described by the Hamiltonian
\cite{YaoPRB2006,LiuNJP2007,YaoPRL2007}
\[
\hat{H}_{k}^{\pm}=(2B_{k}\pm2A_{k},0,D_{k}\pm Z_{k})\cdot\hat
{\boldsymbol{\sigma}}_{k}/2\equiv\mathbf{h}_{k}^{(\pm)}\cdot\hat
{\boldsymbol{\sigma}}_{k}/2.
\]
Typically $|Z_{k}|\gg|B_{k}|\sim|D_{k}|\gg|A_{k}|$, thus the pseudo-spin
dynamics is dominated by its coupling to the central spin. The pseudo-spin
description provides a transparent geometric picture for central spin
decoherence and its control by DD in terms of Bloch vectors, as well as magic
coherence recovery via controlled disentanglement \cite{LiuNJP2007,YaoPRL2007}.

\subsubsection{Non-local and local pair correlations}

\begin{figure}[ptb]
\includegraphics[width=\columnwidth]{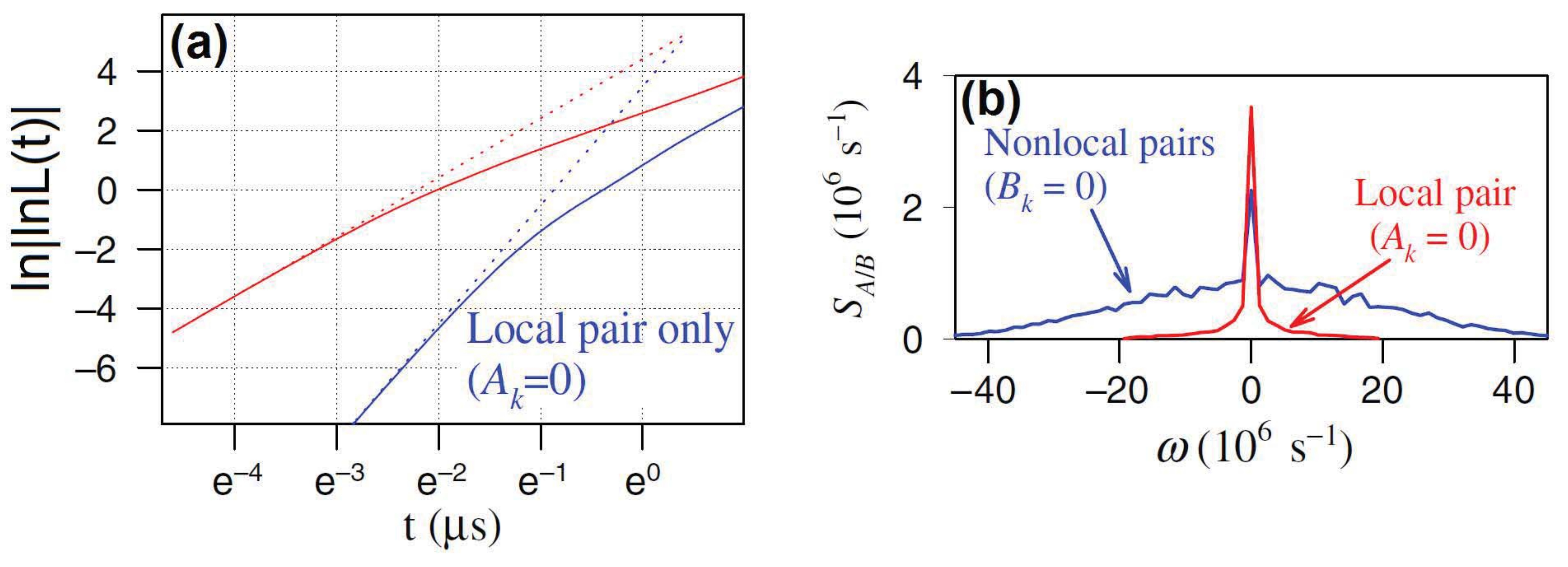}
 \caption{(a)
Non-Markovian-to-Markovian crossover in electron spin decoherence. The dotted
lines are the short-time profile. (b) Excitation spectra for nonlocal and
local nuclear spin pairs. Reproduced from Ref. \cite{LiuNJP2007}.}%
\label{G_MARKOV_CROSSOVER}%
\end{figure}

The pseudo-spins are separated into two groups, corresponding to local pairs
(group $\mathbb{G}_{B}$) with $\mathbf{h}_{k}^{(\pm)}\approx(2B_{k},0,\pm
Z_{k})$ and nonlocal pairs (group $\mathbb{G}_{A}$) with $\mathbf{h}_{k}%
^{(\pm)}\approx\pm(2A_{k},0,Z_{k})$, respectively. Thus the central spin
coherence is factorized as $\left\vert L_{J}(t)\right\vert =\left\vert
L_{A}(t)\right\vert \times\left\vert L_{B}(t)\right\vert $, where $\left\vert
L_{A/B}(t)\right\vert \equiv\prod_{k\in\mathbb{G}_{A/B}}|L_{k}(t)|$. These two
kinds of nuclear spin pairs have qualitatively different contributions to
electron spin decoherence, for both FID and under DD control.

Within the time scale of interest $t\ll1/|B_{k}|,1/|A_{k}|$, the contributions
from local and nonlocal pairs to the FID are
\cite{YaoPRB2006,LiuNJP2007,YaoPRL2007}
\begin{align}
\left\vert L_{A}(t)\right\vert  &  =\prod_{k\in\mathbb{G}_{A}}e^{-2t^{2}%
A_{k}^{2}\operatorname{sinc}^{2}(Z_{k}t)}=e^{-2t\int S_{A}%
(x/t)\operatorname{sinc}^{2}(x)dx},\\
\left\vert L_{B}(t)\right\vert  &  =\prod_{k\in\mathbb{G}_{B}}e^{-(t^{4}%
/2)B_{k}^{2}Z_{k}^{2}\operatorname{sinc}^{4}\frac{Z_{k}t}{2}}=e^{-\frac{1}%
{2}t\int S_{B}(x/t)x^{2}\operatorname{sinc}^{4}\frac{x}{2}dx},
\end{align}
with
\begin{align}
S_{A}(\omega)  &  \equiv\sum_{k\in\mathbb{G}_{A}}\delta(\omega-Z_{k})A_{k}%
^{2},\\
S_{B}(\omega)  &  \equiv\sum_{k\in\mathbb{G}_{B}}\delta(\omega-Z_{k})B_{k}%
^{2},
\end{align}
which are the pseudo-spin excitation spectra [see Fig.
\ref{G_MARKOV_CROSSOVER}(b)]. In the short-time limit ($t\ll1/|Z_{k}|$), the
decoherence caused by non-local pairs is
\begin{equation}
\left\vert L_{k\in\mathbb{G}_{A}}(t)\right\vert \approx e^{-(t/T_{2,A})^{2}},
\label{LGA}%
\end{equation}
which shows Gaussian decay [red lines in Fig. \ref{G_MARKOV_CROSSOVER}(a)] on
a time scale
\begin{equation}
T_{2,A}=\frac{1}{\sqrt{2\sum_{k\in\mathbb{G}_{A}}A_{k}^{2}}}, \label{T2A}%
\end{equation}
while the decoherence caused by local pairs is
\begin{equation}
\left\vert L_{k\in\mathbb{G}_{B}}(t)\right\vert \approx e^{-(t/T_{2,B})^{4}}
\label{LGB}%
\end{equation}
which shows quartic decay [blue lines in Fig. \ref{G_MARKOV_CROSSOVER}(a)] on
a time scale
\begin{equation}
T_{2,B}=\frac{1}{(\sum_{k\in\mathbb{G}_{B}}B_{k}^{2}Z_{k}^{2}/2)^{1/4}}.
\label{T2B}%
\end{equation}
Here the Gaussian decay in Eq. (\ref{LGA}) is actually the expansion of the
power-law decay in Eq. (\ref{LDYN_RDT}) in the short-time limit $t\ll
T_{\mathrm{dyn}}$, and $T_{2,A}$ is equivalent to $T_{\mathrm{dyn}}$, which is
independent of the specific distribution of the HFI coefficients $\{a_{i}\}$
because the differences $a_{i}-a_{j}$ is unimportant due to energy-time
uncertainty in the short-time regime \cite{CoishPRB2010}. On longer time
scales, the sinc function dictates that only pairs with $\omega\in
\lbrack-1/t,1/t]$ contribute significantly, indicative of energy conservation
condition. For sufficiently large $t$ such that $S_{A/B}(\omega)$ can be
regarded as constant $\bar{S}_{A/B}$ within $[-1/t,1/t]$, both $|L_{A}(t)|$
and $|L_{B}(t)|$ show exponential decay [see Fig. \ref{G_MARKOV_CROSSOVER}(a)]
on time scales that depend sensitively on the distribution of the HFI
coefficients (since energy conservation becomes important for long-time
dynamics), in agreement with the ring diagram approximation [see the
discussions after Eq. (\ref{TDYN_RDT})]. The crossover from power-law decay to
exponential decay indicates the crossover from the non-Markovian regime
$(t\ll1/Z_{k})$ to the Markovian regime $(t>1/Z_{k})$. Similar results have
also been derived by a non-Markovian master equation approach
\cite{CoishPRB2010}. For even longer times (which are relevant for a highly
polarized spin bath), the decoherence is determined by the complex structure
of the collective modes of the bath and becomes very sensitive to the
distribution of $\{a_{i}\}$, e.g., exponential decay \cite{CywinskiPRB2009}
and power-law decay \cite{CoishPRB2010,DengPRB2006,DengPRB2008} have been predicted.

Under DD control, the central spin decoherence caused by non-local nuclear
spin pairs are largely suppressed, while the local pairs contributes most to
central spin decoherence \cite{YaoPRB2006,LiuNJP2007}. Under Hahn echo
control, the central spin coherence at the echo time $t=2\tau$ is
\cite{YaoPRB2006,LiuNJP2007,YaoPRL2007}%
\[
L_{\mathrm{H}}(2\tau)\approx\prod_{k\in\mathbb{G}_{B}}e^{-2\tau^{4}Z_{k}%
^{2}B_{k}^{2}\operatorname{sinc}^{4}(Z_{k}\tau/2)}=e^{-2\tau\int S_{B}%
(x/\tau)x^{2}\operatorname{sinc}^{4}(x/2)dx},
\]
which shows quartic decay $e^{-(2\tau/T_{\mathrm{H}})^{4}}$ with coherence
time $T_{\mathrm{H}}=\sqrt{2}T_{2,B}$, which is $\sqrt{2}$ times that of the
FID time. This shows that disturbing the central spin state changes the
bifurcated bath evolution, which in turn changes the central spin decoherence.
At the longer time-scale $1/|Z_{k}|\ll\tau\ll1/|B_{k}|$, the coherence decays
exponentially, indicative of Markovian behavior \cite{YaoPRB2006,LiuNJP2007}.

\subsubsection{Magic coherence recovery}
\label{SEC_MAGIC}

\begin{figure*}[ptb]
\includegraphics[width=1.5\columnwidth]{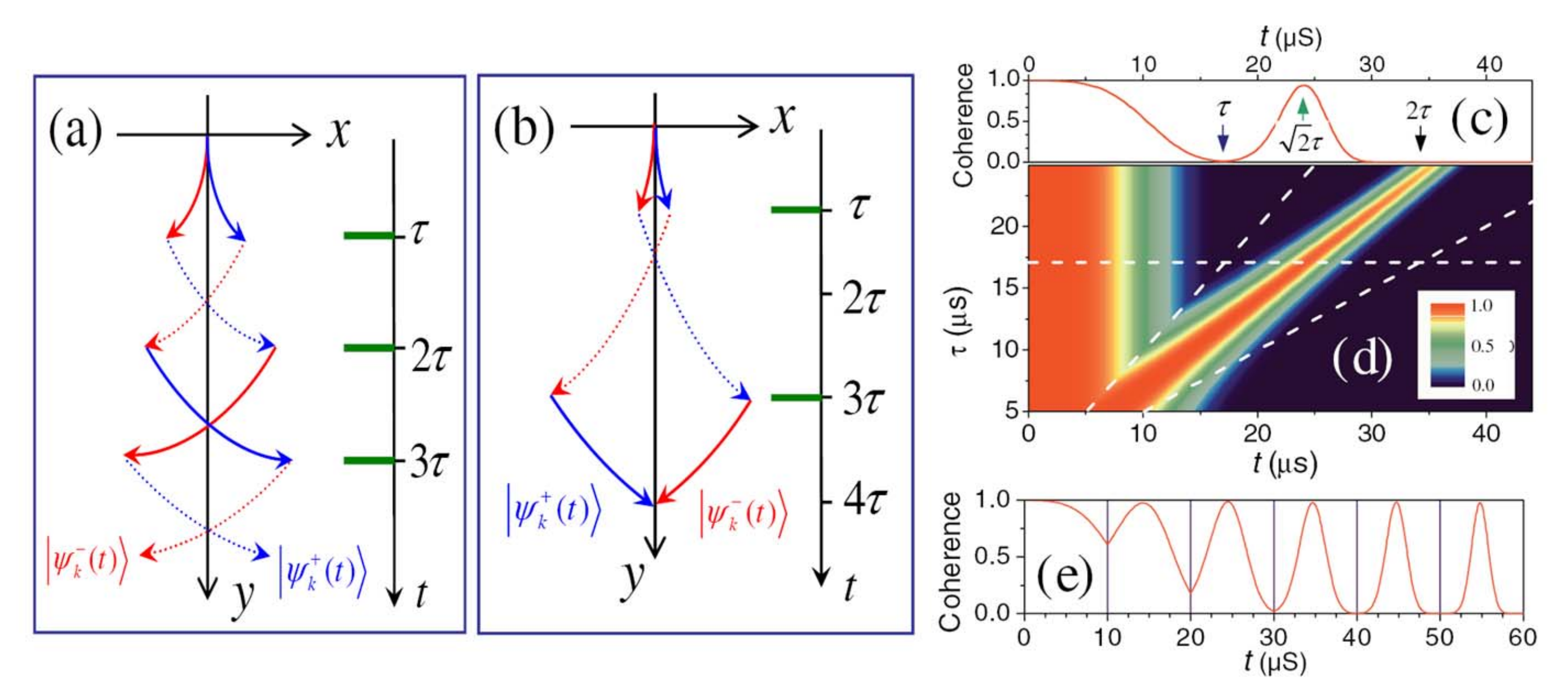}
 \caption{(a)
Bifurcated trajectories of pseudo-spins Bloch vectors under the control of a
sequence of equally spaced $\pi$-pulses. (b) similar to (a) but for the CPMG-2
control. (c) \textquotedblleft True\textquotedblright\ decoherence under Hahn
echo control, with the $\pi$-pulse applied at $\tau=17$ $\mathrm{\mu s}$
(indicated by the blue arrow). (d) Contour plot of \textquotedblleft
true\textquotedblright\ decoherence under Hahn echo vs. evolution time $t$ and
pulse delay time $\tau$. The left tilted dashed line indicates $t=\tau$. The right tilted dashed line indicates the echo time $t=2\tau$. The horizontal line is the cut for the curve
in (c). (e) \textquotedblleft True\textquotedblright\ decoherence under a
sequence of $\pi$-pulses (indicated by purple vertical lines) at intervals of
10 $\mathrm{\mu s}$. Panels (a) and (b) are extracted from Refs.
\cite{LiuNJP2007}, and panels (c)-(e) are extracted from Ref.
\cite{YaoPRL2007}.}%
\label{G_MAGIC_RECOVERY}%
\end{figure*}

The magic recovery of central spin coherence was predicted for an central
electron spin in a nuclear spin bath in the strong field regime (HFI $\ll$
nuclear Zeeman splitting) \cite{YaoPRL2007}. In this case, the noise operator
$\hat{b}=\sum_{j}a_{j}\hat{I}_{j}^{z}$ comes from the isotopic HFI between the
electron spin and the nuclear spins. For nuclear spin-1/2's, the flip-flop $|\uparrow\rangle
_{i}|\downarrow\rangle_{j}\leftrightarrow|\downarrow\rangle_{i}|\uparrow
\rangle_{j}$ of each nuclear spin pair $k\equiv(i,j)$ is mapped to the
precession of the $k$th pseudo-spin $\boldsymbol{\sigma}_{k}$:\ $|\Uparrow
\rangle_{k}\equiv|\uparrow\rangle_{i}|\downarrow\rangle_{j}$ and
$|\Downarrow\rangle_{k}\equiv|\downarrow\rangle_{i}|\uparrow\rangle_{j}$. The
pseudospin field is $\mathbf{h}_{k}^{(\pm)}=\mathbf{h}_{k}^{B}\pm
\mathbf{h}_{k}^{b}/2$ with $\mathbf{h}_{k}^{B}=X_{k}^{B}\mathbf{e}_{x}%
+Z_{k}^{B}\mathbf{e}_{z}$ and $\mathbf{h}_{k}^{b}=Z_{k}^{b}\mathbf{e}_{z}$,
where $X_{k}^{B},Z_{k}^{B}$ are the intrinsic nuclear spin flip-flop amplitude
and energy cost, respectively, due to nuclear dipolar interactions, while
$Z_{k}^{b}$ is the HFI induced correction to the energy cost.

The bifurcated evolution $|J\rangle\rightarrow|J_{\pm}(t)\rangle$ of the
pseudospin starting from a pure state (say $|J\rangle=|\Uparrow\rangle$) can
be mapped to Bloch vectors $\boldsymbol{\sigma}_{\pm}(t)$ and the central spin
decoherence is determined by their distance $d(t)=|\boldsymbol{\sigma}%
_{+}(t)-\boldsymbol{\sigma}_{-}(t)|$ (see the Appendix). When $Z_{k}^{B}=0$,
the fields $\mathbf{h}_{k}^{(\pm)}=(X_{k}^{B},0,\pm Z_{k}^{b})$ leads to
$t^{2}$ increase of $d(t)$ in the short-time limit, while the application of a
$\pi$ pulse at time $\tau$ reverse the evolution direction and gives rise to
coherence recovery at the magic time $t_{\mathrm{mag}}=\sqrt{2}\tau$ instead
of the echo time $t_{\mathrm{d}}=2\tau$ (see the Appendix). Here the presence
of $Z_{k}^{B}$ does not change the $t^{2}$ increase of $d(t)$, so it shows the
same magic coherence recovery [see Fig. \ref{G_MAGIC_RECOVERY}(b) and (c)].
More generally, under an arbitrary DD characterized by the modulation function
$s(t)$, the distance%

\begin{equation}
d^{(1)}(t)\propto\int_{0}^{t}t^{\prime}s(t^{\prime})dt^{\prime} \label{DJ1}%
\end{equation}
vanishes (and hence coherence recovery occurs) at the magic time
$t_{\mathrm{mag}}$ as determined by $\int_{0}^{t_{\mathrm{mag}}}ts(t)dt=0$,
e.g., $t_{\mathrm{mag}}=\sqrt{N(N+1)}\tau$ for the DD consisting of $N$
equally spaced $\pi$-pulses $\tau_{k}=\tau$ [see Fig. \ref{G_MAGIC_RECOVERY}%
(a) and (d)]. By contrast, in the absence of intrinsic bath dynamics
($\mathbf{h}_{k}^{B}=0$), the distance $d(t)\propto t$. Under DD control, the
distance
\begin{equation}
d^{(0)}(t)\propto\int_{0}^{t}s(t^{\prime})dt^{\prime} \label{DJ0}%
\end{equation}
vanishes at the echo time $t_{\mathrm{d}}=2\tau$, corresponding to the
elimination of inhomogenous dephasing at the echo time.

Equations (\ref{DJ1}) and (\ref{DJ0}) are reminiscent of the Taylor expansion
of the classical random phase $\tilde{\varphi}(t_{\mathrm{d}})=\int
_{0}^{t_{\mathrm{d}}}s(t)\tilde{b}(t)dt\approx\sum_{n}\tilde{b}_{n}\int
_{0}^{t_{\mathrm{d}}}s(t)t^{n}dt$ based on $\tilde{b}(t)=\sum_{n}\tilde{b}%
_{n}t^{n}$. For a pure initial state of the bath (no classical analog), the
lowest-order term $d^{(0)}(t)$ is absent, so $d^{(1)}(t)=0$ gives rise to
magic coherence recovery at $t_{\mathrm{mag}}$, suggesting that elimination of
the coupling to the environment is not an necessary condition for the recovery
of coherence. For a thermal initial state of the bath, since the time-averaged
coupling between the central spin and the bath is nonzero at the magic time in
the first order, the rapid inhomogeneous dephasing will prevent magic
coherence recovery from being observed. Direct observation of magic coherence
recovery is possible once the inhomogeneous nuclear spin distribution is
narrowed, e.g., a projective measurement of the noise operator $\hat{b}$ could
be used to limit the nuclear spin configurations by post-selection
\cite{GiedkePRA2006,KlauserPRB2006,StepanenkoPRL2006}.

\subsubsection{Anomalous decoherence effects}

\label{SEC_STRONG_SLOW}

\begin{figure}[ptb]
\includegraphics[width=\columnwidth]{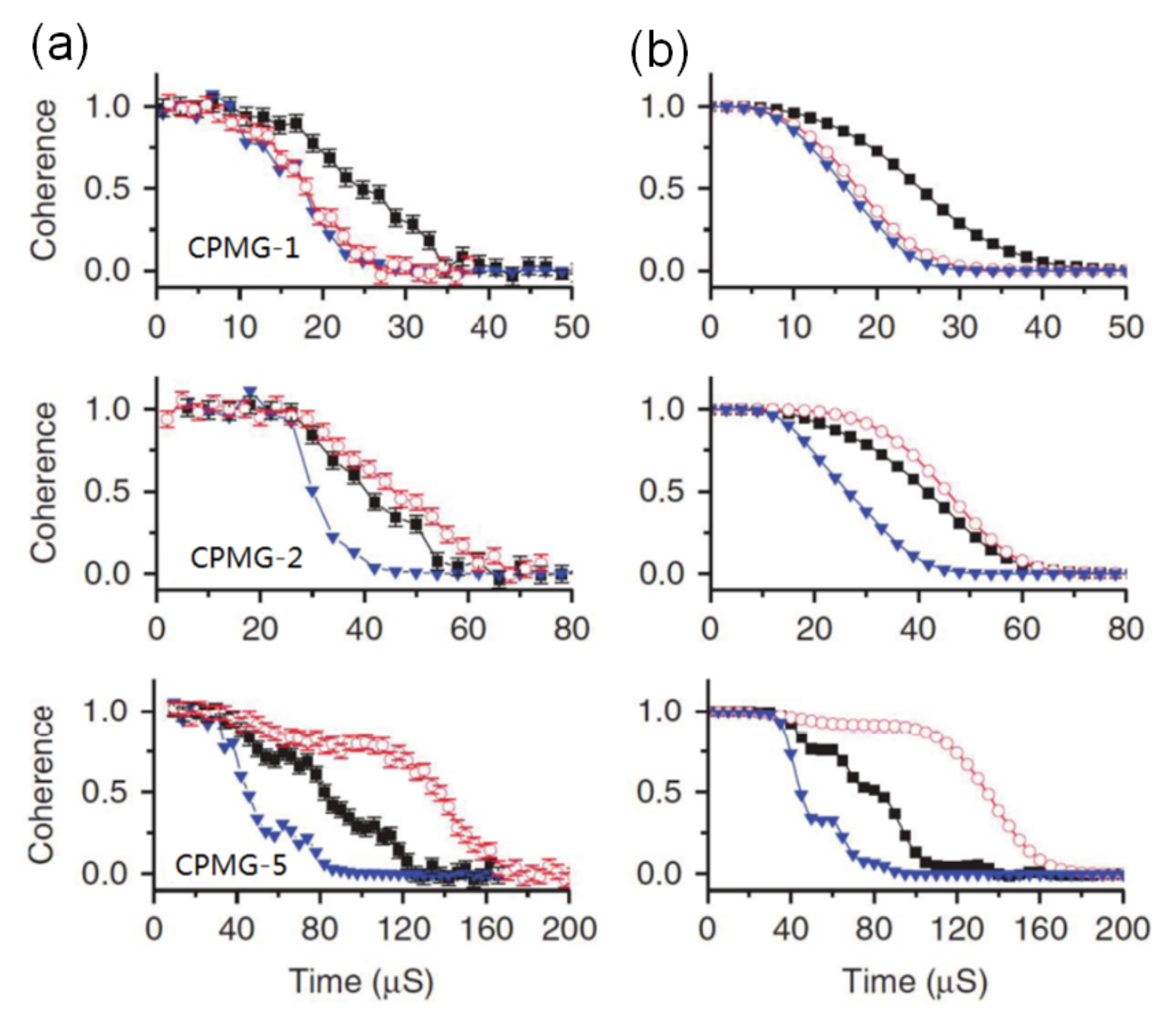}
 \caption{(a)
Measured single (black line with square symbols) and double (red line with
circle symbols) quantum coherence, under the control of different numbers of
equally spaced pulses (CPMG-1, CPMG-2 and CPMG-5, from top to bottom). The
scaled single quantum coherence $|L_{0,+1}|^{4}$ (blue line with triangle
symbols) is also shown for comparison. (b) The theoretical results, plotted in
the same format as in (a). Adapted from Ref. \cite{HuangNatCommun2011}.}%
\label{G_ANOMALOUS_DECOHERENCE}%
\end{figure}

An important feature of classical decoherence theories is that different
processes coupled to the same noise source have similar decoherence behaviors
and stronger noises cause faster decoherence. However, this is not the case in
the quantum picture, since stronger coupling to the environment allows DD
control to strongly manipulate the environmental dynamics to recover the lost
coherence. For example, the spin-1 electronic state of the NV center in
diamond with eigenstates $\{|m\rangle\}$ ($m=0,\pm1$) is subjected to noises
from the $^{13}$C nuclear spin bath. Surprisingly, under DD control, the
double transition $|+1\rangle\leftrightarrow|-1\rangle$ could have longer
coherence time than the single transition $|0\rangle\leftrightarrow
|\pm1\rangle$, even though the noise amplitude for the former is twice that
for the latter \cite{ZhaoPRL2011,HuangNatCommun2011}. This anomalous
decoherence effect can be understood from the manipulation of pseudospin
evolutions via DD control of the central spin.

In the semiclassical noise picture, the nuclear spin bath can be described as
a random fluctuating local field \cite{KuboJPSJ1954,AndersonJPSJ1954}. For
Gaussian noise \cite{CywinskiPRB2008,MaPRB2015,WitzelPRB2014}, the central
spin decoherence for the transition $|m\rangle\leftrightarrow|n\rangle$ is
$L_{m,n}(t)=e^{-(m-n)^{2}\langle\tilde{\varphi}^{2}(t)\rangle/2}$, which obeys
the scaling relation
\begin{equation}
\label{SCALING}|L_{+1,-1}(t)|=|L_{0,\pm1}(t)|^{4},
\end{equation}
We can see that decoherence of double transition $L_{+1,-1}(t)$ decays in the
same way as that of single transitions $L_{0,\pm1}(t)$, but is faster. The
scaling relation in Eq. (\ref{SCALING}) remains valid when the electron spin
is subjected to arbitrary DD control. However, numerical calculations in the
quantum picture shows that under DD control with more and more $\pi$ pulses,
the classical scaling relation in Eq. (\ref{SCALING}) is violated more and more
significantly, and finally the double quantum coherence even decays slower
than the single quantum coherence [see Fig.\ \ref{G_ANOMALOUS_DECOHERENCE}(b)].

This counterintuitive effect can be understood by analyzing the microscopic
nuclear spin bath evolution $\hat{U}_{m}(t)\equiv e^{-i\hat{H}_{m}t}$
conditioned on the central spin state, where $\hat{H}_{m}\equiv\hat{H}%
_{B}+m\hat{b}$. Under a moderate magnetic field $(\gtrsim0.1$ T) along the N-V
symmetry axis ($z$ axis), the flip of individual nuclear spins is suppressed
by the large nuclear Zeeman splitting, so the elementary excitation of the
nuclear spins are the flip-flop of nuclear spin pairs, which can be mapped to
the precession of non-interacting pseudo-spins with the effective Hamiltonian
conditioned on the electron spin state,%
\[
\hat{H}_{m}^{\mathrm{eff}}=\sum_{k}\mathbf{h}_{k}^{(m)}\cdot\hat
{\boldsymbol{\sigma}}_{k}=\sum_{k}(\mathbf{h}_{k}^{B}+m\mathbf{h}_{k}%
^{b})\cdot\hat{\boldsymbol{\sigma}}_{k}.
\]
where $\mathbf{h}_{k}^{b}=Z_{k}\mathbf{e}_{z}$ comes from the HFI
($Z_{k}\equiv{_{k}}\langle \Uparrow|\hat{b}|\Uparrow\rangle_{k}-{_{k}}%
\langle\Downarrow|\hat{b}|\Downarrow\rangle_{k}$) and $\mathbf{h}_{k}%
^{B}=X_{k}\mathbf{e}_{x}$ is from the nuclear dipolar interaction
($X_{k}\equiv2{_{k}}\langle \Uparrow|\hat{H}_{B}|\Downarrow\rangle_{k}$), so
the coupling to the central spin dominates the bath dynamics ($h_{k}^{b}\gg
h_{k}^{B}$). According to Eq. (\ref{LH_SINGLESPIN}), the Hahn echo of electron
spin coherence for the transition $|m\rangle\leftrightarrow|n\rangle$ is%
\[
L_{m,n}^{\mathrm{H}}(2\tau)=\prod_{k}\left[  1-2\sin^{2}\Theta_{k}^{m,n}%
\sin^{2}\frac{h_{k}^{(m)}\tau}{2}\sin^{2}\frac{h_{k}^{(n)}\tau}{2}\right]  ,
\]
where $\Theta_{k}^{m,n}$ is the angle between $\mathbf{h}_{k}^{(m)}$ and
$\mathbf{h}_{k}^{(n)}$. The decay in the short-time limit is
\[
L_{m,n}^{\mathrm{H}}(2\tau)\approx\prod_{k}e^{-|\mathbf{h}_{k}^{(m)}%
\times\mathbf{h}_{k}^{(n)}|^{2}\tau^{4}/8}=\prod_{k}e^{-(m-n)^{2}%
|\mathbf{h}_{k}^{B}\times\mathbf{h}_{k}^{b}|^{2}\tau^{4}/8},
\]
which obeys the classical scaling Eq. (\ref{SCALING}). At a longer time-scale,
however, the strong coupling to the central spin makes the two fields of
$L_{+1,-1}^{\mathrm{H}}(2\tau)$ nearly antiparallel $(\sin\Theta_{k}%
^{+1,-1}\ll1)$, while that of the single quantum coherence are nearly
perpendicular $(\sin\Theta_{k}^{0,\pm1}\approx1)$. Consequently, the long-time
decay of the double quantum coherence is much smaller than that of the single
quantum coherence, thus violating Eq. (\ref{SCALING}) at longer times.
Application of more $\pi$ pulses prolongs the electron spin coherence time and
makes this long-time behavior more pronounced.

This anomalous decoherence has been experimentally observed by Huang
\textit{et al.} in type-IIa diamond at room temperature
\cite{HuangNatCommun2011}. In the experimental setup, the magnetic field is
weak, so the electron spin decoherence is mainly caused by the single $^{13}$C
nuclear spin dynamicis, quite similar to the pseudospin dynamics dicussed above.

\subsection{Multi-spin correlation effects}

\begin{figure}[ptb]
\includegraphics[width=\columnwidth,clip]{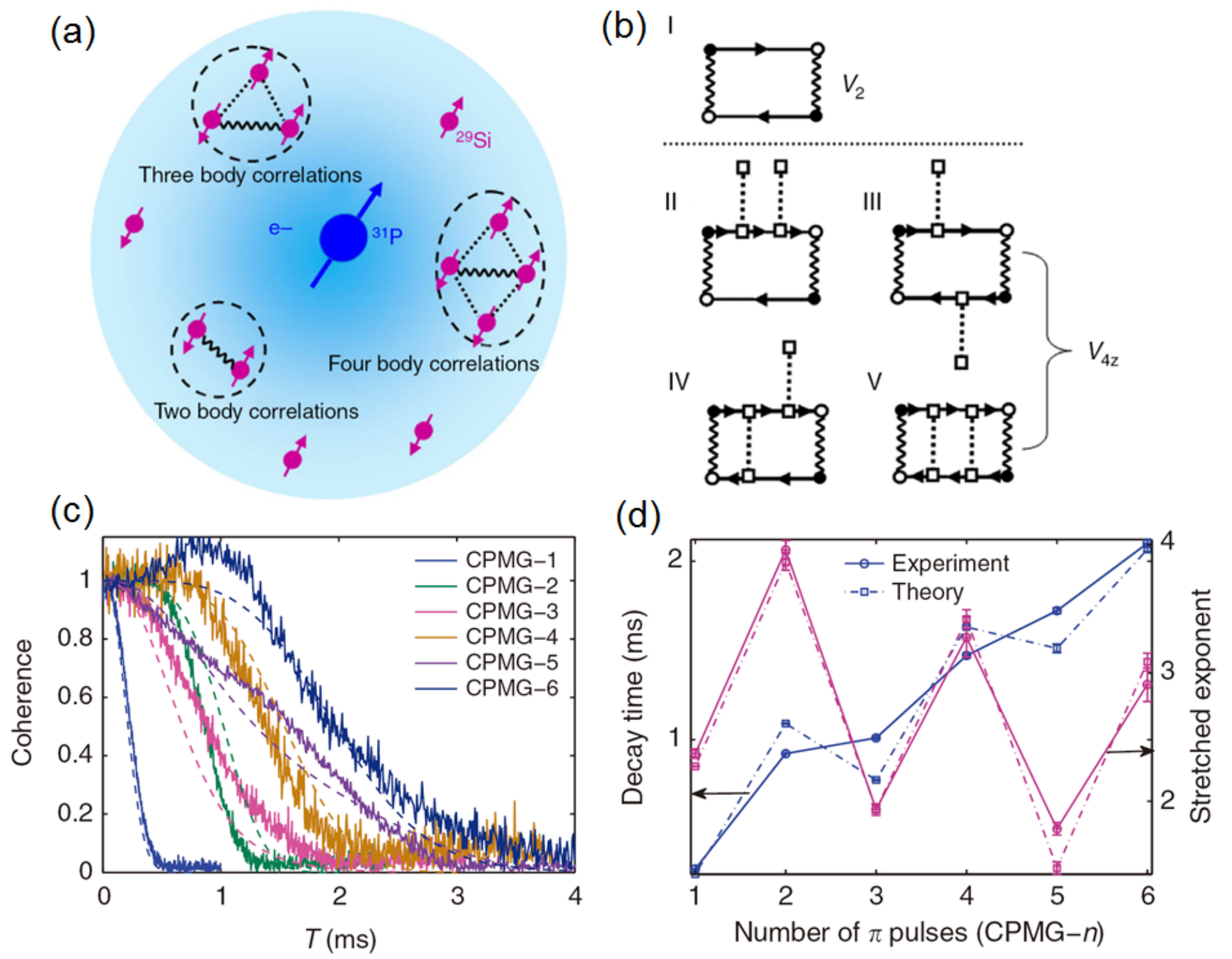}
 \caption{Revealing
many-body correlations in nuclear spin baths by central spin decoherence. (a)
Electron spin of a phosphorous donor in silicon interacts with a bath of
$^{29}$Si nuclear spin-1/2's possessing various many-body processes. (b)
topologically inequivalent connected Feynman diagrams corresponding to
different many-body processes in the nuclear spin bath: (I) $V_{2}$ -
second-order pairwise flip-flop, (II-V) $V_{4z}$ - fourth-order pairwise
flip-flop dressed by diagonal interactions. (c) Measured (solid lines) and
calculated (dashed lines) coherence of the P donor electron spin in the
natural $^{29}$Si nuclear spin bath under CPMG control. (d) Comparisons of the
experimental (solid lines) and theoretical (dashed line) decay times
$T_{\mathrm{SD}}$ (blue) and stretched exponents $n$ (magenta) of the central
spin decoherence under the CPMG control in (c). Extracted from Ref.
\cite{MaNatCommun2014}.}%
\label{G_LCE_NC}%
\end{figure}The effects of multi-spin correlations on central spin decoherence
become pronounced when the coherence time is prolonged to be comparable or
longer than the inverse of typical nuclear-nuclear interaction, which can be
realized by applying multi-pulse DD control
\cite{ZhaoPRB2012,WitzelPRL2007,MaNatCommun2014} or tuning the external
magnetic field near some optimal working points
\cite{MaPRB2015,BalianPRB2014,BalianPRB2015} (also called ``clock'' transitions
\cite{WolfowiczNatNano2013} where the central spin is insensitive to the
magnetic noise in the first order). The CCE method can explicitly show the
contributions of different multi-spin clusters in the nuclear spin bath to
central spin decoherence, providing an intuitive tool to identify the
underlying nuclear spin processes. For NV centers in diamond and donor spins
in silicon, the CCE-2 calculations (truncated up to the clusters with two
nuclear spins) always give converged results for Hahn echo of spin coherence
\cite{ZhaoPRB2012,BalianPRB2014}, indicating the pairwise flip-flop processes
dominate the central spin decoherence. For central spin decoherence under
multi-pulse DD control \cite{ZhaoPRB2012} or near the optimal working points
in silicon \cite{MaPRB2015,BalianPRB2015} , the CCE-6 calculations (truncated
up to the clusters with six nuclear spins) are always needed to give converged
results, indicating that the multi-spin correlations contributes significantly
to central spin decoherence.

More interestingly, recent studies show that DD control of the central spin
can selectively suppress or amplify certain many-body processes in the nuclear
spin bath \cite{MaNatCommun2014}. In this case, LCE provides a systematic and
transparent way to visualize the gradual development of different many-body
processes in a nanoscale spin bath, by analysing the individual influence of
each LCE diagram on the central spin decoherence \cite{SaikinPRB2007}. For
example, consider a central electron spin in a relatively large nuclear spin
bath with a strong external magnetic field and the HFI between the central
spin and bath spins much larger than the nuclear-nuclear interactions, such as
shallow donors in silicon (e.g. Si:P and Si:Bi) and electron spin in
semiconductors (e.g. GaAs and InAs quantum dots). For CPMG-$N$ (or UDD-$N$)
control of the central spin with odd $N$, the second-order pairwise flip-flop
diagram ($V_{2}$ term in Fig. \ref{G_LCE_NC}(b)) dominates the central spin decoherence and
almost fully reproduces the exact decoherence calculated from CCE, while for
CPMG-$N$ control with even $N$, the effects of the second-order pairwise
flip-flop diagram are cancelled and the fourth-order flip-flop diagrams ($V_{4z}$ terms in
Fig. \ref{G_LCE_NC}(b)), corresponding to renormalized pairwise flip-flop
dressed by the diagonal interactions (or pairwise flip-flop processes of two
nuclear spins renormalized the dipolar diagonal interaction with the other
nuclear spins in the bath), dominates the decoherence. This even-odd effect
indicates that the second-order flip-flop [$V_{2}$ term Fig. \ref{G_LCE_NC}%
(b)] and fourth-order flip-flop processes [$V_{4z}$ term in Fig.
\ref{G_LCE_NC}(b)] can be selectively detected by applying an appropriate
number of DD pulses, as has been theoretically predicted and experimentally
observed recently in Si:P system \cite{MaNatCommun2014}. Actually, a similar
even-odd effect has been noticed before in cluster expansion calculations
\cite{WitzelPRL2007} (without analyzing the underlying microscopic processes):
in the presence of an even (odd) number of DD pulses, the decoherence scale as
$\ln L=O(\lambda^{4})$ [$\ln L=O(\lambda^{2})$] with respect to the dipolar
interaction strength $\lambda$ between bath spins. In the experiment
\cite{MaNatCommun2014}, the measured decoherence $e^{-(t/T_{\mathrm{SD}})^{n}%
}$ caused by $^{29}$Si nuclei has a stretching factor $n$ oscillating between
about $2$ (for odd $N$) and $4$ (for even $N$), as shown in Figs.
\ref{G_LCE_NC}(c) and (d), indicating the detection of either the second-order
flip-flop processes or fourth-order flip-flop processes. The different
signatures of the many-body processes in the bath under DD control of the
central spin, in particular the even-odd effect in the number of DD control
pulses, provide a useful approach to studying many-body physics in the nuclear
spin bath.

\section{Summary and outlook}

\label{SEC_OUTLOOK} Central electron spin decoherence in nanoscale nuclear
spin baths is a critical issue for quantum technologies. In recent years,
quantum pictures and quantum many-body theories have been established and have
provided a quantitative description and unprecedented understanding for the
central spin decoherence under many experimental conditions (such as DD
control and moderate to strong magnetic fields). Accompanying the great
progresses in prolonging the central spin coherence time through various DD
schemes, the coherent evolution of the central spin in turn serves as an
ultrasensitive probe for weak signals
{{\cite{ChernobrodJAP2005,DegenAPL2008,TaylorNatPhys2008,HallPRL2009,HallPNAS2010}
}}and many-body dynamics in the environments
\cite{QuanPRL2006,ChenNJP2013,WeiPRL2012,PengPRL2015,WeiSciRep2014,WeiSciRep2015}
with nanoscale resolution.

When the semi-classical noise model and especially the noise filter
description are valid \cite{MaPRB2015}, central spin decoherence under DD
control has been used to reconstruct the environmental noise spectra
\cite{AlvarezPRL2011,BylanderNatPhys2011,BarGillNatCommun2012,CywinskiPRA2014}%
, which in turn can be used to design optimal quantum control for protecting
the quantum coherence and quantum gates \cite{WitzelPRB2014}. In particular,
the decay of the central spin coherence on very long time scales (up to
seconds \cite{MaPRB2015}) allows studying the low-energy excitations in the
environment, since as the evolution time $t$ increases, the noises that cause
significant central spin decoherence have frequencies $\sim1/t$.

Central spin decoherence under DD control has also been widely used for
quantum sensing of single nuclear spins
\cite{ColeNanotechnol2009,ZhaoNatNano2011}. When the period of the DD control
matches the transition frequencies of the target nuclear spin(s)
\cite{ZhaoNatNano2011}, the noises from the target nuclear spins are
resonantly amplified, causing enhanced central spin decoherence (manifested as
a sharp coherence dip when sweeping the DD period). Several groups have
adopted the DD scheme to successfully detect single $^{13}$C nuclear spins
\cite{ZhaoNatNano2012,KolkowitzPRL2012,TaminiauPRL2012} and $^{13}$C clusters
\cite{ShiNatPhys2014} in diamond. Shallow NV centers near the surface have
also been used to sense the NMR of single protein molecules
\cite{ShiScience2015} and nano-scale NMR of nuclear species
\cite{MaminScience2013,StaudacherScience2013} on diamond surfaces. Recently,
there are also new proposals and concepts for quantum sensing, such as using
multiple NV spins as the quantum sensor \cite{MaPRApp2016}, distinguishing
nuclear spins of different species by sweeping the DD pulse number
\cite{MaarXiv2015a}, and design of multi-dimensional DD to distinguish the
nuclear spin correlations in single molecules \cite{MaarXiv2015c,BossPRL2016}.

Another promising direction is to employ the central spin decoherence to
reveal the many-body physics and thermodynamic properties of the environment,
since in some cases the central spin decoherence caused by the environment is
directly related to the partition function of the environment. It has been
found that central spin coherence shows sharp decay when the environment is
tuned near a quantum critical point \cite{QuanPRL2006,ChenNJP2013}. For a
central spin homogenously coupled to a ferromagnetic Ising model, the central
spin coherence vanishes at times corresponding to the Lee-Yang zeros of the
partition function of the Ising model \cite{WeiPRL2012,PengPRL2015}. Moreover,
central spin decoherence has extended the phase transitions in the environment
to the complex plane of physical parameters \cite{WeiSciRep2014} and enabled
thermodynamic holograph of the partition function of the environment
\cite{WeiSciRep2015}.

\section{Acknowledgements}

We acknowledge the support by Hong Kong RGC, CUHK Vice Chancellor's One-off
Discretionary Fund, the NSFC (Grant No. 11274036 and No. 11322542), the
MOST (Grant No. 2014CB848700), NSFC program for 'Scientific Research Center' (Program No. U1530401), and the computational support from the Beijing Computational Science Research Center (CSRC).

\appendix

\section{Bloch vector representation of single spin dynamics}

\label{APP Bloch}

\begin{figure}[ptb]
\includegraphics[width=\columnwidth]{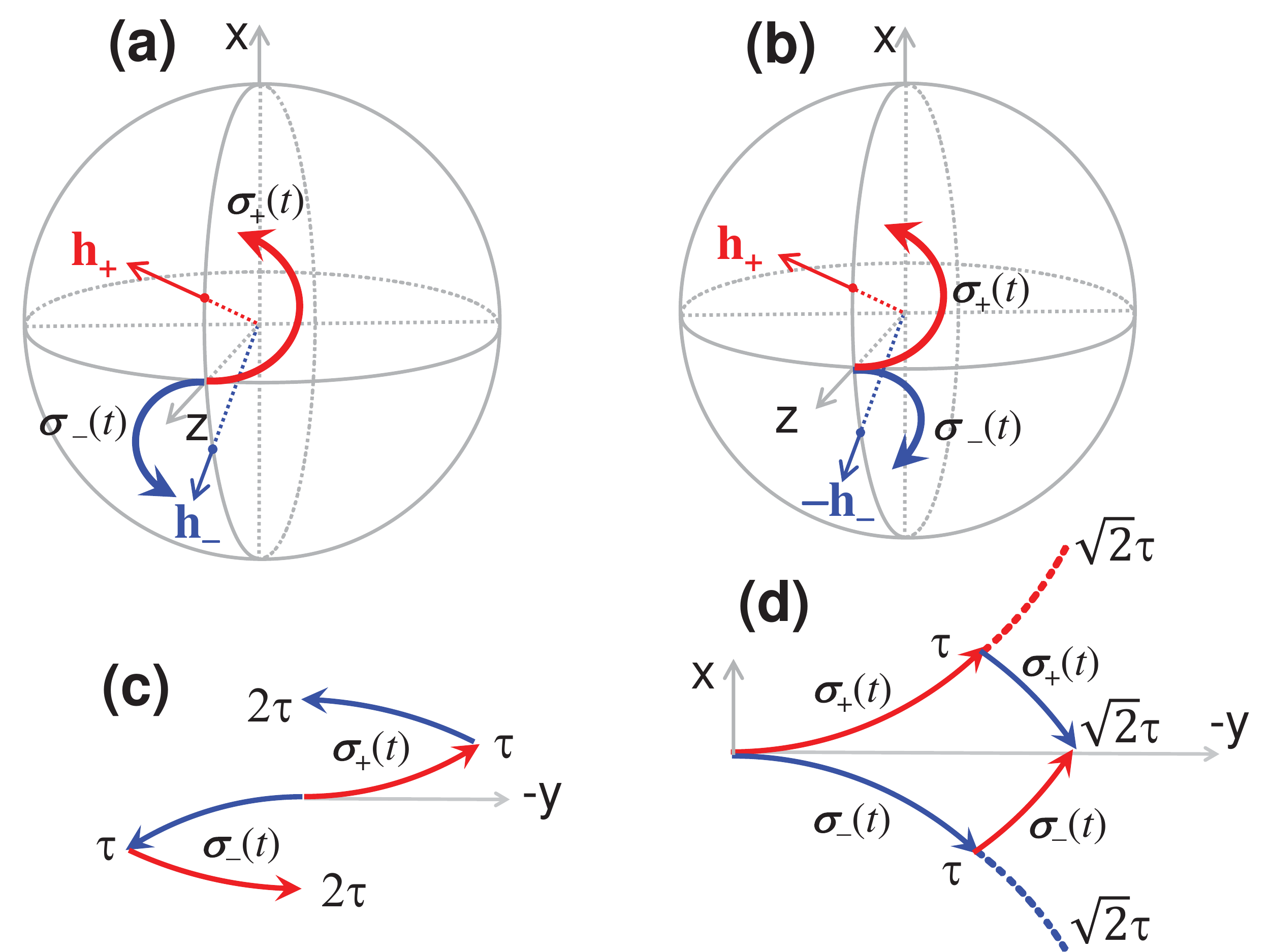}
\caption{Larmor
precession of the Bloch vectors $\boldsymbol{\sigma}_{\pm}(t)$ of the two bath
pathways around (a) $\mathbf{h}_{\pm}=(\pm X,0,Z)$ and (b)\ $\mathbf{h}_{\pm
}=(X,0,\pm Z)$. (c) and (d) are the corresponding projection of the Bloch
vectors in the $xoy$ plane.}%
\label{G_PSEUDOSPIN}%
\end{figure}

We consider that the bath consists of a single spin-1/2, which starts from a
pure spin-up state $|J\rangle=|\Uparrow\rangle$ along the $z$ axis and
bifurcates into two pathways $|J_{\pm}(t)\rangle=e^{-i\hat{H}_{\pm}t}%
|J\rangle$ with
\[
\hat{H}_{\pm}=\hat{H}_{B}\pm\hat{b}/2=(\mathbf{h}_{B}\pm\frac{\mathbf{h}_{b}%
}{2})\cdot\frac{\boldsymbol{\hat{\sigma}}}{2}.
\]
The pathways can be mapped to the Bloch vectors $\boldsymbol{\sigma}_{\pm
}(t)\equiv\langle J_{\pm}(t)|\hat{\boldsymbol{\sigma}}|J_{\pm}(t)\rangle$,
which start from $\mathbf{e}_{z}$ at $t=0$ and then undergo Larmor precession
around the fields $\mathbf{h}_{\pm}$ on a unit sphere. The central spin
decoherence $|L(t)|^{2}=1-d^{2}(t)/4$ is determined by the distance
$d(t)=|\boldsymbol{\sigma}_{+}(t)-\boldsymbol{\sigma}_{-}(t)|$ between the
Bloch vectors \cite{YaoPRB2006,LiuNJP2007}. To visualize the bifurcated bath
evolution, we consider two special cases: (A) $\mathbf{h}_{\pm}=(\pm X,0,Z)$
and$\ $(B) $\mathbf{h}_{\pm}=(X,0,\pm Z)$. In either case, the bath spin
precesses with angular frequency $h=\sqrt{X^{2}+Z^{2}}$ on a circle of radius
$\sin\theta=X/h$.

For case (A), the FID%
\[
L(t)=1-2\sin^{2}\theta\sin^{2}\frac{ht}{2}\overset{t\ll1/h}{\longrightarrow
}e^{-X^{2}t^{2}/2}%
\]
shows Gaussian decay in the short-time limit, corresponding to a linear
increase of the distance $d(t)\approx2Xt$ with time [Fig. \ref{G_PSEUDOSPIN}%
(a) and (c)]. At $t=\pi/h$, the distance is maximal $d_{\max}=2\sin(2\theta)$
and the coherence is minimal: $L_{\min}=\cos(2\theta)$. Under Hahn echo
control, the distance in the short-time limit is $d_{\mathrm{H}}%
(t)\approx2X(\tau-(t-\tau))$, so central spin decoherence is minimized at the
echo time
\begin{equation}
t_{\mathrm{d}}=2\tau. \label{TD_ECHO}%
\end{equation}

For case (B), the FID \cite{YaoPRB2006,LiuNJP2007}%
\begin{align*}
L(t)  &  =1-2\cos^{2}\theta\sin^{2}\frac{ht}{2}-i\cos\theta\sin(ht)\\
&  \overset{t\ll1/h}{\longrightarrow}e^{-iZt(1-X^{2}t^{2}/6)-Z^{2}X^{2}%
t^{4}/8}%
\end{align*}
exhibits $t^{4}$ decay in the short-time limit, corresponding to quadratic
increase of the distance $d(t)\approx XZt^{2}$ with time [Fig.
\ref{G_PSEUDOSPIN}(b) and (d)]. At $t=\pi/h$, the distance is maximal
$d_{\max}=2\sin(2\theta)$ and the coherence is minimal: $L_{\min}=\cos
(2\theta)$. Under Hahn echo control, the distance in the short-time limit is
$d_{\mathrm{H}}(t)\approx\lbrack(\tau^{2}-(t^{2}-\tau^{2})]XZ$, so central
spin decoherence is minimized at the \textit{magic} time:
\begin{equation}
t_{\mathrm{mag}}=\sqrt{2}\tau. \label{T_MAGIC}%
\end{equation}

The different coherence recovery times [Eqs. (\ref{TD_ECHO}) and
(\ref{T_MAGIC})] follow from the different time dependences of the Bloch
vector distances: $d(t)\propto t$ for case (A) and $d(t)\propto t^{2}$ for
case (B). For case (A) [Fig. \ref{G_PSEUDOSPIN}(c)], the Bloch vectors
$\boldsymbol{\sigma}_{\pm}(t)$ move in opposite directions with almost
constant velocity $X$. After the $\pi$ pulse, both Bloch vectors reverse their
velocities, so minimal distance occurs at $2\tau$. For case (B) [Fig.
\ref{G_PSEUDOSPIN}(d)], both Bloch vectors move away from the $-y$ axis
quadratically with time, e.g., the distance of each Bloch vector from the $-y$
axis reaches $\tau^{2}$ at $t=\tau$. If there were no $\pi$ pulses at $\tau$,
then evolution from $\tau$ to $\sqrt{2}\tau$ would double the distance to
$2\tau^{2}$. Now the $\pi$ pulse reverse the evolution direction of both Bloch
vectors, so $\boldsymbol{\sigma}_{\pm}(t)$ both return to the $-y$ axis at
$\sqrt{2}\tau$. Such coherence recovery at a \textquotedblleft
magic\textquotedblright\ time (i.e., different from the echo time) was first
predicted in Ref. \cite{YaoPRL2007,LiuNJP2007} and will be discussed in more
detail in Sec. \ref{SEC_MAGIC}.

%

\end{document}